\documentclass[aps,preprint,groupedaddress,nofootinbib,showpacs,floatfix]{revtex4}
\usepackage{amsmath}
\usepackage{graphicx}
\usepackage{amssymb}
\usepackage{hyperref}
\usepackage{color}

\newcommand{\beq}{\begin{equation}}
\newcommand{\eeq}{\end{equation}}
\newcommand{\bea}{\begin{eqnarray}}
\newcommand{\eea}{\end{eqnarray}}

\newcommand{\ttb}{t\bar t}
\newcommand{\RNum}[1]{\uppercase\expandafter{\romannumeral #1\relax}}

\newcommand{\gsim}{\lower.7ex\hbox{$\;\stackrel{\textstyle>}{\sim}\;$}}
\newcommand{\lsim}{\lower.7ex\hbox{$\;\stackrel{\textstyle<}{\sim}\;$}}

\begin{document}
	
\preprint{
{\vbox {
\hbox{\bf MSUHEP-19-014}
}}}
\vspace*{0.2cm}

\title{Updating and Optimizing Error PDFs in the Hessian Approach. Part II}

\author{Tie-Jiun Hou$^{(a,1)}$, Zhite Yu$^{(b,2)}$, Sayipjamal Dulat$^{(c,3)}$, Carl Schmidt$^{(b,4)}$, C.-P. Yuan$^{(b,5)}$}
	
\email{$^{1}$tjhou@msu.edu, $^{2}$yuzhite@msu.edu, $^{3}$sdulat@hotmail.com, $^{4}$schmidt@pa.msu.edu  $^{5}$yuan@pa.msu.edu}

\affiliation{$^{(a)}$Department of Physics, College of Sciences, Northeastern University, Shenyang 110819, China \\
	$^{(b)}$Department of Physics and Astronomy, Michigan State University, East Lansing, MI 48824 U.S.A. \\
	$^{(c)}$School of Physics Science and Technology, Xinjiang University, Urumqi, Xinjiang 830046 China
}

\date{\today}

\begin{abstract}

In an earlier publication, we introduced the software package, {\tt \texttt{ePump}} (error PDF Updating Method Package), that can be used to update or optimize a set of parton distribution functions (PDFs), including the best-fit PDF set and Hessian eigenvector pairs of PDF sets (i.e., error PDFs), and to update any other set of observables, in the Hessian approach. Here, we validate the {\tt \texttt{ePump}} program with a detailed comparison against a full global analysis, and we demonstrate the potential of  {\tt \texttt{ePump}}  by presenting selected phenomenological applications relevant to the Large Hadron Collider. For example, we use the package to estimate the impact of the recent LHC data of the measurements of $W$, $Z$ boson and top quark pair differential distributions on the CT14HERA2 PDFs.

\end{abstract}

\maketitle

\tableofcontents

\section{Introduction}

An understanding of uncertainties due to parton distribution functions (PDFs) is crucial to precision studies of the standard model, as well as to searches for new physics beyond the standard model at hadron colliders, such as the CERN Large Hadron Collider (LHC).
As extensively discussed in Ref.~\cite{Schmidt:2018hvu}, a technique for estimating the impact of new data on the PDFs, without performing a full global analysis, is extremely useful.
(See also Refs.~\cite{Paukkunen:2014zia,Paukkunen:2013grz,Camarda:2015zba}.)
For this purpose, we have developed a software package, {\tt \texttt{ePump}} (error PDF Updating Method Package), which can be used to obtain both the updated best-fit PDF and updated eigenvector PDFs from an earlier global analysis.
The package can also directly update the predictions for experimental observables and their PDF uncertainties without requiring the use of the updated PDFs to re-calculate the theory predictions.
Finally, an alternative use of the package is to optimize a given set of Hessian PDFs for a particular set of observables, so that a reduced number of error PDFs can be used, while maintaining the PDF uncertainty on the observables to any desired precision.

In Ref.~\cite{Schmidt:2018hvu} some examples were given comparing the results of {\tt \texttt{ePump}} with a full global analysis, as well as several phenomenological analyses using {\tt \texttt{ePump}}. In addition, an exercise using \texttt{ePump} was performed in Ref.~\cite{Willis:2018yln} to show how to assess the potential of precision measurement of triple differential distributions of high-mass (up to sub-TeV) Drell-Yan pairs to reduce the PDF induced errors in predicting the cross section of an extra $Z^\prime$ boson with mass greater than a few TeVs produced at the LHC. In this work we provide further checks and more details of the validation of {\tt \texttt{ePump}} against the full global analysis machinery, and we provide more examples of using {\tt \texttt{ePump}} to update current PDFs with new LHC data.

In a global analysis of experimental data, the PDFs are defined as a function of a number of fitting parameters, and in turn, the global $\chi^2$ and the theoretical prediction for any
observable are also functions of the parameters.  The crucial approximations used by \texttt{ePump} are these:  1) The global $\chi^2$ is a quadratic function of the parameters around its global minimum.  2) All other relevant quantities (including theoretical predictions of new observables used in the update, as well as the PDFs themselves) are linear functions of the parameters.
It is these simplifying assumptions that allow \texttt{ePump} to obtain updated best-fit PDFs and error PDFs, and to update the predictions and uncertainties for any other observable, in
just a few seconds of CPU time.  Note that these approximations are the exact same as those used to calculate the PDF uncertainty for any observable in the Hessian method.
However, the impact of these approximations must still be considered when interpreting the results from \texttt{ePump}.  In addition, subtleties in the calculations of PDF uncertainties, such
as the use of dynamical tolerances and Tier-2 penalties, could potentially induce further discrepancies between the predictions of \texttt{ePump} versus a full global analysis.  Thus, it is
useful to validate \texttt{ePump} against a full global analysis in as many distinct applications as possible.

The paper is structured as follows. In Sec.~\ref{section:CT14}, we perform the aforementioned validation of \texttt{ePump}.
To do this we start with a base best-fit and error PDF set, obtained from a global analysis using the CT14HERA2 parametrization.  The data included is the CT14HERA2 data sets minus some subset of the data.  We then use \texttt{ePump} to update the PDFs by adding back the excluded data sets, and we compare with the standard CT14HERA2 PDFs.
If the Hessian approximations were exact, we should find that the \texttt{ePump} predictions reproduce exactly the CT14HERA2 PDFs.
Thus, we can test how well the approximations work for different classes of data sets.
In Ref.~\cite{Schmidt:2018hvu} results were shown for this exercise where the jet data was included by \texttt{ePump}.  In this paper we present more details of this check with jet data, and also present additional checks with Deeply-Inelastic Scattering (DIS) and Drell-Yan data.  In each of these cases, we will see that the updated PDFs obtained from \texttt{ePump} are very close to the global-fit results, i.e., CT14HERA2 PDFs in this case.
Furthermore, in Sec.~\ref{subsection:mJ}, we show, as an example, how to use \texttt{ePump} to directly update the theoretical predictions of the Higgs boson production cross
section $\sigma(gg\rightarrow h)$ from gluon fusion  in proton-proton collider, including its uncertainties induced by the updated error PDFs.

The speed of \texttt{ePump} makes it very useful to perform analyses to investigate the influence of multiple data sets on the PDFs that otherwise might require many different time-consuming global fittings.
In Sec.~\ref{section:}, we demonstrate how to use \texttt{ePump} to quickly identify the experimental data sets that constrain the CT14HERA2 PDFs most stringently. We find that among all of the 33 data sets included in the CT14HERA2 fits, less than half of them are necessary to effectively constrain the CT14HERA2 PDF errors.
Detailed information on the impact of those individual data sets to constrain the CT14HERA2 PDFs, such as which parton flavors and at which $x$ values, will also be discussed.

Of course, one of the main uses for a tool such as \texttt{ePump} is to quickly assess the impact of new data sets prior to updating with a full global analysis.
In Sec.~\ref{section:new data} we provide two detailed examples of this by using \texttt{ePump} to update the CT14HERA2 PDFs with some recent LHC data.
First, we examine the impact from the LHC top quark pair ($\ttb$) production data provided by ATLAS and CMS Collaborations. Second, we examine the impact from the ATLAS 7 TeV data on $W$ and $Z$ productions~\cite{Aad:2011dm}.
We find that while the $\ttb$ data can provide potential constraints on the $g$-PDF, its impact is quite minimal after we have included the inclusive high transverse momentum ($p_T$) jet production data from the Tevatron and the LHC in the same fit.
On the other hand, we find a large impact on the quark PDFs, particularly in the small-$x$ region, when updated by adding
the ATLAS 7 TeV $W$ and $Z$ data~\cite{Aad:2011dm}.
This large deviation of the updated PDFs from the original CT14HERA2 PDFs suggests that the \texttt{ePump} result should only be trusted qualitatively in this case, and for quantitative results with this data set a full global fit is required.
This conclusion is further supported by examining the magnitude of the two measures $\tilde{d}^0$ and $d^0$, introduced in  Ref.~\cite{Schmidt:2018hvu}, which give the distance between the original and updated PDFs in the parameter space, relative to the updated and original errors, respectively. For the ATLAS 7 TeV $W$ and $Z$ data the value of $\tilde{d}^0=1.49$ indicates that the original best-fit PDF was far outside the error band for the updated PDFs, so that the new best fit obtained by \texttt{ePump} is more likely to be affected by nonlinearities in the dependence of the observables and the PDFs on the fitting parameters.
This, in turn, could produce results that differ from the true global fit.

Finally, concluding remarks are given in Sec.~\ref{section:summary}.	

\section{Validation of \texttt{ePump} using data sets in CT14HERA2}\label{section:CT14}

The data sets used for PDF global fitting in CT14HERA2~\cite{Hou:2016nqm} consist of the HERA Run I+II combined data~\cite{Abramowicz:2015mha}, 15 other sets of DIS data, 14 sets of Drell-Yan data, and 4 sets of jet production data, as listed in Table I and II of Ref.~\cite{Dulat:2015mca}.
Here, we will take the CT14HERA2 PDFs, including the best-fit and error sets, as the full global fit result to compare against the results of \texttt{ePump}.
We shall see how well \texttt{ePump} reproduces the best-fit PDFs and uncertainty of CT14HERA2 for different classes of experimental data.
The test goes as follows.
First, we perform a full global analysis with the CT14HERA2 parametrization, using all of the CT14HERA2 data except for a particular subset of data.
For instance, when we perform the global analysis with the jet data excluded, we obtain a new set of best-fit and error PDFs, called CT14HERA2mJ.
We then use \texttt{ePump} to update CT14HERA2mJ by treating the excluded jet data as ``new'' data, with the updated PDFs called CT14mJeAll.
A comparison between the CT14mJeAll and the CT14HERA2 best-fit and error PDFs can then be used to show how well \texttt{ePump} reproduces the full global analysis for this subset of data.
Note that since \texttt{ePump} depends on quadratic and linear approximations, we should not expect perfect agreement.
In addition, in the \texttt{ePump} prediction, there are assumptions in how the Tier-2 penalties from the new data affect the updated error PDFs, and therefore the uncertainties in the updated PDFs.
However, as we shall see, the updated best-fit PDFs and their uncertainties from \texttt{ePump} are pretty close to those from the full global analysis.

We shall perform this analysis three times by removing different subsets of data from CT14HERA2:  1) excluding all of the DIS data except the HERA I+II combined data (CT14HERA2mD),
2) excluding all Drell-Yan data (CT14HERA2mY), and 3) excluding all Jet data (CT14HERA2mJ).
We then add the excluded data back with \texttt{ePump} and compare the updated PDFs with the CT14HERA2 PDFs.
To be precise, for the CT14HERA2 parametrization there are 27 parameters, corresponding to 54 error PDFs.
In addition, two gluon extreme sets ({\it i.e.}, eigen-PDF sets 55 and 56) were introduced via the Lagrangian Multiplier method in the CT14HERA2 fit to enlarge the uncertainty in the $g$-PDF in the small-$x$ region.
For our  CT14HERA2mD, CT14HERA2mY, and CT14HERA2mJ fits, we did not produce these extra gluon extreme sets.
Thus, everywhere in this section we shall exclude the two gluon extreme sets also in the CT14HERA2 PDF errors, in order to have a truer comparison.
For convenience we summarize our notations in this paper here:
\begin{itemize}
\item CT14HERA2mD, CT14HERA2mY, and CT14HERA2mJ are the base sets as described above, to be used by \texttt{ePump}.
\item  The letter ``e" followed by a data set name indicates that the PDFs are obtained from \texttt{ePump} by adding the
given data set as ``new" data to the base set.  For example, in Sec.~\ref{section:impact Jet} the PDFs CT14mJeCDF are obtained from \texttt{ePump}
by adding the CDF inclusive jet data to the base set CT14HERA2mJ.
\item  The letters ``eAll'' indicate that PDFs are obtained from \texttt{ePump} by adding back all of data that was excluded in the base set as ``new" data.
Thus, these sets are the \texttt{ePump} approximation to be compared with the full CT14HERA2 PDF set.
\item  The suffixes ``.54" or ``.52" (as for example, in CT14HERA2.54) are used to indicate that the error bands are obtained with 54 or 52 eigen-PDFs, respectively,
rather than with the full 56 eigen-PDFs.
\end{itemize}
Finally, we note that we always show symmetric error bands in this paper.  As described in Ref.~\cite{Schmidt:2018hvu}, the symmetric Hessian error bands are invariant under a change of the eigen-PDF basis (unlike the asymmetric errors) and therefore are more reliable when assessing the impact of new data on the PDF errors when using \texttt{ePump}.

\subsection{CT14HERA2 excluding all DIS data (except HERA Run I+II combined): CT14HERA2mD} \label{subsection:mD}

\begin{figure}[h]
	\includegraphics[width=0.45\textwidth]{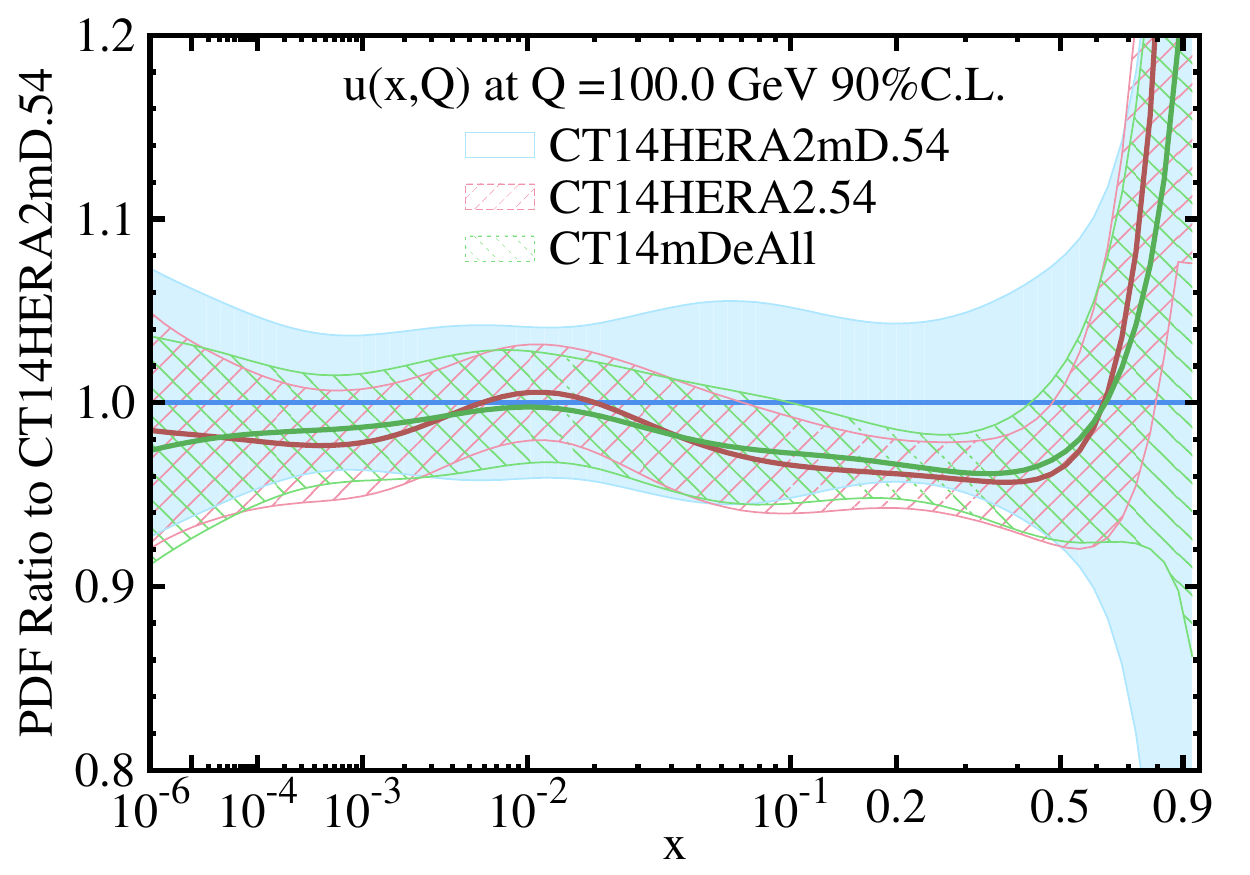}
	\includegraphics[width=0.45\textwidth]{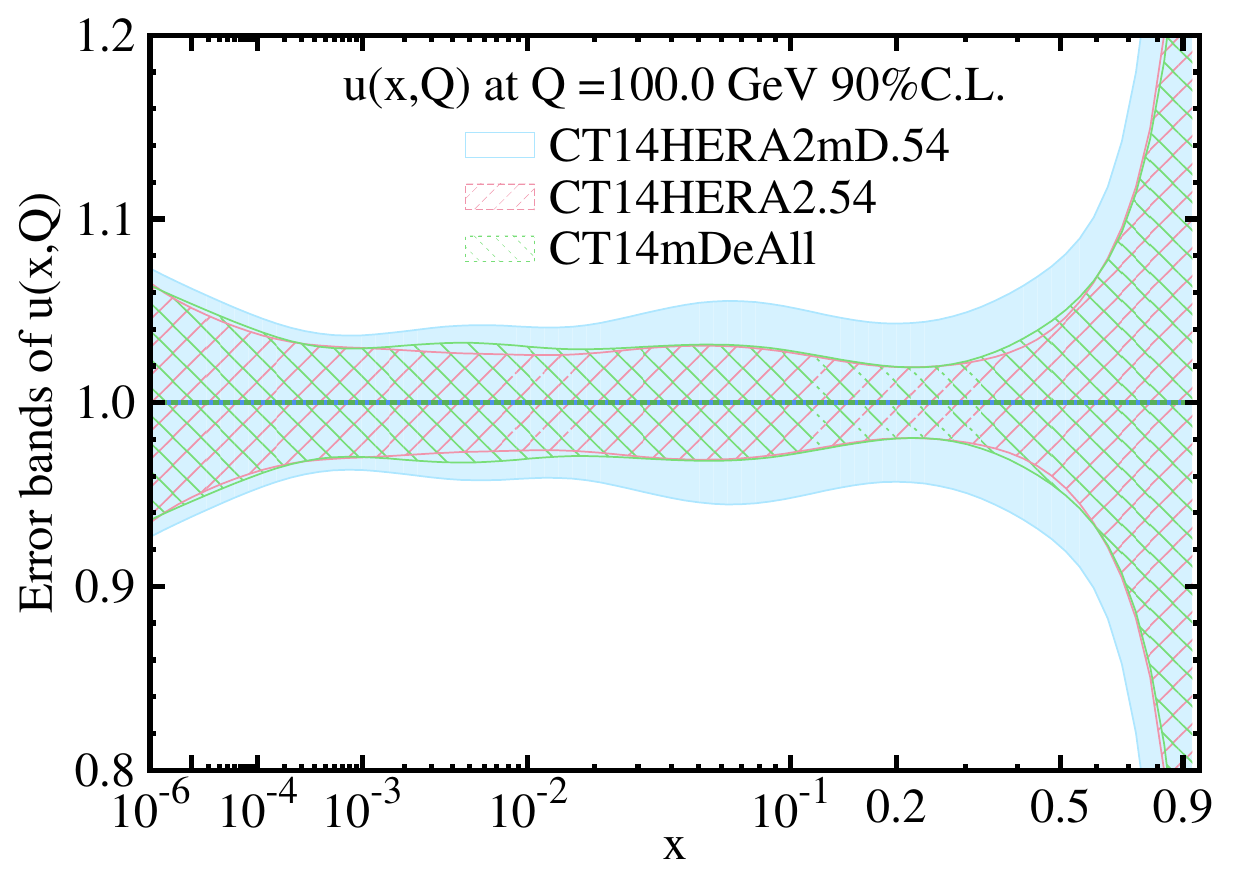}\\
	\includegraphics[width=0.45\textwidth]{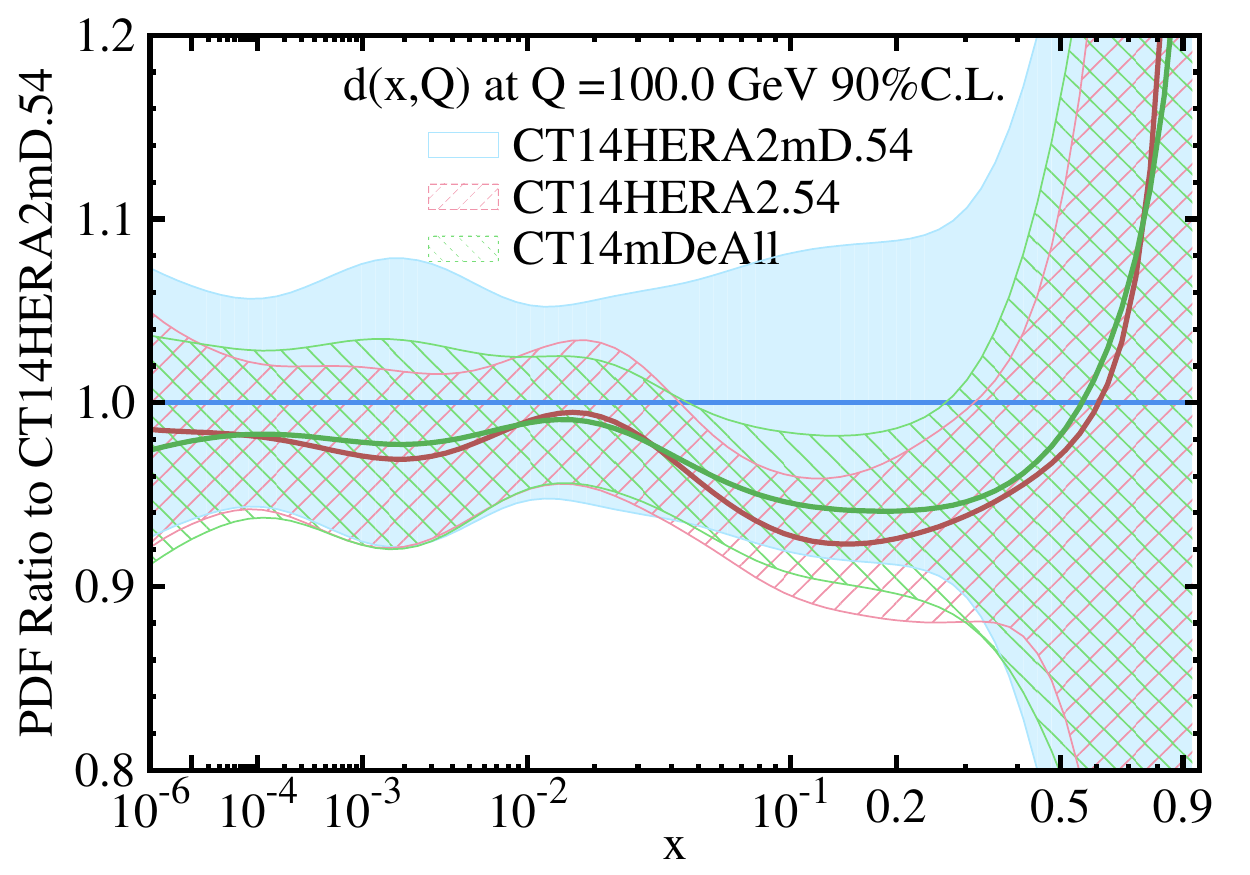}
	\includegraphics[width=0.45\textwidth]{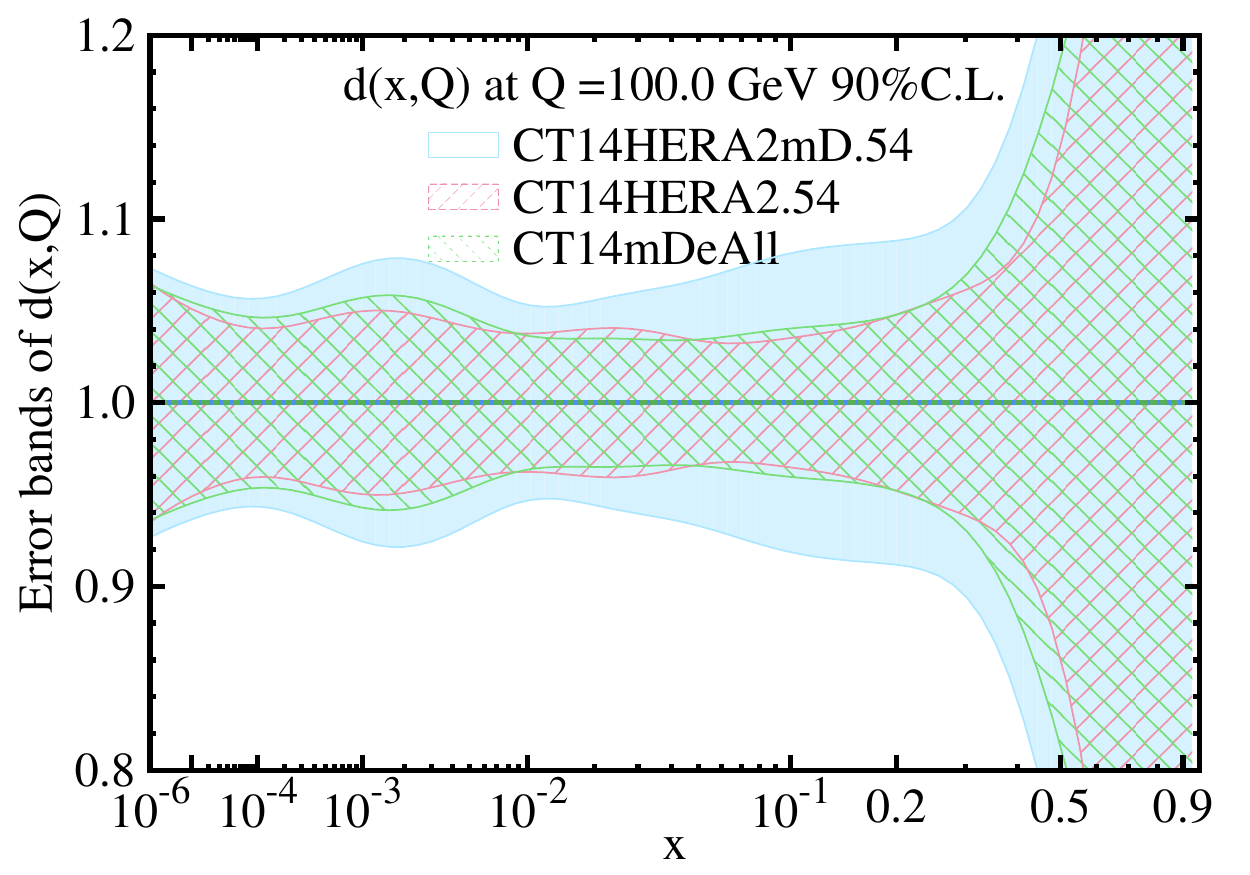}
	\caption{
		Comparison of \texttt{ePump}-updated $u$ and $d$ PDFs, at $Q=100~\rm{GeV}$.  Left panel: the PDF ratios over the best-fit of the base CT14HERA2mD. Right panel: the error bands relative to their own best-fit.}
	\label{Fig:Dud}
\end{figure}

There are 3287 data points in total in the CT14HERA2 fit~\cite{Dulat:2015mca,Hou:2016nqm}.
Among these, the DIS experiments contribute 2381 data points, of which 1120 data points are from the precision HERA Run I+II combined neutral current and charged current data.
If we remove all of the DIS data from the CT14HERA2 fit, this only leaves 906 data points for the reduced (non-DIS) global fit, with the 2381 DIS data points to be added in by \texttt{ePump} as ``new'' data.
In this instance, we may not expect \texttt{ePump} to reproduce the full CT14HERA2 fit results well, since there are more data points (2381) in the ``new'' data than in the ``old'' data (906).
As a consequence, in many regions of PDF parameter space the ``new'' DIS data constrain the PDFs much more than the ``old'' non-DIS data.  As discussed in Ref.~\cite{Schmidt:2018hvu}, if the updated fit moves too far from the original one in the parameter space, the linear and quadratic approximations used in \texttt{ePump} may break down.
Furthermore, it is difficult to obtain a well-converged global fit if all of the DIS data is removed, because too many fitting parameters are left unconstrained.
If one insists on removing all of the DIS data (including HERA Run I+II), some of the 27 (without extreme sets) or 28 (with extreme sets) parameters must be fixed before doing the fit. Then one can get a set of global-fit PDFs with fewer parameters than CT14HERA2, and can still use \texttt{ePump} to update them. However, since some parameters that should be constrained by DIS data are already fixed, the update obtained by adding the DIS data using \texttt{ePump} will not fully reflect the impact of the new data,  and the comparison with CT14HERA2 becomes less meaningful.

Therefore, in the present analysis we choose to keep the HERA Run I+II combined data in the original base fit, allowing us to use the full 27 free parameters.
These data provide important information on the decomposition of parton flavors inside the proton, and therefore provide sufficient constraints on the PDFs for a reasonable base set to update with \texttt{ePump}.
Our base fit, CT14HERA2mD, is then obtained from a global fit to a total of 2026 data points, which include all non-DIS data and the HERA I+II combined data, and exclude all other DIS data.
The remaining DIS data contains 1261 data points, which will be taken as ``new'' data to update the CT14HERA2mD PDFs by \texttt{ePump}.  The updated PDFs, named CT14mDeAll, can then be compared to the CT14HERA2 PDFs.

In Fig.~\ref{Fig:Dud}, we compare the \texttt{ePump}-updated PDFs (CT14mDeAll) to the base PDFs (CT14HERA2mD) and the true global-fit PDFs (CT14HERA2).
It can be seen that the update with \texttt{ePump} yields very similar results as the true global fit. Given the quadratic and linear approximations in \texttt{ePump}, and the number of ``new'' data points (1261) compared to the number of ``old" data points (2026), the results are extremely satisfactory.
Moreover, these well-approximated updated PDFs were calculated in just a few seconds of CPU time.
So, prior to a full global fit, one can quickly obtain a first look at the impact of the new data using \texttt{ePump}. As is expected, the DIS data provide important information on the
$u$ and $d$ PDFs, whose error bands have shrunk by almost one half with the inclusion of the non-HERA DIS data. It is also evident that the $u$ quark PDF is constrained more than the $d$ quark PDF.    This is easily understood by the fact that the electric charge of the $u$ quark is twice that of the $d$ quark, and so it contributes more to the cross section in low energy DIS neutral current processes.

\subsection{CT14HERA2 excluding all Drell-Yan data: CT14HERA2mY} \label{subsec:mY}

\begin{figure}[h]
	\includegraphics[width=0.45\textwidth]{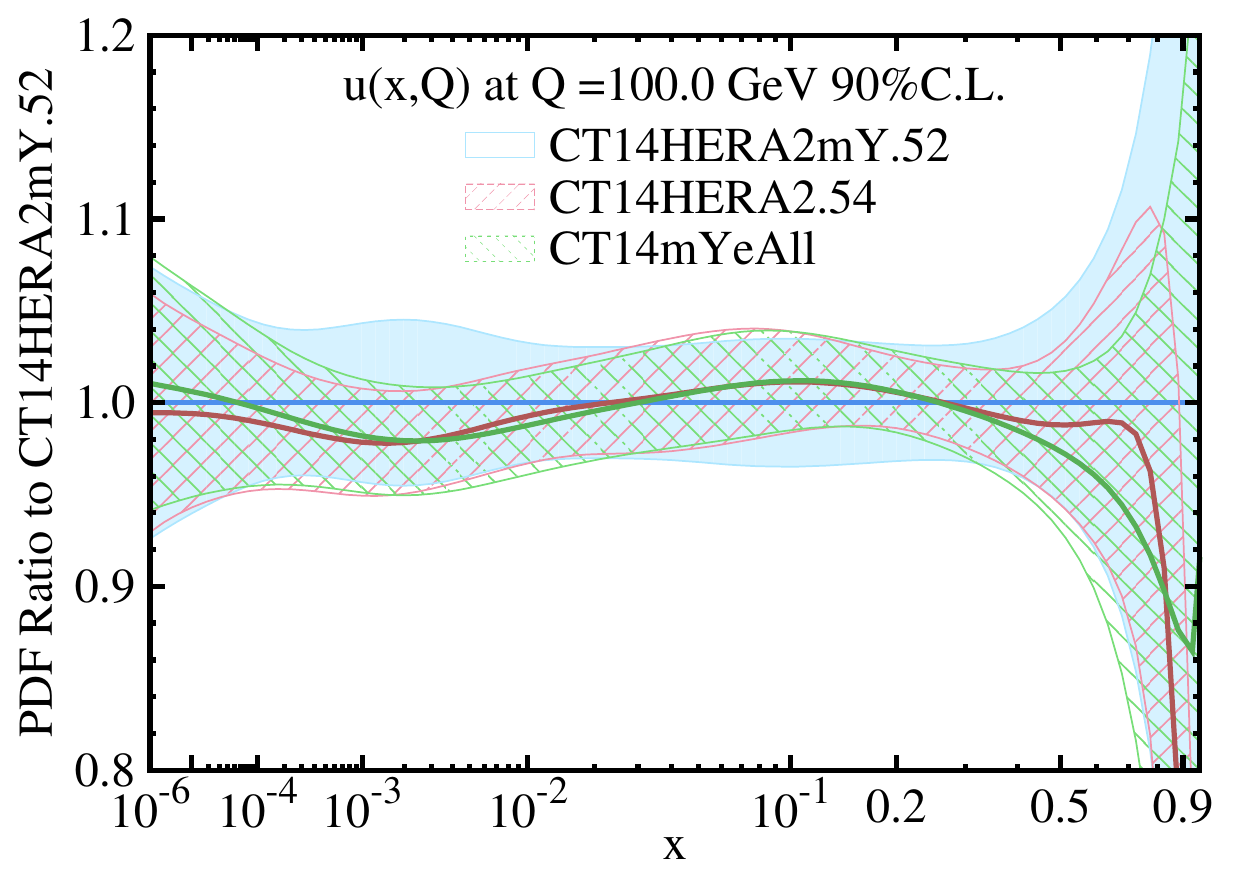}
	\includegraphics[width=0.45\textwidth]{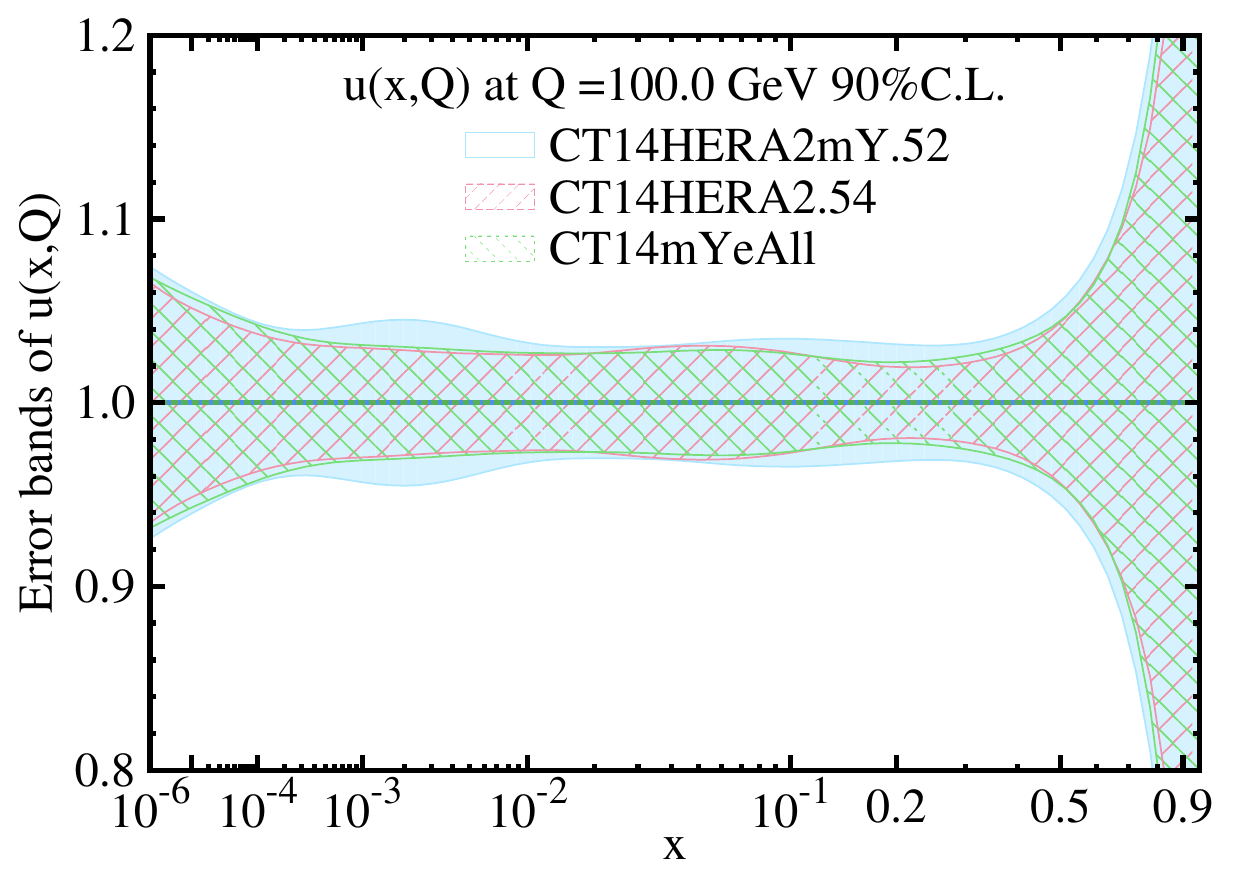}\\
	\includegraphics[width=0.45\textwidth]{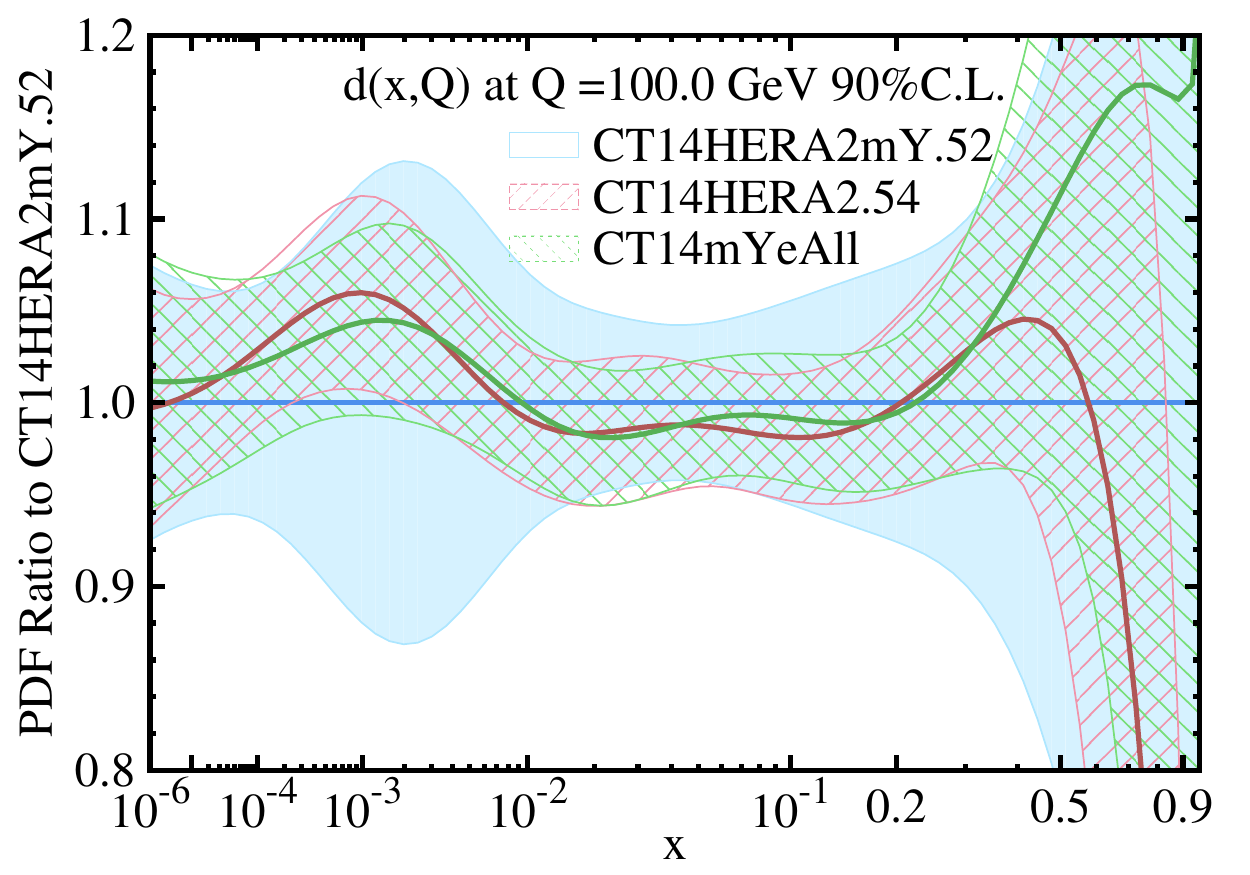}
	\includegraphics[width=0.45\textwidth]{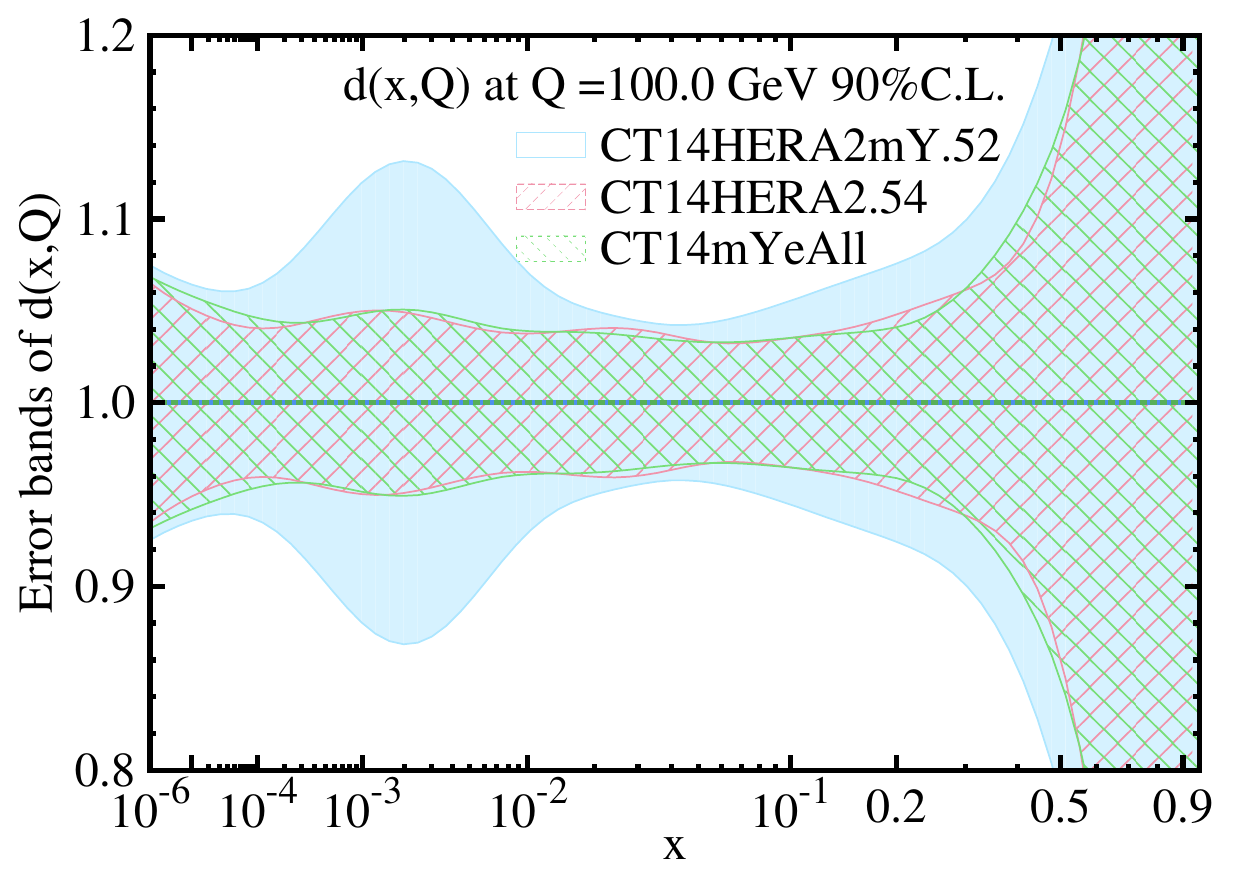}
	\caption{Comparison of \texttt{ePump} updated $u$ and $d$ PDFs, at $Q=100~\rm{GeV}$. Left panel: the PDF ratios over the best-fit of the base CT14HERA2mY. Right panel: the error bands relative to their own best-fit.
	}
	\label{Fig:DYud}
\end{figure}

Here we perform a similar study for the Drell-Yan data in CT14HERA2. The global fit to all the DIS and jet data, and excluding the Drell-Yan data, is named CT14HERA2mY, which contains 2786 data points.
It is worth mentioning that the global fit for CT14HERA2mY with 27 parameters are not very well converged, for the same reason explained in Sec.~\ref{subsection:mD}. Fortunately, we only need to fix one parameter to get a good fit. Thus we are left with enough (26) free parameters to test \texttt{ePump} and the results are still meaningful, as we will see in the following.
The \texttt{ePump}-updated PDFs, obtained by adding back all the Drell-Yan data, which contains 501 data points, to the CT14HERA2mY fit, are shown in Fig.~\ref{Fig:DYud}, together with CT14HERA2 (after removing the two extreme $g$-PDF sets).
Again, we see that \texttt{ePump} yields a result very similar to the true global fit CT14HERA2, with only a small difference for $x$ less than about 0.4, which is negligible compared to the size of the error band.
In the large $x$ region, the PDFs are small and there is little experimental data to constrain them, so they are determined by analytic extrapolation and depend strongly on the non-perturbative parameterization forms assumed at the PDF initial scale (which is 1.3 GeV in the CT14HERA2 fit).  Therefore, we are not concerned by
the differences in the best-fit PDFs at $x$ greater than about 0.4, which are nevertheless still well within the error bands.

\begin{figure}[h]
	\includegraphics[width=0.45\textwidth]{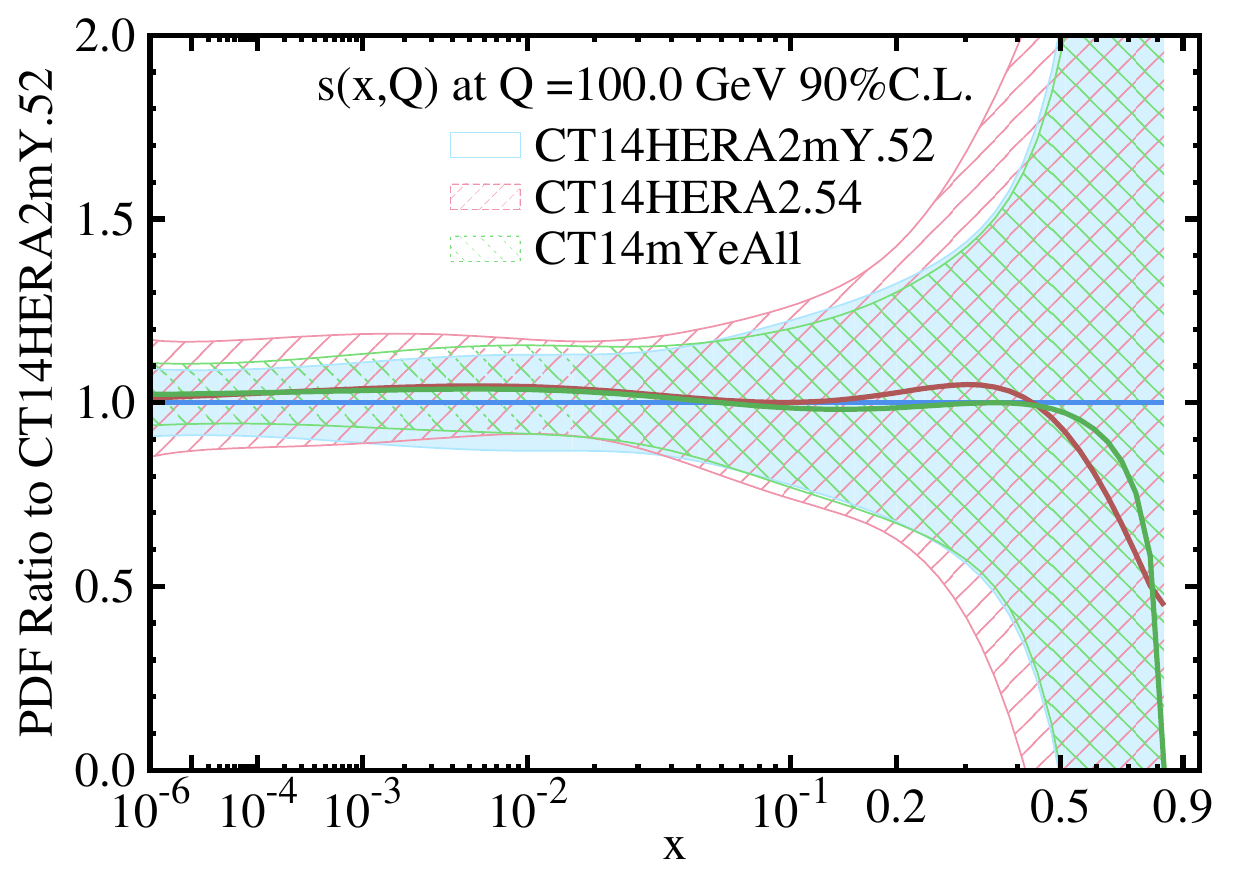}
	\includegraphics[width=0.45\textwidth]{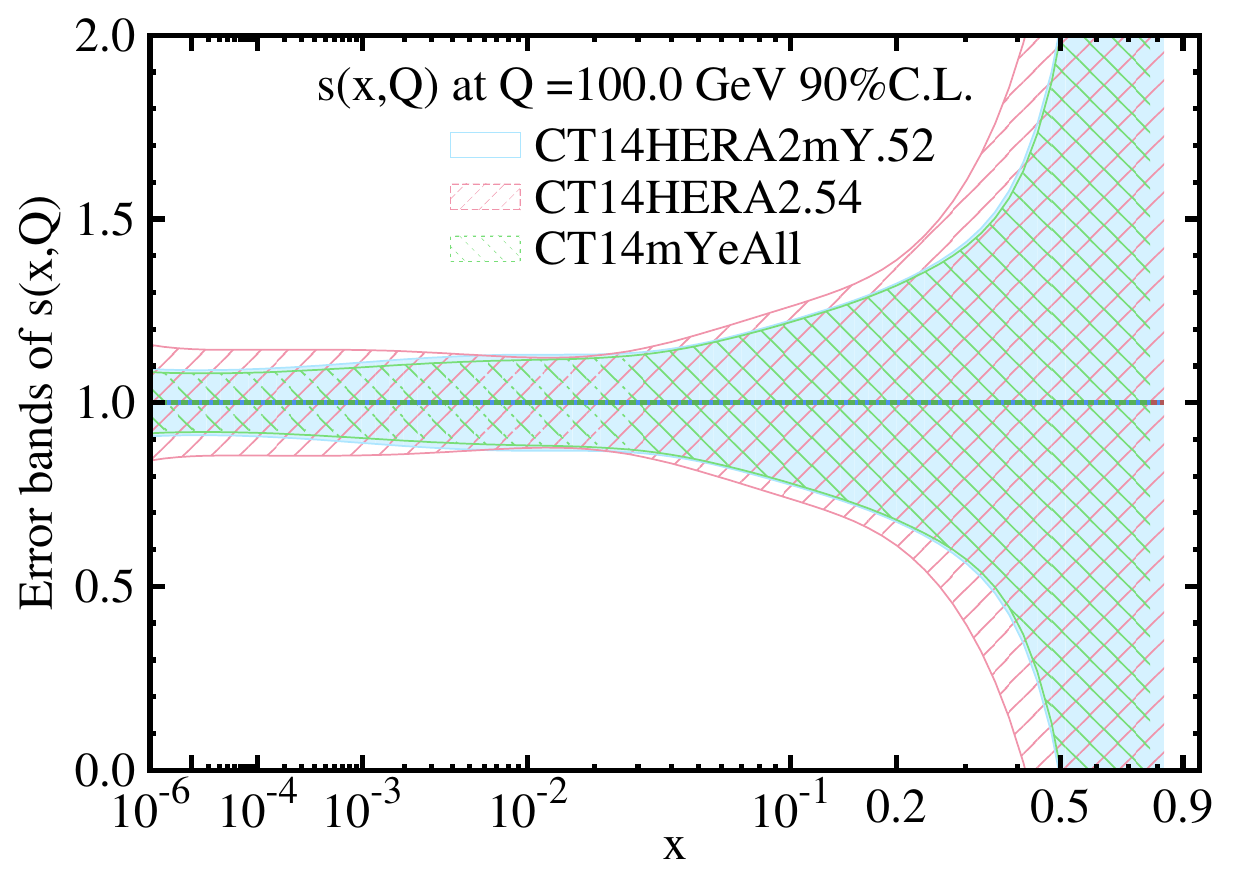}
	\caption{Same as Fig.~\ref{Fig:DYud}, but for $s$-PDF.
	}
	\label{Fig:DYs}
\end{figure}

The updated $s$-PDF is shown in Fig.~\ref{Fig:DYs}, where one finds a dramatic difference in the $s$-PDF uncertainty between \texttt{ePump} and the global-fit.
The CT14HERA2 error band of the $s$-PDF increases for almost all $x$ values when the Drell-Yan data is added, while the \texttt{ePump}-updated CT14mYeAll error band stays the same as the base CT14HERA2mY.  An increase in the PDF error band found in a global fit usually indicates the presence of some tension between the new and the old data. Due to the quadratic approximation in \texttt{ePump}, it can never produce an increase in the size of the error band, but rather in most cases it will reduce the error.  This can be inferred directly from Eq.~(20) in Ref.~\cite{Schmidt:2018hvu} by the positivity of $\lambda^{(r)}$. Thus, we find that when there is strong tension between the new and old data sets, it will not be revealed by an enlargement of the \texttt{ePump}-updated PDF error bands, in contrast to that of a true global fit. We shall discuss other methods to explore possible tension between different data sets with \texttt{ePump} later.

\subsection{CT14HERA2 excluding all jet data: CT14HERA2mJ}\label{subsection:mJ}

\begin{figure}[h]
\includegraphics[width=0.45\textwidth]{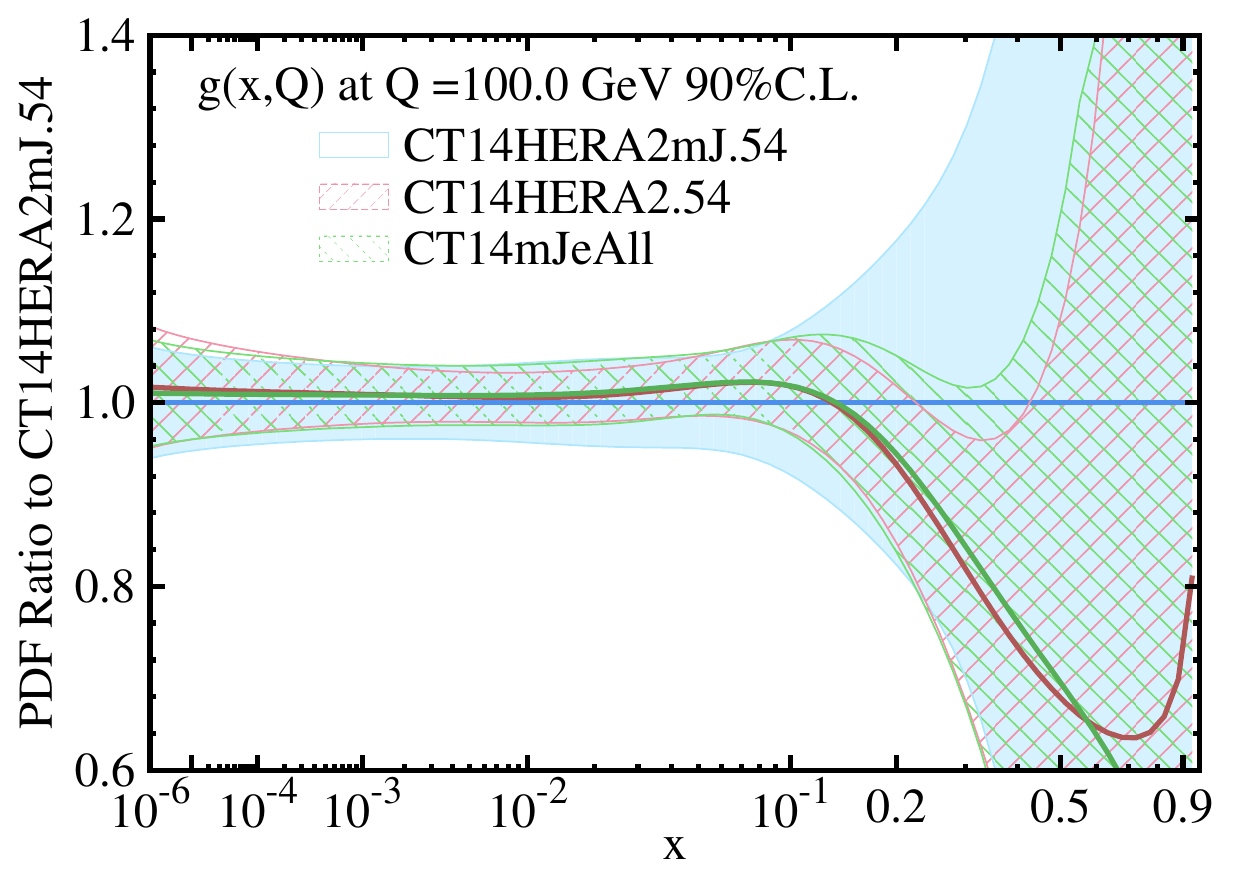}
\includegraphics[width=0.45\textwidth]{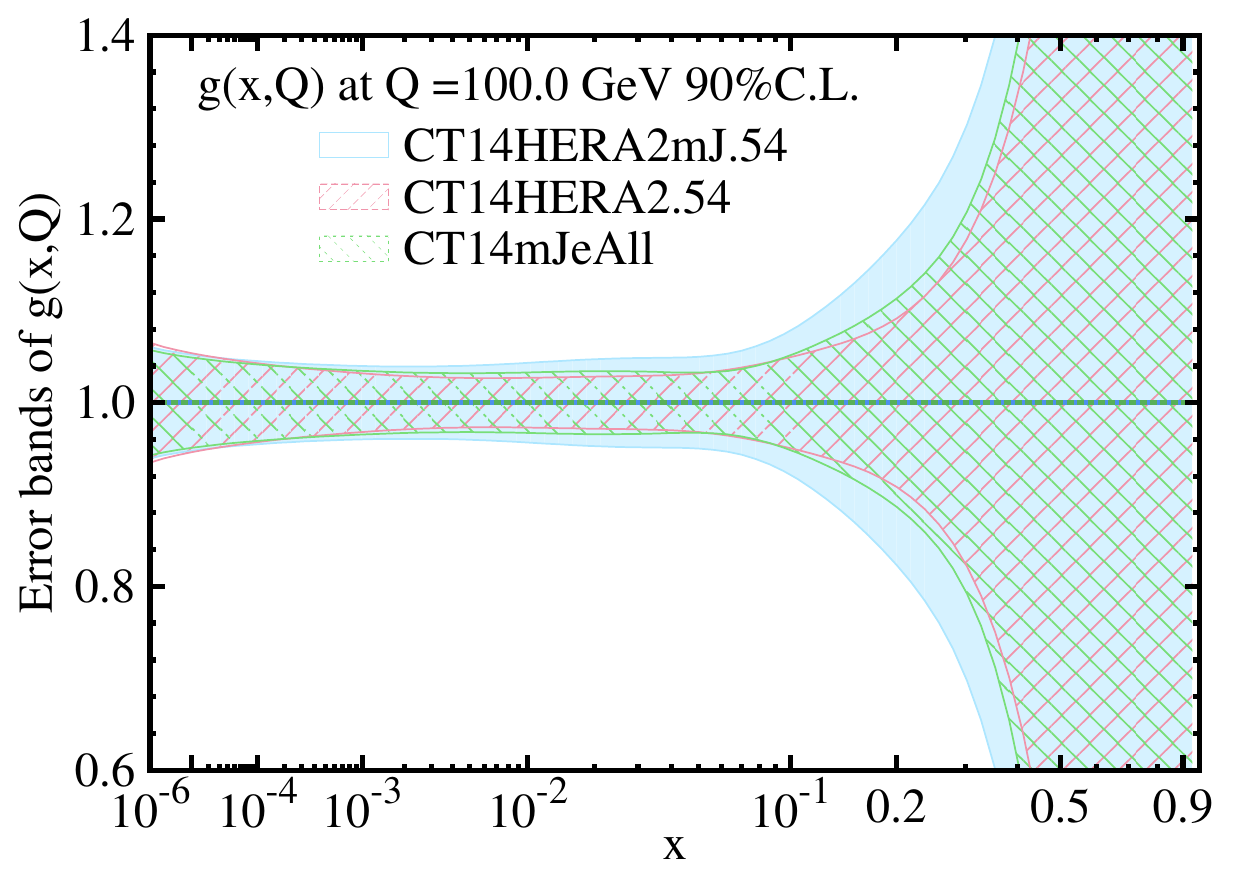}
\caption{
Comparison of \texttt{ePump}-updated $g$-PDFs, at $Q=100~\rm{GeV}$. Left panel: the PDF ratios over the best-fit of the base CT14HERA2mJ. Right panel: the error bands relative to their own best-fit.	
}
\label{Fig:Jg}
\end{figure}

\begin{figure}[h]
\includegraphics[width=0.45\textwidth]{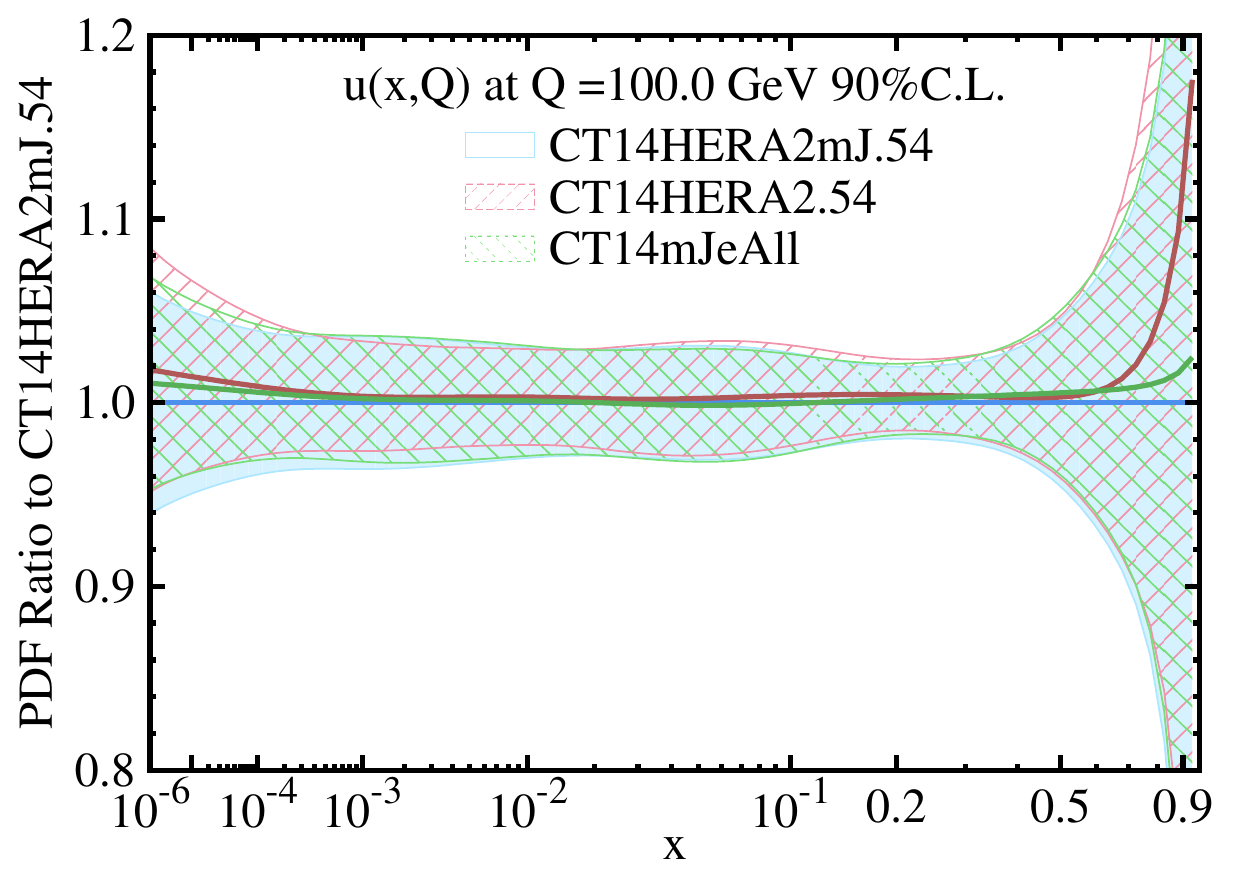}
\includegraphics[width=0.45\textwidth]{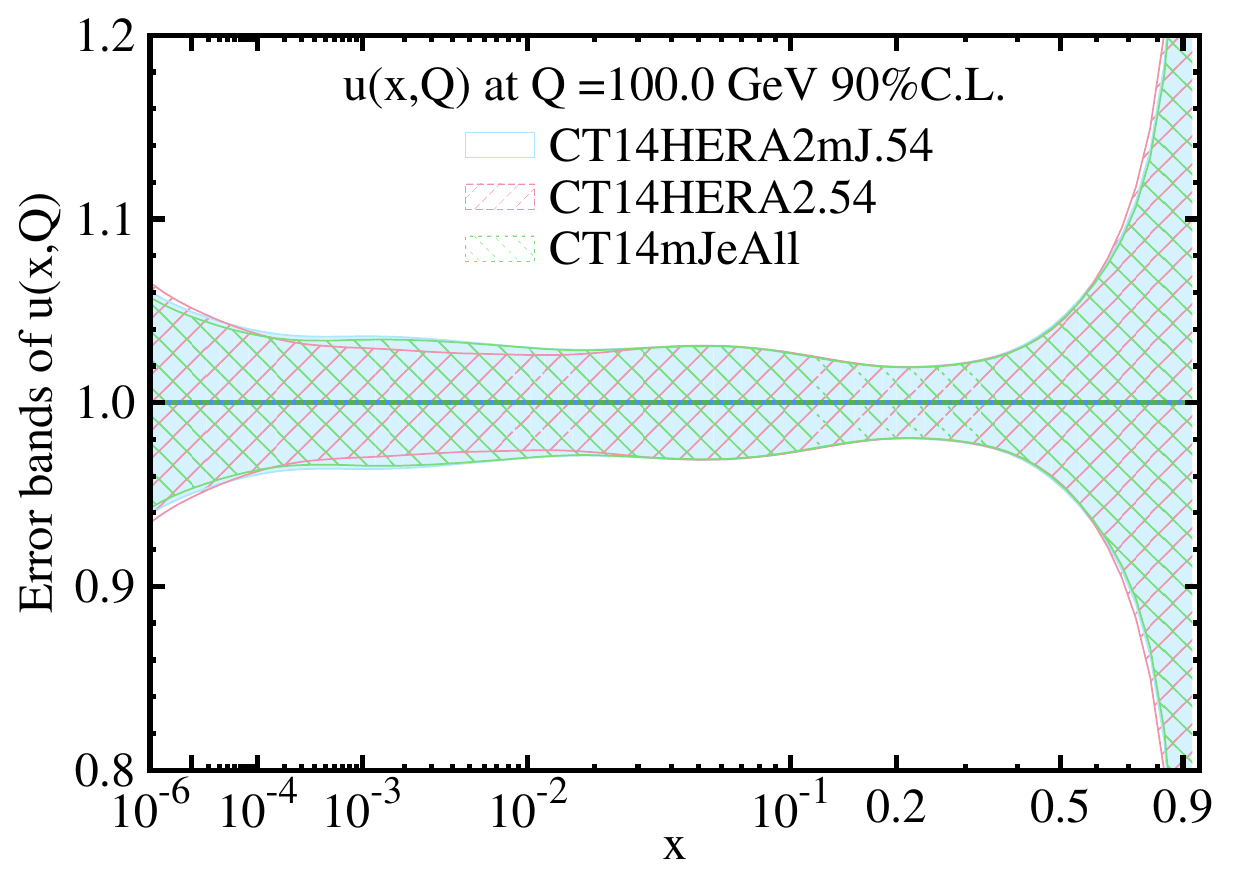}\\
\includegraphics[width=0.45\textwidth]{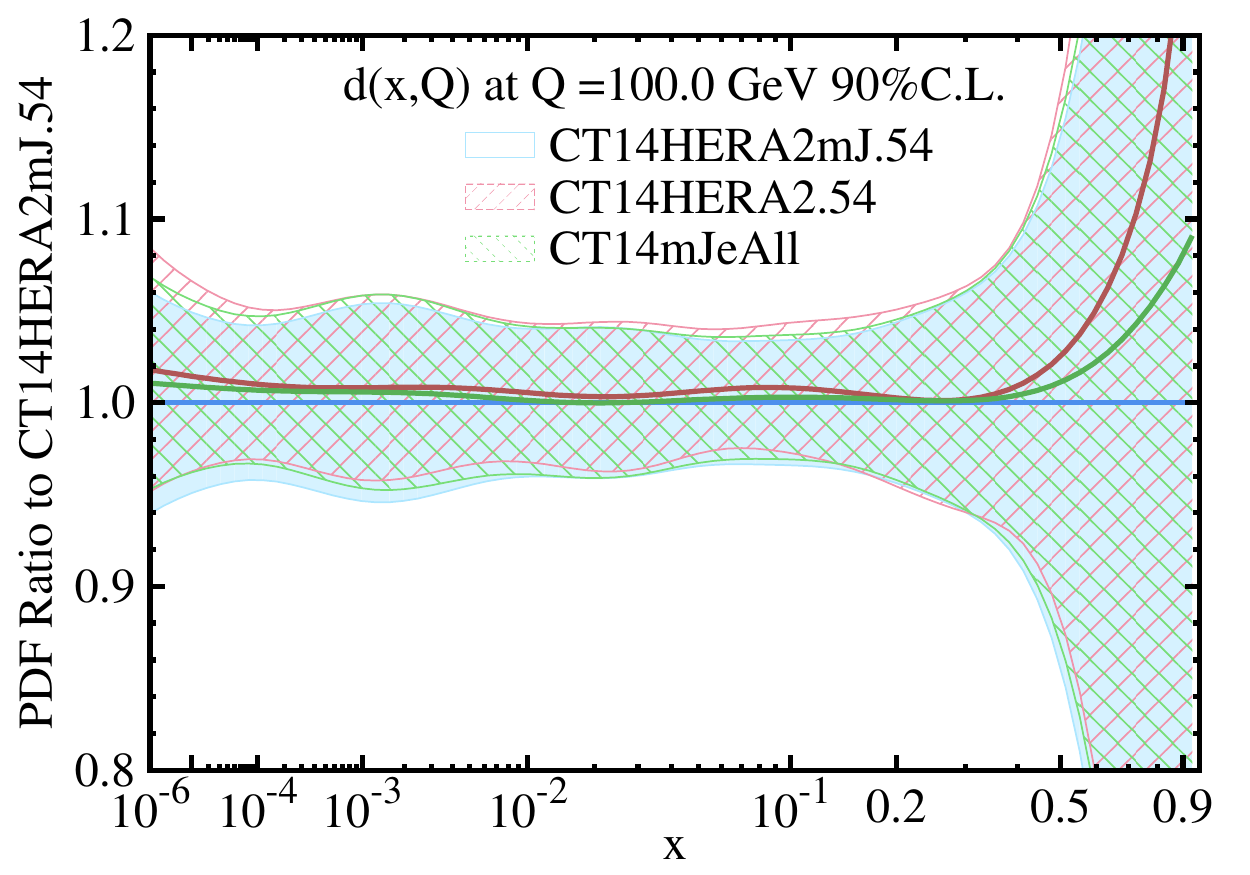}
\includegraphics[width=0.45\textwidth]{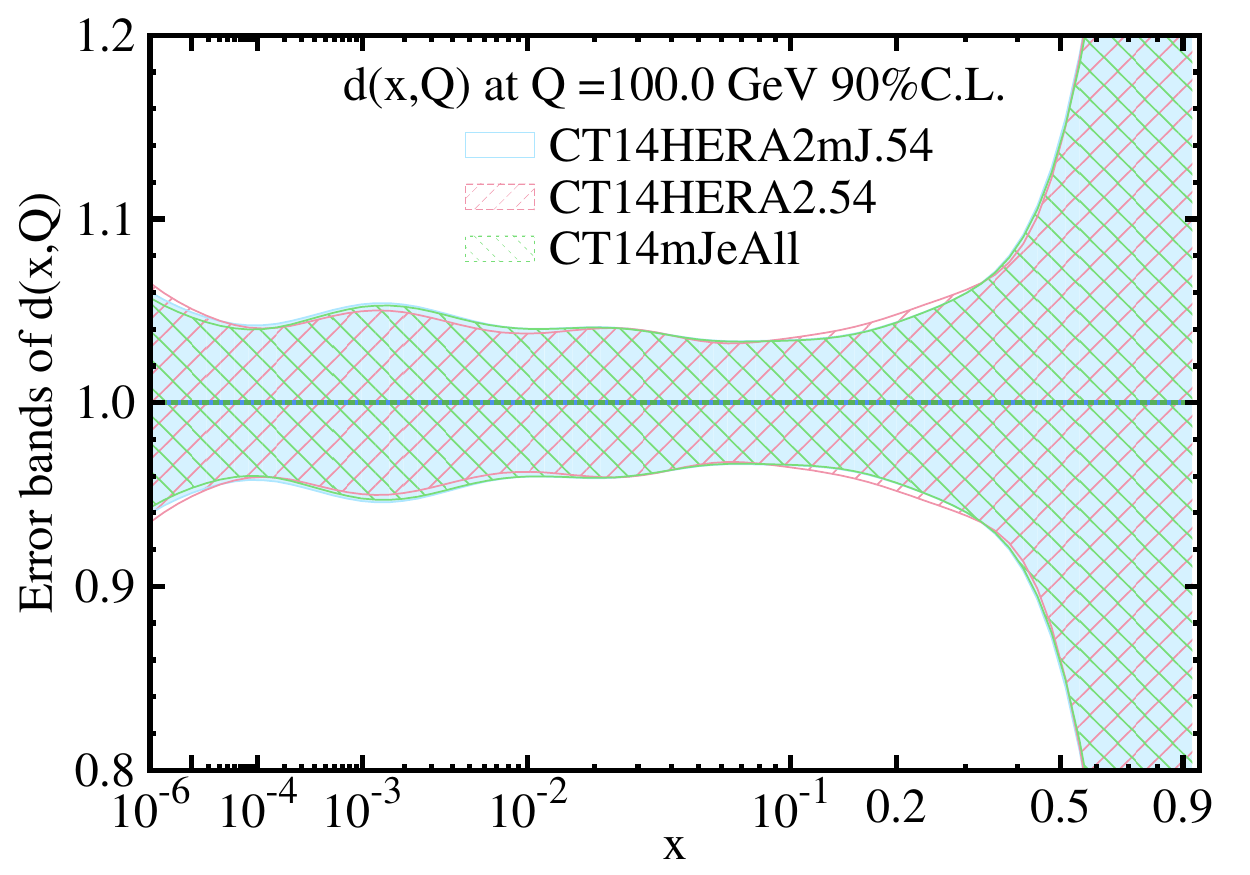}
\caption{Same as Fig.~\ref{Fig:Jg}, but for $u$ and $d$ PDFs.
}
\label{Fig:Jud}
\end{figure}

CT14HERA2 contains four sets of inclusive jet production data:
CDF~\cite{Aaltonen:2008eq} and D$\O$~\cite{Abazov:2008ae} at the Tevatron Run-2, and ATLAS 7 TeV~\cite{Aad:2011fc} and CMS 7
TeV ~\cite{Chatrchyan:2012bja} at the LHC.
We denote the global fit to all the CT14HERA2 data, minus the four jet data sets, as CT14HERA2mJ, which contains 2882 data points.\footnote{CT14HERA2mJ contains 27 free parameters. We do not need to fix any parameters.}
The \texttt{ePump}-updated PDFs, obtained by adding back the four jet data sets as ``new" data, which contain 405 data points, are designated as CT14mJeAll.
As shown in Fig.~\ref{Fig:Jg}, the jet data mainly constrain the $g$-PDF, with little effect on the $u$ and $d$ PDFs, cf. Fig.~\ref{Fig:Jud}.
The agreement between the \texttt{ePump}-updated PDF and the CT14HERA2 global fit is quite satisfactory.
The $g$-PDF is modified and the error band is increasingly reduced as $x$ grows from 0.01 to 0.3.
At large $x$ values, the difference in the best-fit PDFs is not significant due to the large error band size.
Again, the somewhat larger CT14HERA2 PDF error band of $d$ PDF at $x$ in the range from 0.1 to 0.4, as compared to CT14mJeAll, indicates some tension caused by adding the jet data to the rest of the CT14HERA2 data in the global fit, which \texttt{ePump} is unable to see.

The Higgs boson cross section is strongly dependent on the $g$-PDF, so an interesting question to ask is:
What is the impact of the jet data included in the CT14HERA2 fit on the prediction of the
Higgs boson production cross section
$\sigma(gg\rightarrow h)$ at the LHC?
As explained in Section \RNum{2}.D of Ref.~\cite{Schmidt:2018hvu},
\texttt{ePump} can not only update the PDFs but also physical observables within a few seconds of CPU time.
Table~\ref{Tab:Jh} shows the result of the comparison between the \texttt{ePump}-updated prediction and the CT14HERA2 prediction for various LHC center-of-mass energies. After adding jet data into the fits, the central values of $\sigma(gg\rightarrow h)$ become slightly larger and the errors become smaller. Using \texttt{ePump}, the jet data reduce the error of  $\sigma(gg\rightarrow h)$  by about 20\%, while the true global fit results in about a 30\% reduction. This difference may be due to the linear and quadratic approximations in \texttt{ePump}, or it could be due to the effect of the new data on the Tier-2 penalty, which is only treated
on average in \texttt{ePump}. Nevertheless, one can use \texttt{ePump} to quickly estimate the impact of some ``new'' data to updated PDFs and physical observables.
For instance, in this case, we could conclude from \texttt{ePump} updating that including jet data in the fit will lead to a more precise result of $\sigma(gg\rightarrow h)$ with its  uncertainty reduced by about 20\%.
\begin{table}[h]
\begin{tabular}{c|c|c|c}
\hline
\hline
$\sqrt S $(TeV) & CT14HERA2mJ& CT14mJeAll & CT14HERA2\\
\hline
7 & 14.52 $\pm$ 0.62 & 14.68 $\pm$ 0.50 & 14.60 $\pm$ 0.43\\
\hline
8 & 18.37 $\pm$ 0.80 & 18.59 $\pm$ 0.63 & 18.49 $\pm$ 0.53\\
\hline
13 & 42.1 $\pm$ 2.0 & 42.7 $\pm$ 1.5 & 42.5 $\pm$ 1.2\\
\hline
14 & 47.5 $\pm$ 2.3 & 48.2 $\pm$ 1.8 & 48.0 $\pm$ 1.4\\
\hline
\hline
\end{tabular}
\caption{Theoretical predictions of $\sigma(gg\rightarrow h)$ in pb at the LHC, for various center-of-mass energies, based on different PDF sets. CT14mJeAll predictions were obtained directly from \texttt{ePump} after adding jet data to update the CT14HERA2mJ PDFs.
Here, the listed CT14HERA2 PDF errors do not include the contribution from the two extreme $g$-PDF sets.}
\label{Tab:Jh}
\end{table}

\section{Impact of individual CT14HERA2 data sets on PDFs} \label{section:}

\texttt{ePump} can be used to quickly assess the impact of individual data sets on constraining the PDFs in a global analysis.
In this section, we will demonstrate this by using \texttt{ePump} to assess the data sets used in the CT14HERA2 global analysis.

\subsection{The impact of jet data in the CT14HERA2 fit}\label{section:impact Jet}

\begin{figure}[h]
	\centering
	\includegraphics[width=0.45\textwidth]{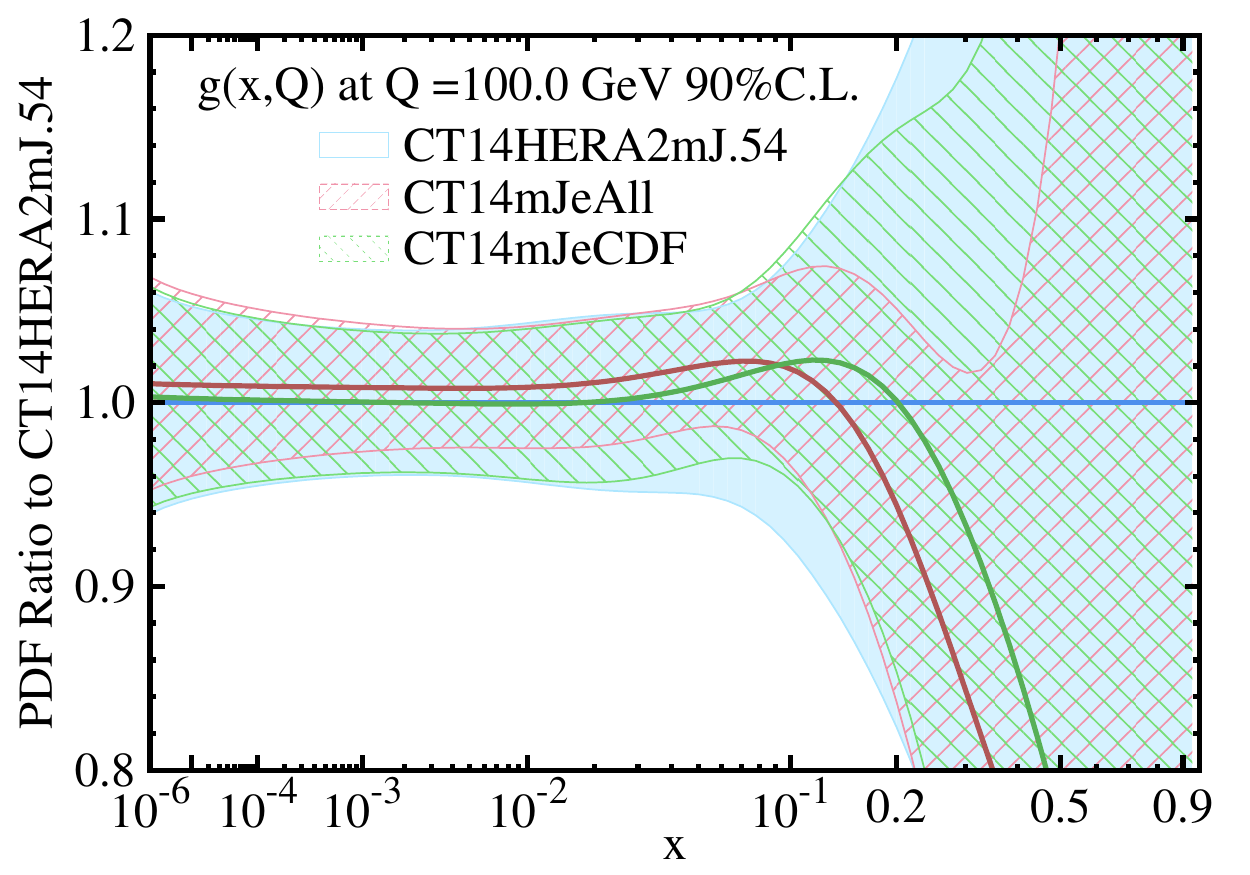}
	\includegraphics[width=0.45\textwidth]{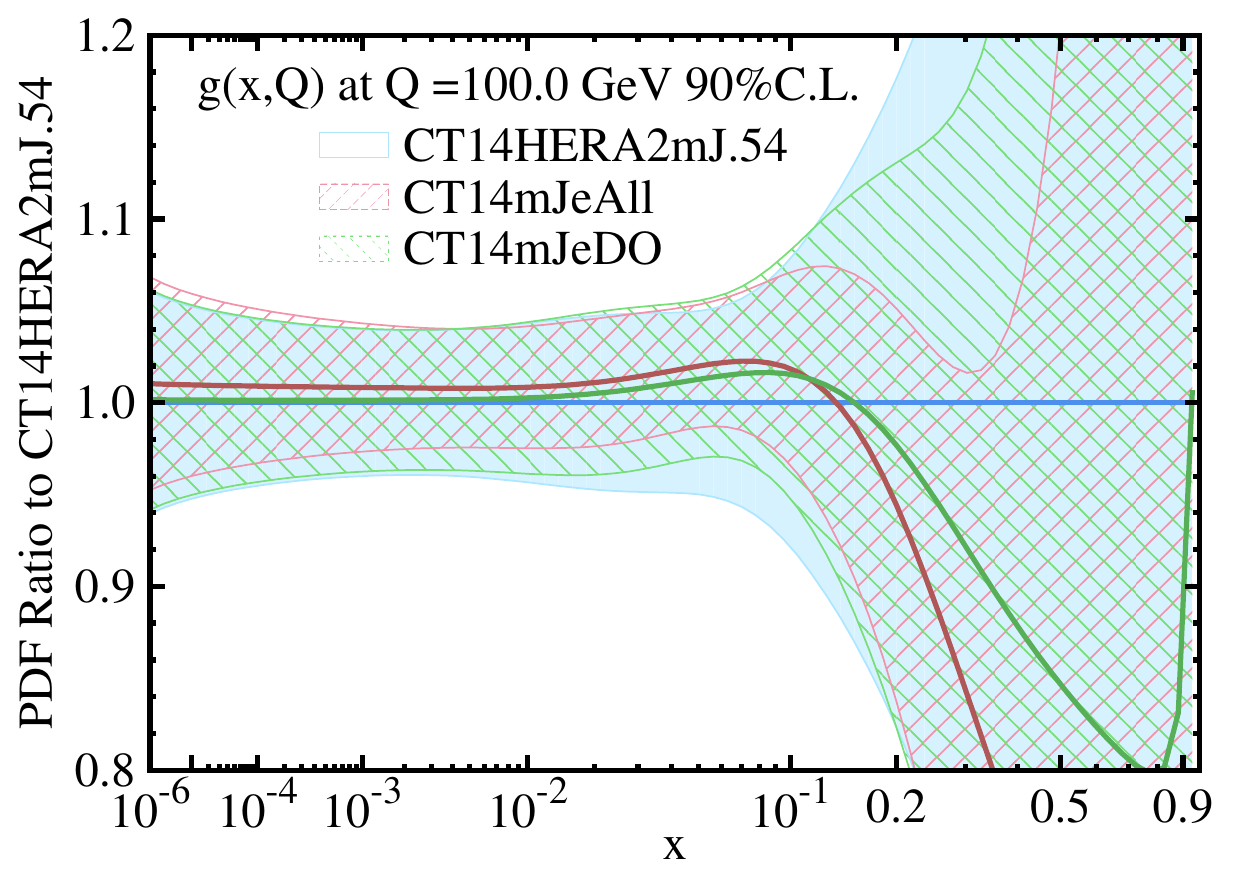}\\
	\includegraphics[width=0.45\textwidth]{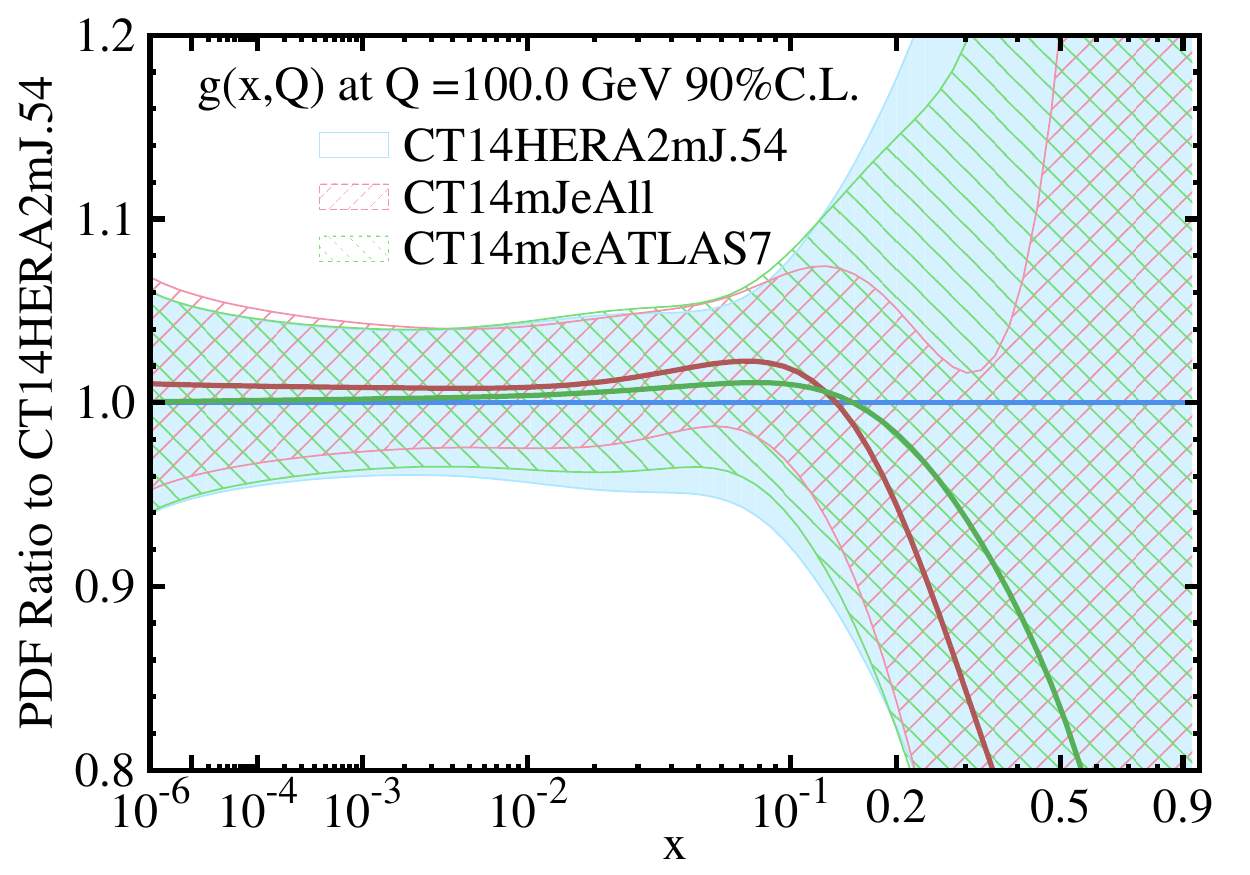}
	\includegraphics[width=0.45\textwidth]{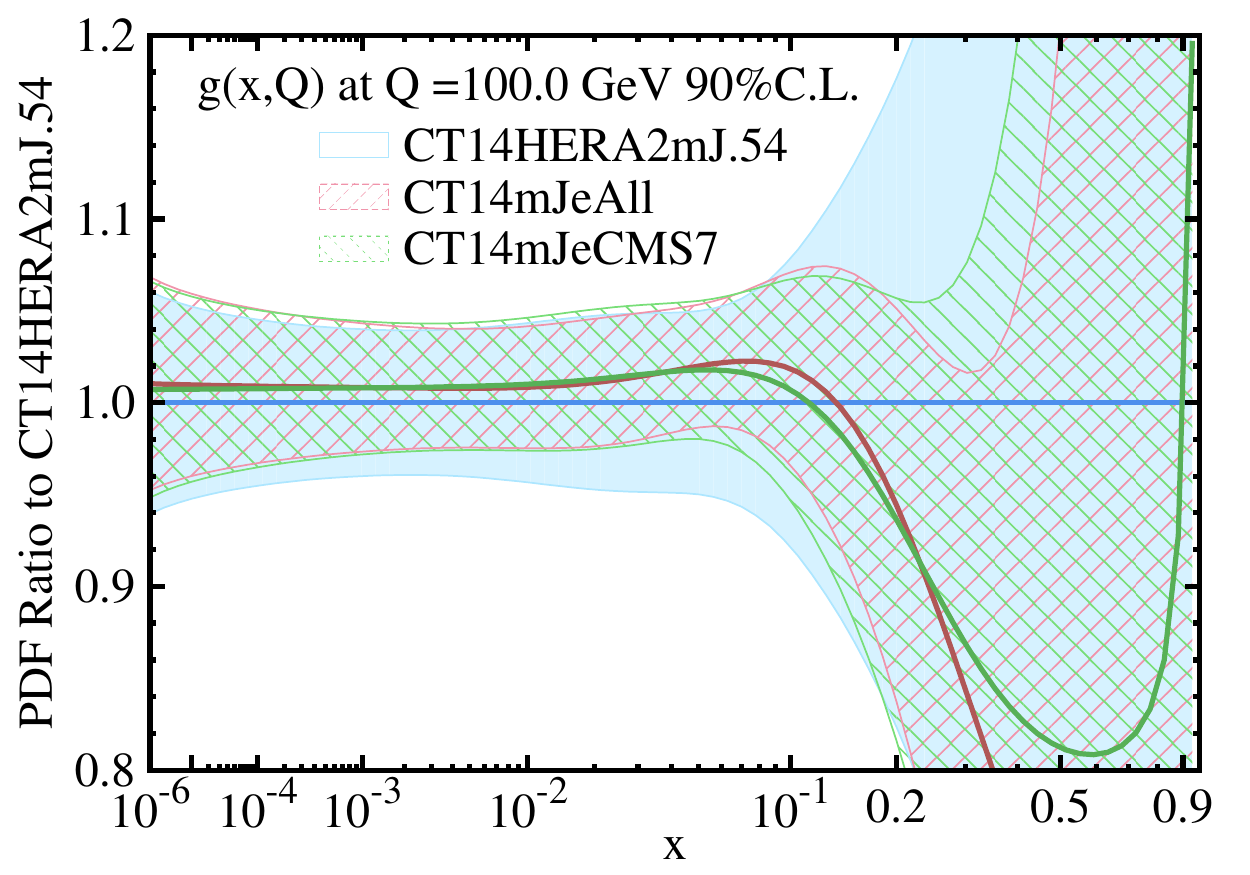}
	\caption{Impact of individual jet data sets on the $g$-PDF, using \texttt{ePump} to add the data sets to CT14HERA2mJ, which are then compared with CT14mJeAll. The curves are the ratios to the best-fit of CT14HERA2mJ, and they can be used to
identify the relative impact of each jet data set on the $g$-PDF.}
	\label{Jg1}
\end{figure}

As already noted in Sec.~\ref{subsection:mJ}, the jet data mainly constrain the $g$-PDF and have little effect on other flavors. From Fig.~\ref{Fig:Jg}, we see that the jet data prefer a larger $g$-PDF at $x=10^{-2} \sim 10^{-1}$ and smaller $g$-PDF at $x=0.2 \sim 0.4$. The error band is reduced by a fairly large amount in the range of $x=10^{-2} \sim 0.2$, by about 1/4 to 1/3.

In order to see the impact of individual jet data in the CT14HERA2 fit, we use \texttt{ePump} to add each jet data set individually to CT14HERA2mJ. The results are shown in Fig.~\ref{Jg1}, with CT14mJeAll shown together in the same graph for comparison. It can be seen that the four jet data sets produce the same qualitative effects on the $g$-PDF, but quantitatively the CMS 7 TeV jet data~\cite{Chatrchyan:2012bja} yields the result that is most similar to CT14mJeAll.
It increases the $g$-PDF slightly at small $x$, with maximum pull upward around $x\sim 0.1$, but pulls it downward sharply above $x\sim 0.2$.
While all of the jet data sets reduce the error band, the CMS 7 TeV jet data reduces it the most and is the closest to the all-jet result.
The others reduce the errors by a distinctly smaller amount. From this we can draw the conclusion that the CMS jet data has the dominant impact on the $g$-PDF, among all the jet data included in CT14HERA2.
It is worth noting that in the range of $x=0.1\sim 0.2$, the CDF
Run-2 inclusive jet data set leads to a harder $g$-PDF than the others, while the D$\O$ Run-2,
ATLAS 7 TeV and CMS 7 TeV jet data
yield similar results for the $g$-PDF.

\subsection{The impact of Drell-Yan data in the CT14HERA2 fit}\label{section: impact DY}

\begin{figure}[h]
\includegraphics[width=0.45\textwidth]{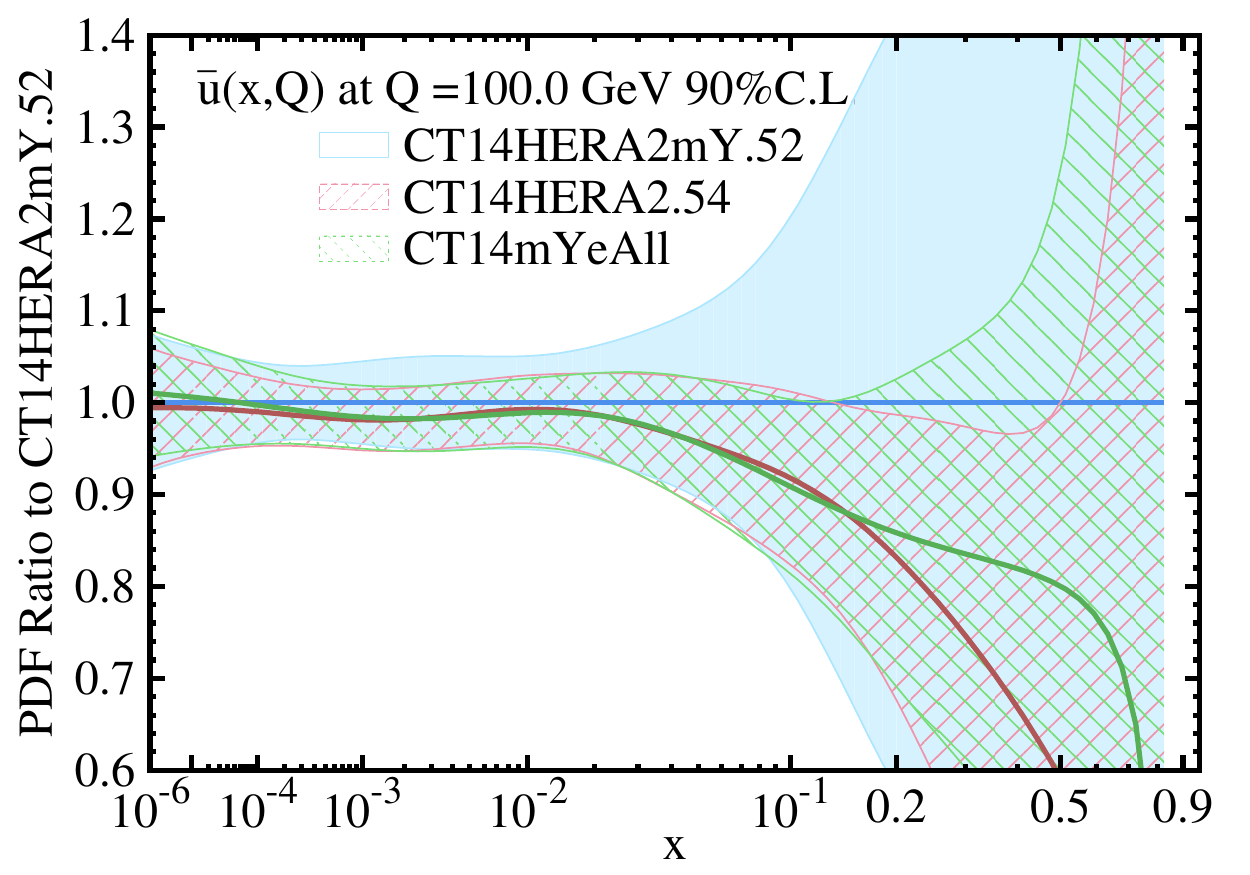}
\includegraphics[width=0.45\textwidth]{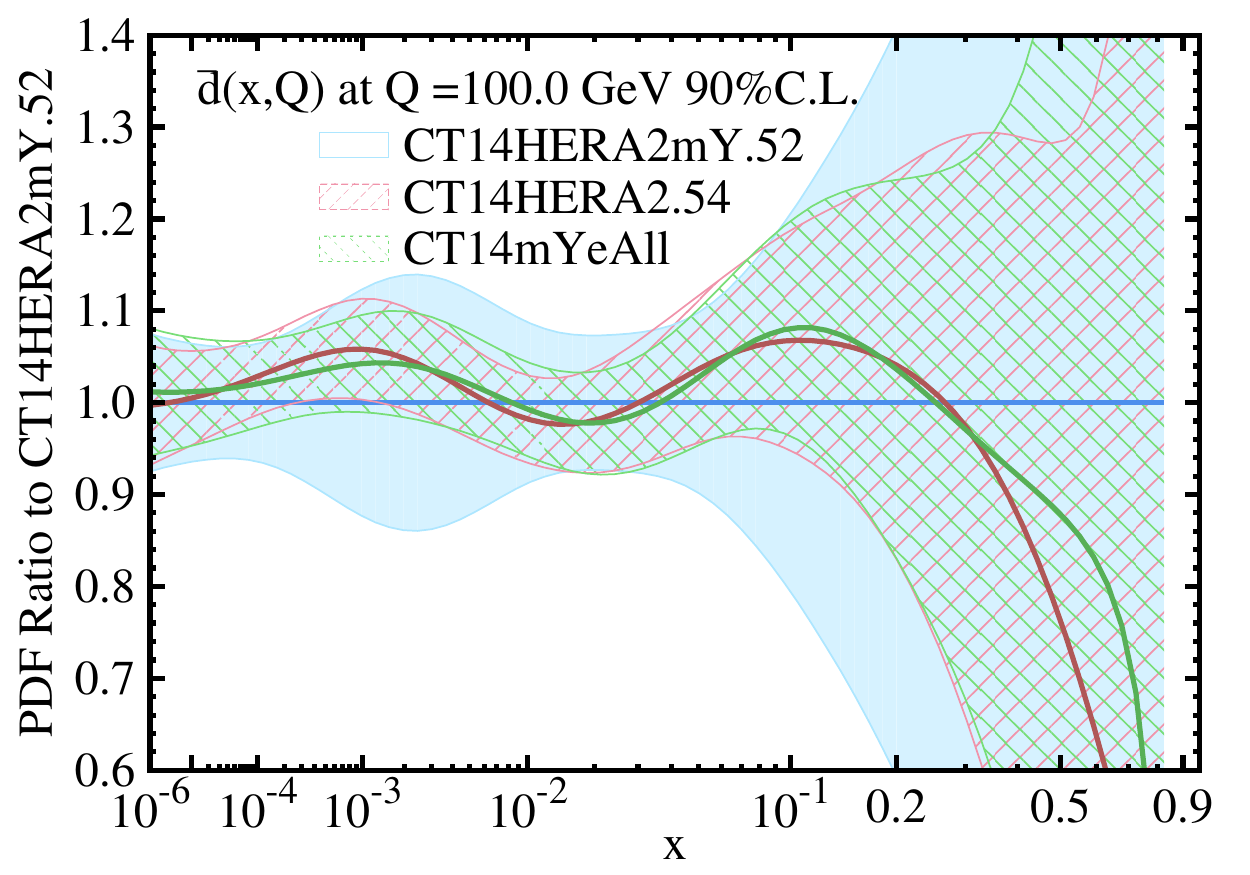}
\caption{Same as Fig.~\ref{Fig:DYud}, but for $\bar u$ and $\bar d$ PDFs.
}
\label{Fig:DYubardbar}
\end{figure}

\begin{figure}[h]
\includegraphics[width=0.45\textwidth]{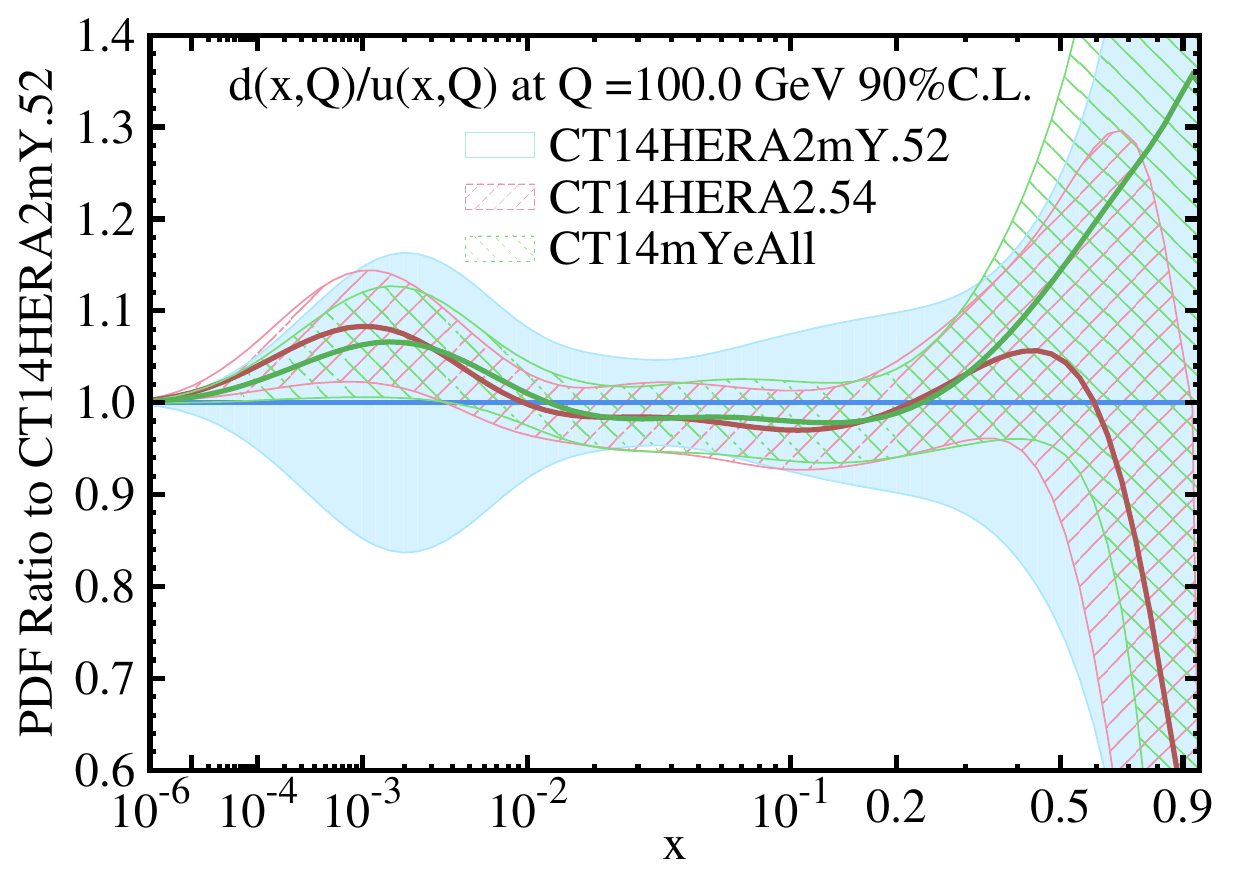}
\includegraphics[width=0.45\textwidth]{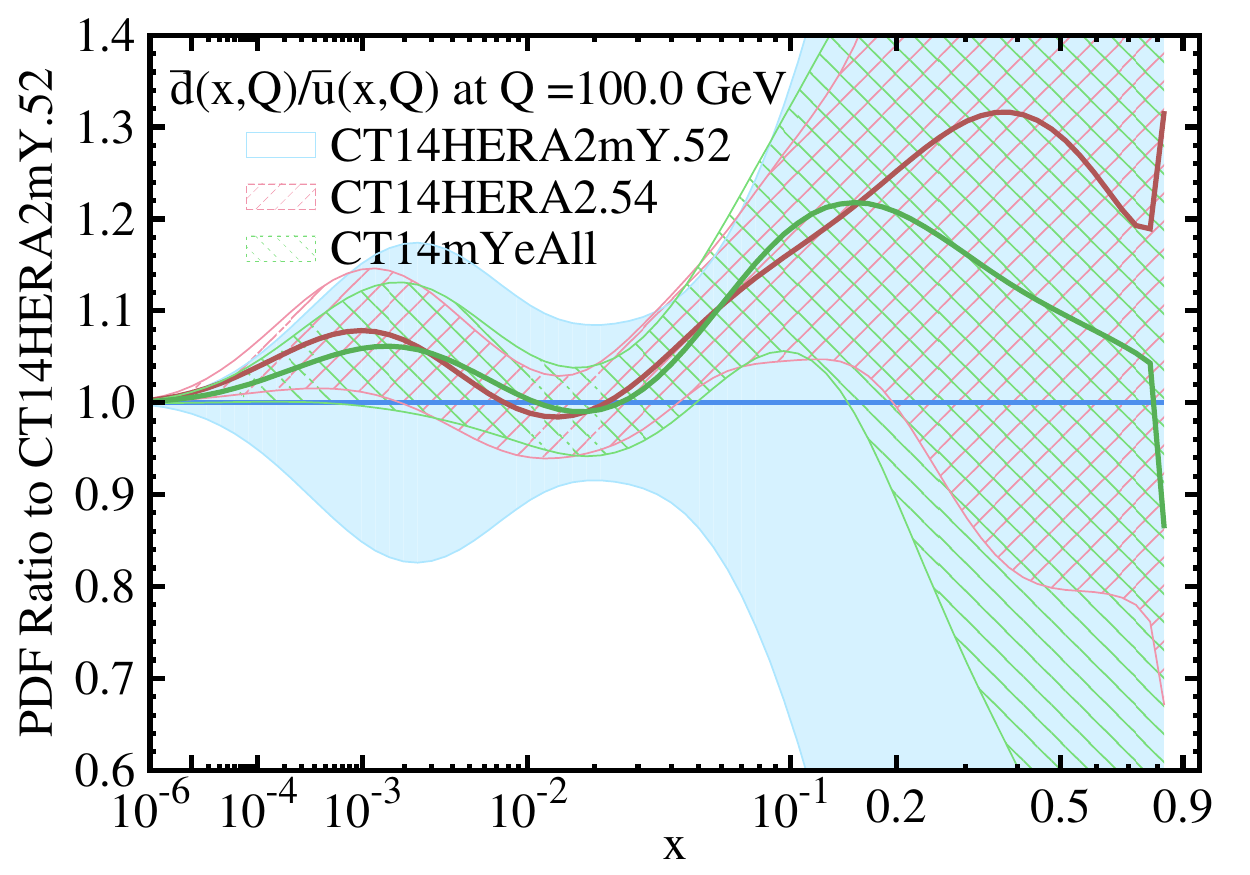}
\caption{Same as Fig.~\ref{Fig:DYud}, but for
	$d/u$ and $\bar d/\bar u$ PDF ratios.}
\label{Fig:DYratio}
\end{figure}

The impact of Drell-Yan data on $u$, $d$ and $s$-PDFs has been shown in Figs.~\ref{Fig:DYud} and \ref{Fig:DYs} in Sec.~\ref{subsec:mY}. The impact on other flavor PDFs can be found in
Figs.~\ref{Fig:DYubardbar} and \ref{Fig:DYratio}. It is found that Drell-Yan data mainly constrain $u$, $d$, $\bar u$ and $\bar d$ PDFs, with little effect on $g$ and $s$-PDFs. Drell-Yan data not only reduce the uncertainties of $u$, $d$, $\bar u$ and $\bar d$ a lot, but also change the best-fit PDFs dramatically. It is worth noting that for most values of $x$, the Drell-Yan data pulls the $u$ and $d$ PDFs in opposite directions, cf.  Fig.~\ref{Fig:DYud}. Around $x=10^{-3}$, the $u$ PDF is decreased while the $d$ PDF is increased, and at  $x \sim 0.1$, $u$ becomes larger while $d$ becomes smaller. This feature is also visible for the $\bar u$ and $\bar d$ PDFs and is a characteristic of Drell-Yan data.

\begin{figure}[ht]
	\centering
	\includegraphics[width=0.45\textwidth]{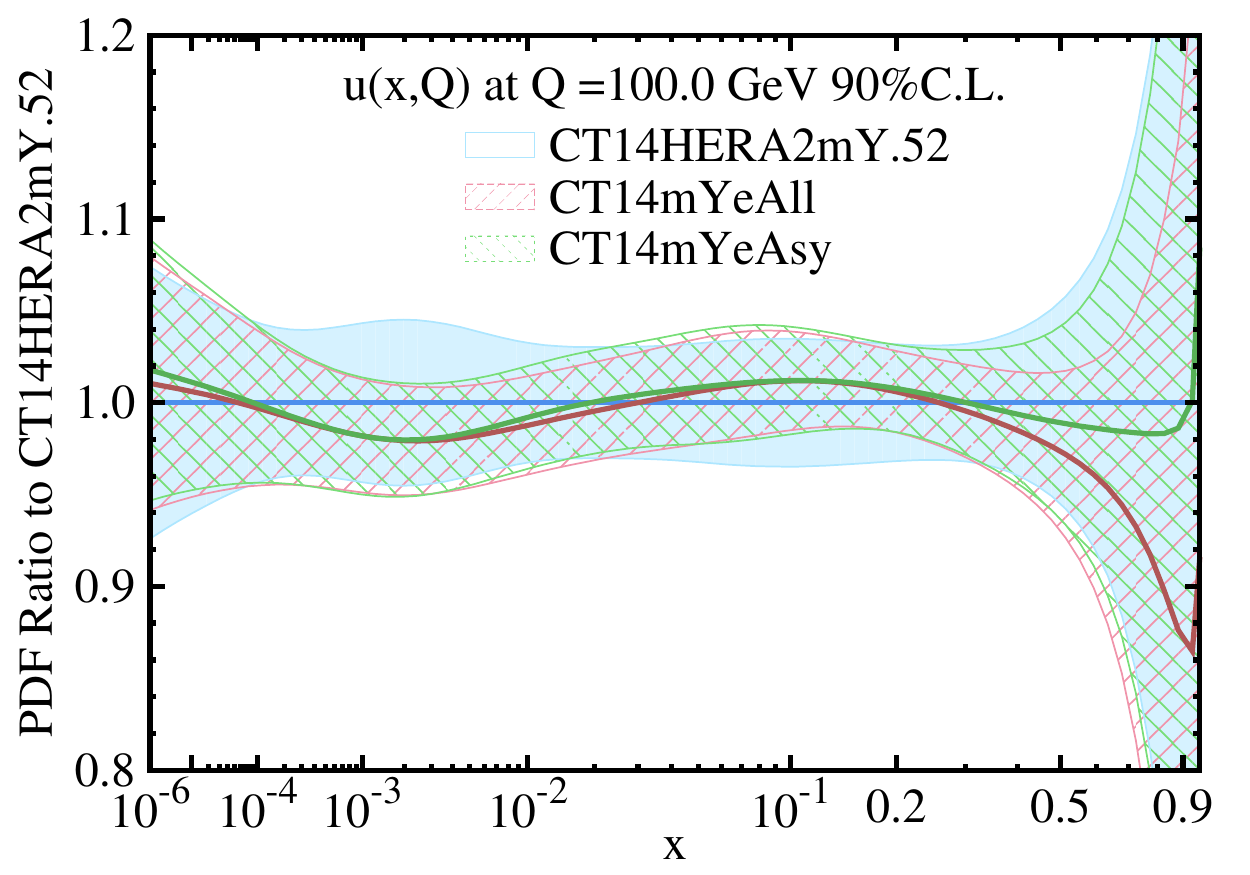}
	\includegraphics[width=0.45\textwidth]{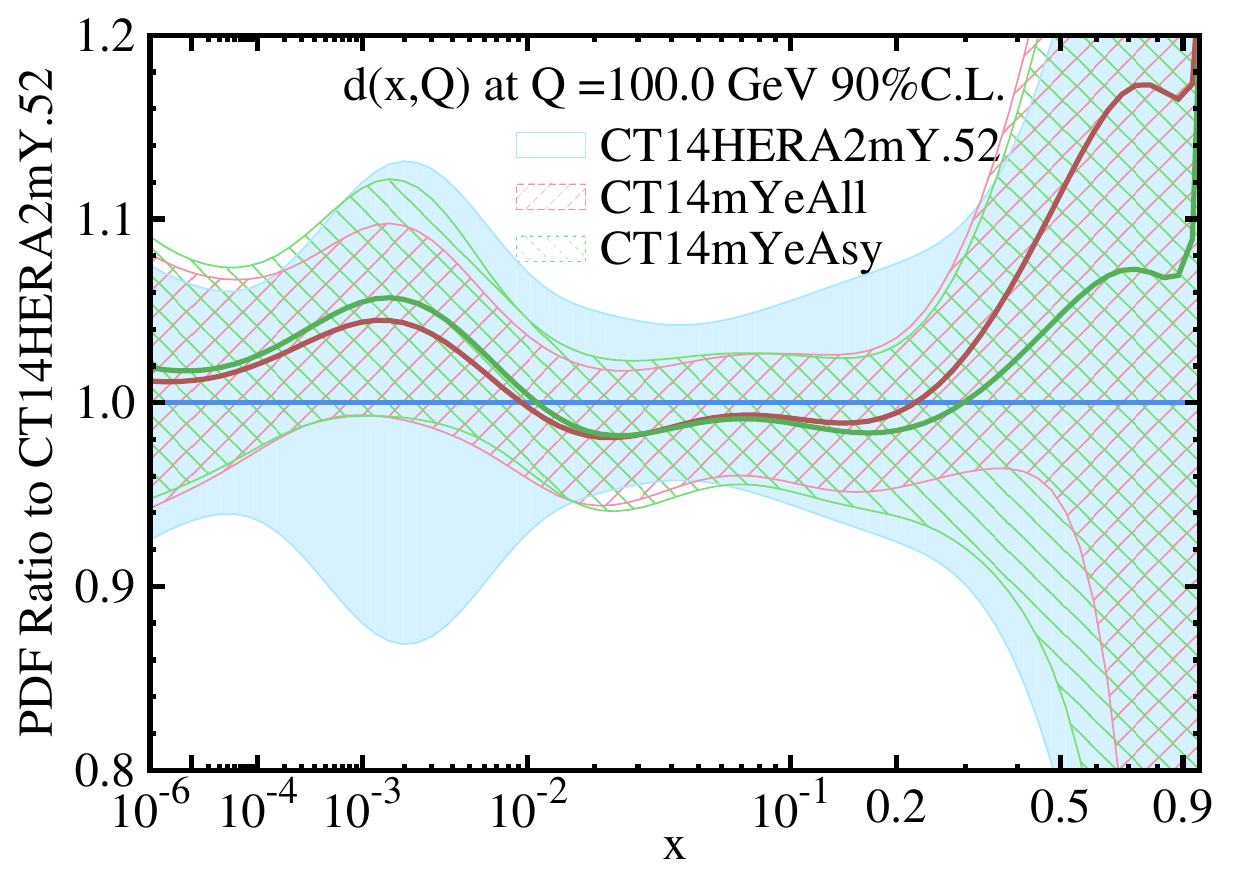}
	\caption{Comparison of CT14mYeAsy and CT14mYeAll for $u$ and $d$ PDFs at $Q=100~{\rm GeV}$. CT14mYeAsy is obtained by adding CMS 7 TeV $\mu$ asymmetry data, CMS 7 TeV electron asymmetry data, ATLAS 7 TeV $WZ$ data and D$\O$ Run2 $\mu$ asymmetry data to CT14HERA2mY, using \texttt{ePump}. The PDF ratios are over the best-fit of CT14HERA2mY.}
	\label{DY asy u d}
\end{figure}

\begin{figure}[h]
	\centering
	\includegraphics[width=0.45\textwidth]{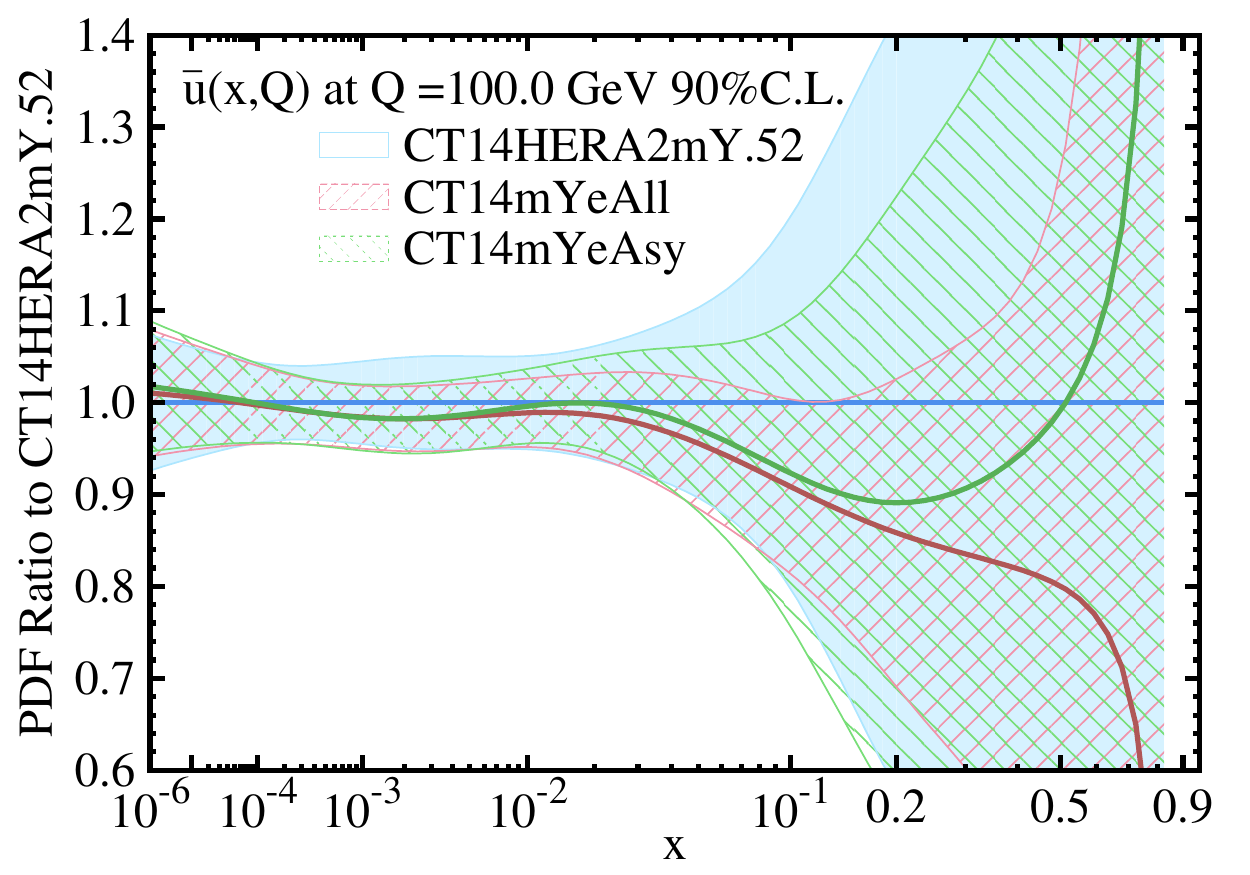}
	\includegraphics[width=0.45\textwidth]{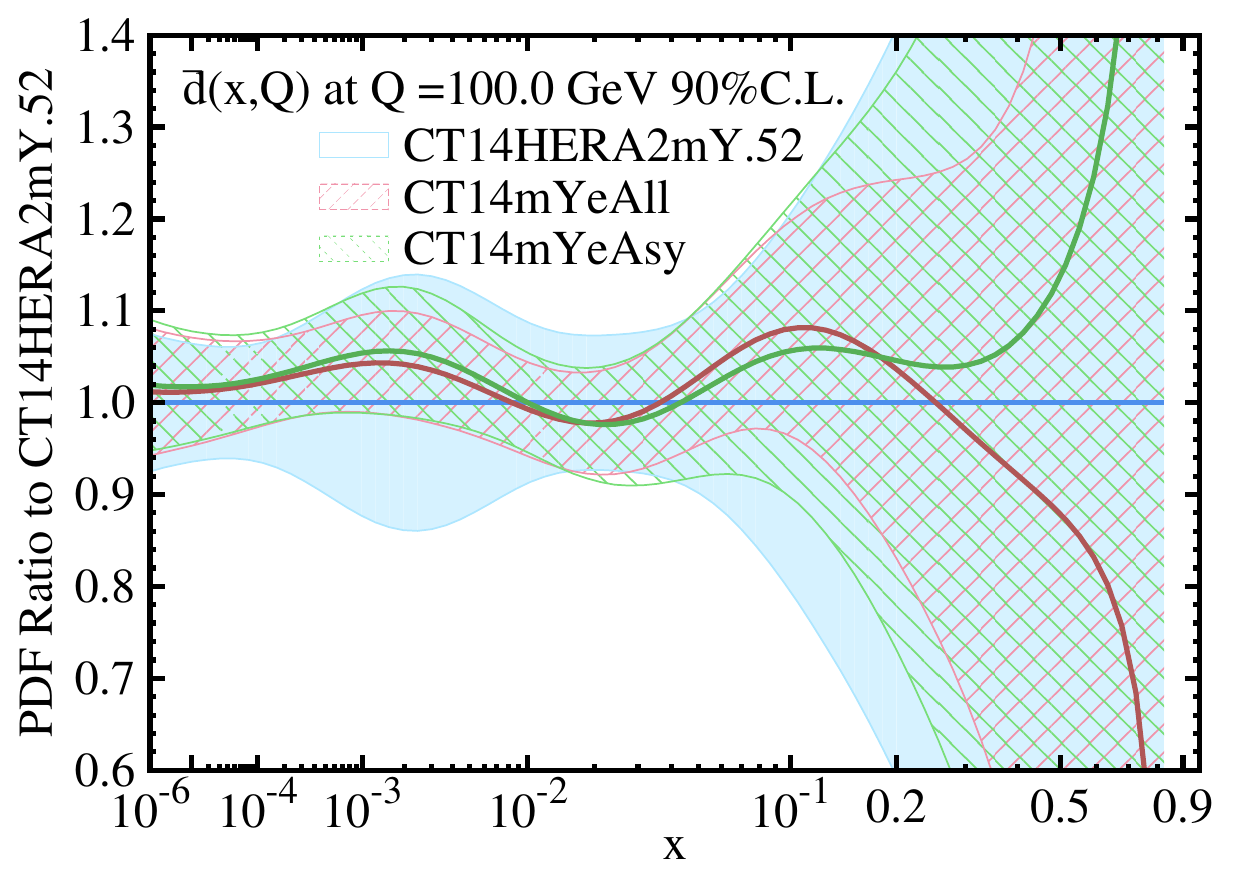}
	\caption{
		Same as Fig.~\ref{DY asy u d}, but for
		 $\bar u$ and $\bar d$ PDFs.}
	\label{DY asy ubar dbar}
\end{figure}

\begin{figure}[h]
	\centering
	\includegraphics[width=0.45\textwidth]{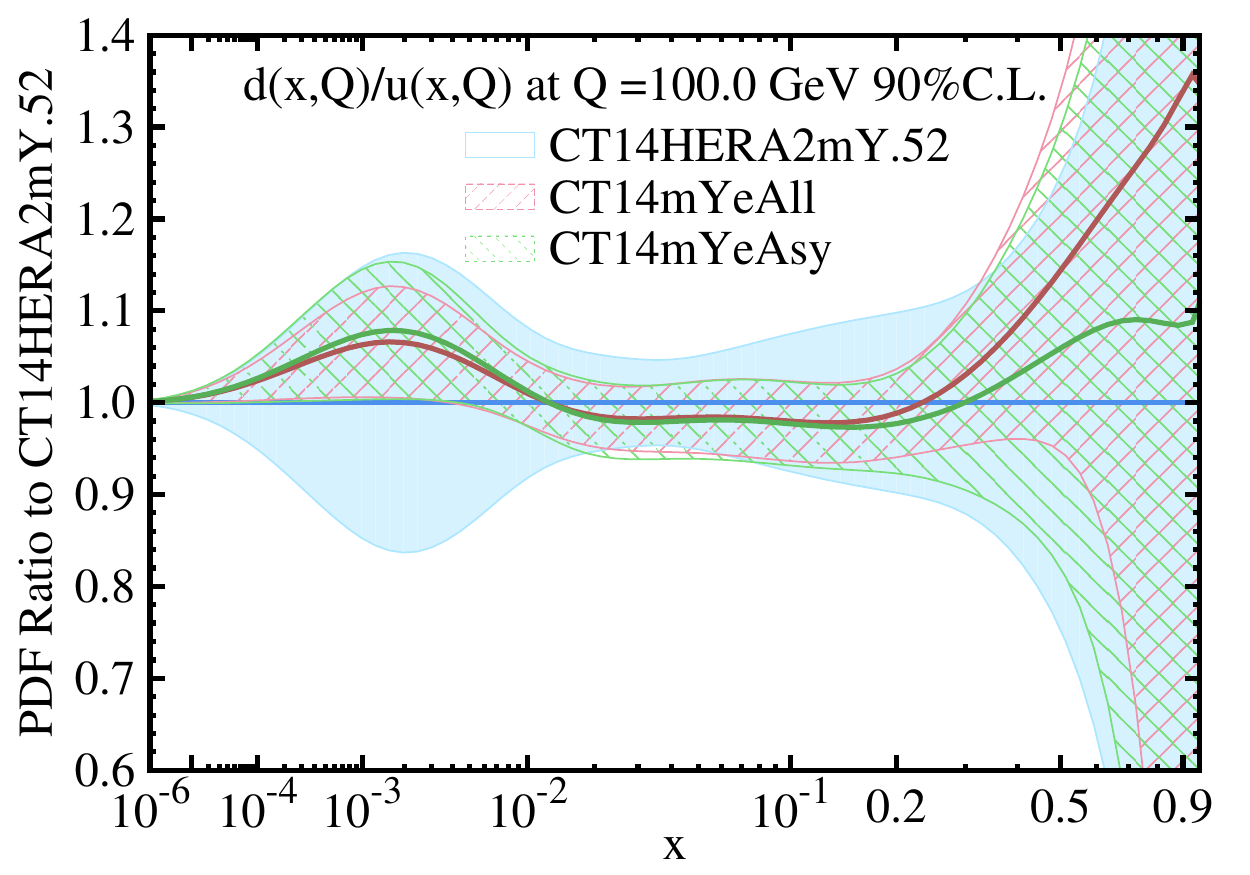}
	\includegraphics[width=0.45\textwidth]{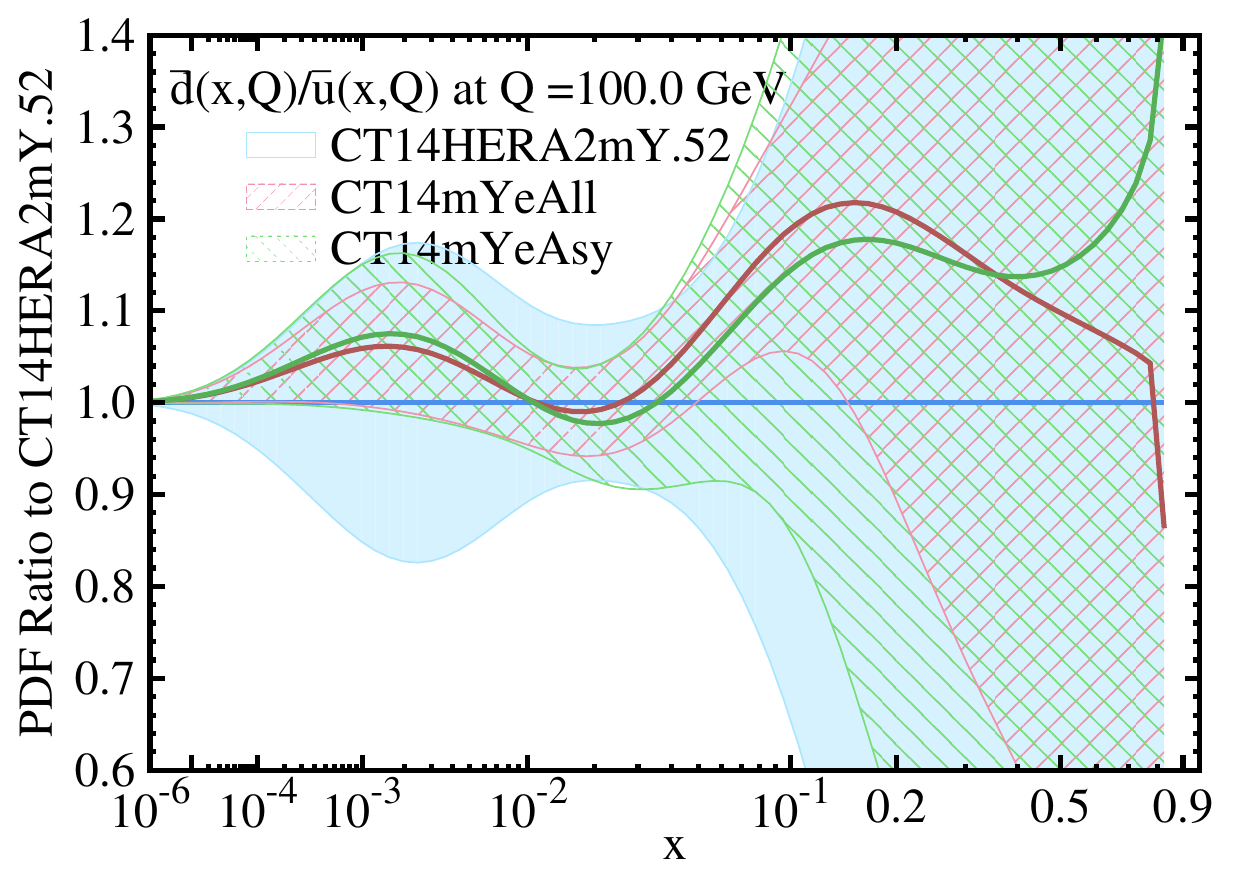}
	\caption{		Same as Fig.~\ref{DY asy u d}, but for
	 $d/u$ and $\bar d/\bar u$ PDF ratios.}
	\label{DY asy du and dbarubar}
\end{figure}

The fact that the $u$ and $d$ PDFs or the $\bar u$ and $\bar d$ PDFs get pulled in opposite directions indicates that the fitting program is directly modifying the differences between them.  Namely, it is $(u-d)$ and $(\bar u-\bar d)$, or rather, the ratios $d/u$ and $\bar d/\bar u$ that are directly probed by the Drell-Yan data. A reasonable conjecture is that this is due to $W^{\pm}$ charge asymmetry data measured at the Tevatron and the LHC.
To check this, we use \texttt{ePump} to add each Drell-Yan data set individually to CT14HERA2mY, and compare with CT14mYeAll, which we have already shown is very close to CT14HERA2.
We find that although most data sets give a similar trend, only the CMS 7 TeV $\mu$ asymmetry data~\cite{Chatrchyan:2013mza}, the CMS 7 TeV electron asymmetry data~\cite{Chatrchyan:2012xt}, the ATLAS 7 TeV $WZ$ data~\cite{Aad:2011dm} and the D$\O$ Run2 $\mu$ asymmetry data~\cite{Abazov:2007pm} have an appreciable impact.
This result is as expected, since most of them are lepton charge asymmetry data.
We can use \texttt{ePump} to add just these four charge asymmetry data sets to CT14HERA2mY. The result, called CT14mYeAsy, is shown in Figs.~\ref{DY asy u d}, \ref{DY asy ubar dbar} and \ref{DY asy du and dbarubar}, compared with CT14mYeAll.
The results from just including the lepton charge asymmetry data are pretty close except in the large $x$ region, where the relative PDF errors are large and the approximations used by \texttt{ePump} are unreliable~\cite{Schmidt:2018hvu}.

\begin{figure}[h]
	\includegraphics[width=0.45\textwidth]{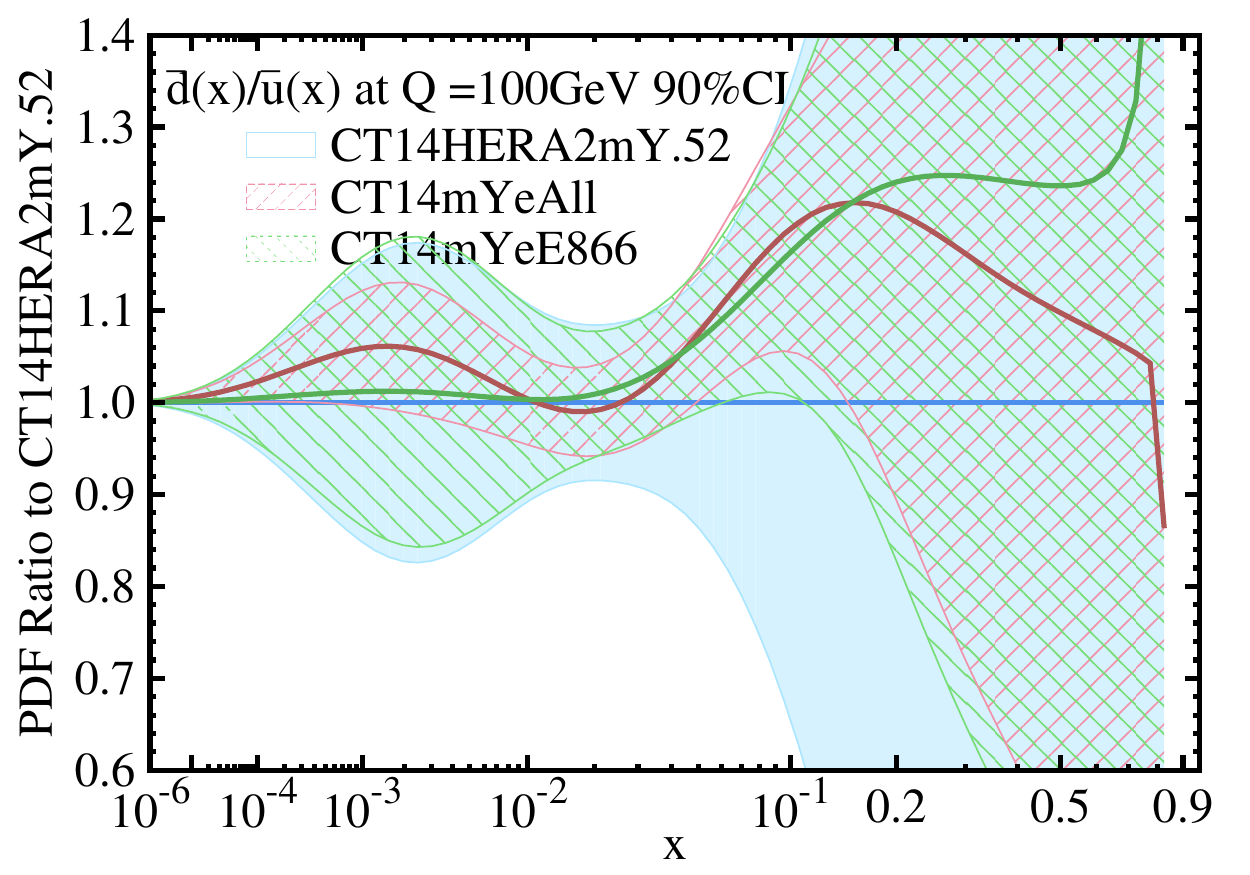}
	\includegraphics[width=0.45\textwidth]{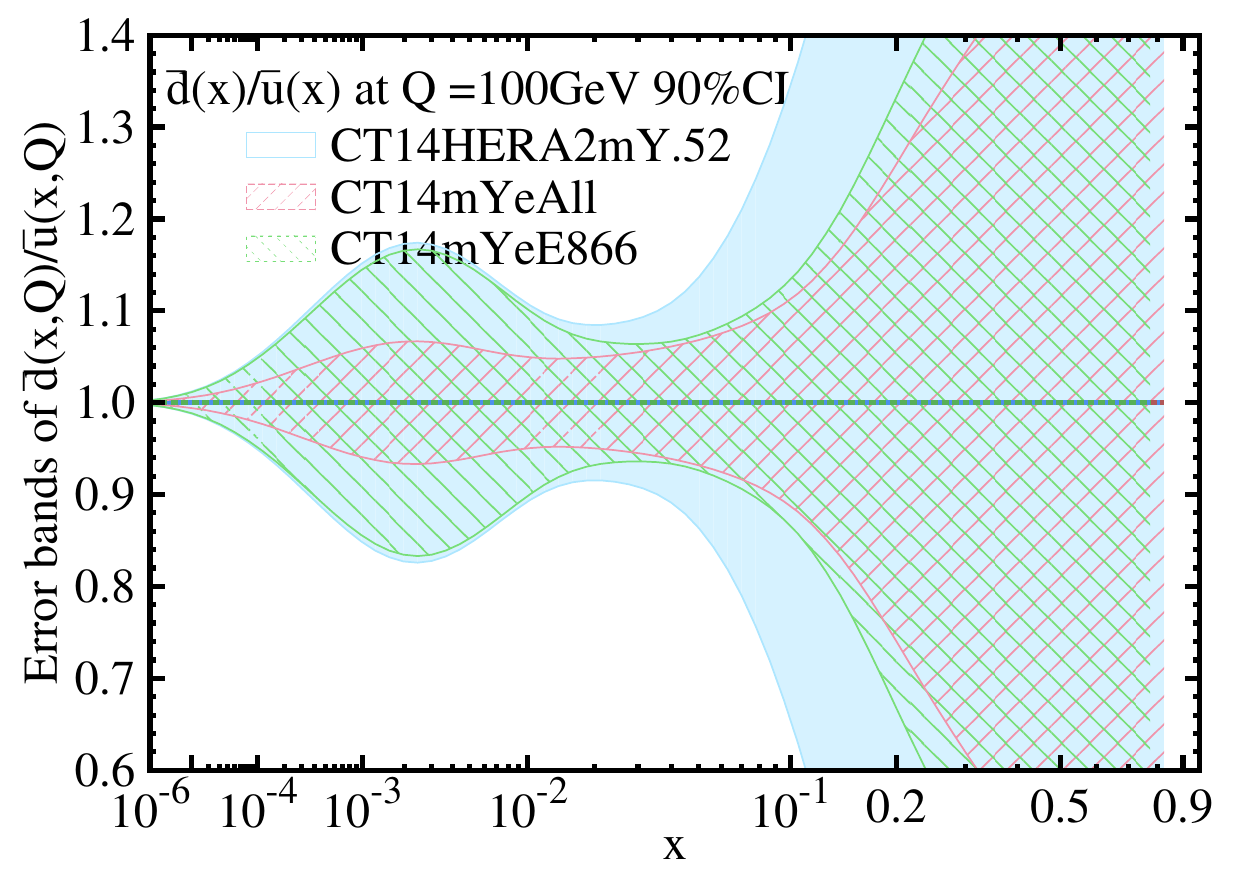}\\
	\includegraphics[width=0.45\textwidth]{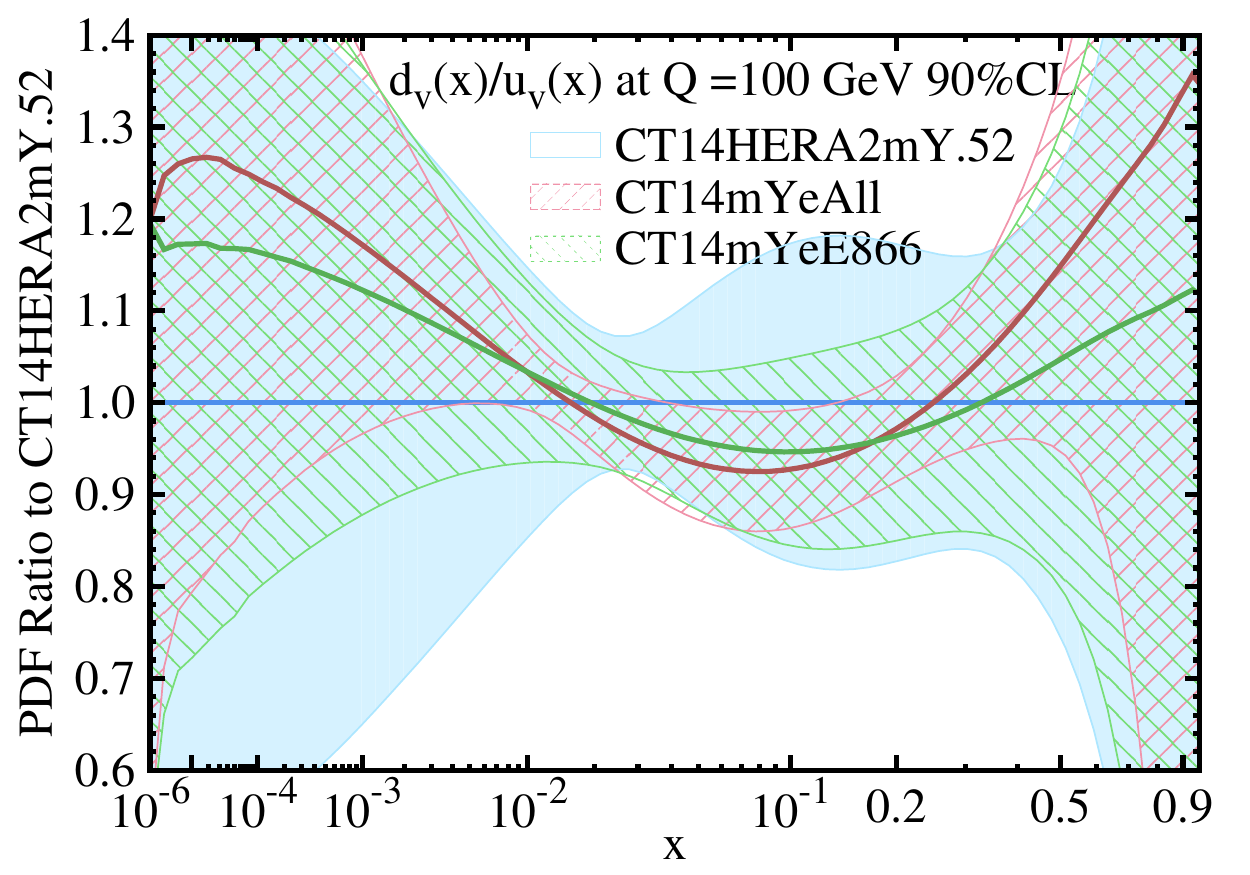}
	\includegraphics[width=0.45\textwidth]{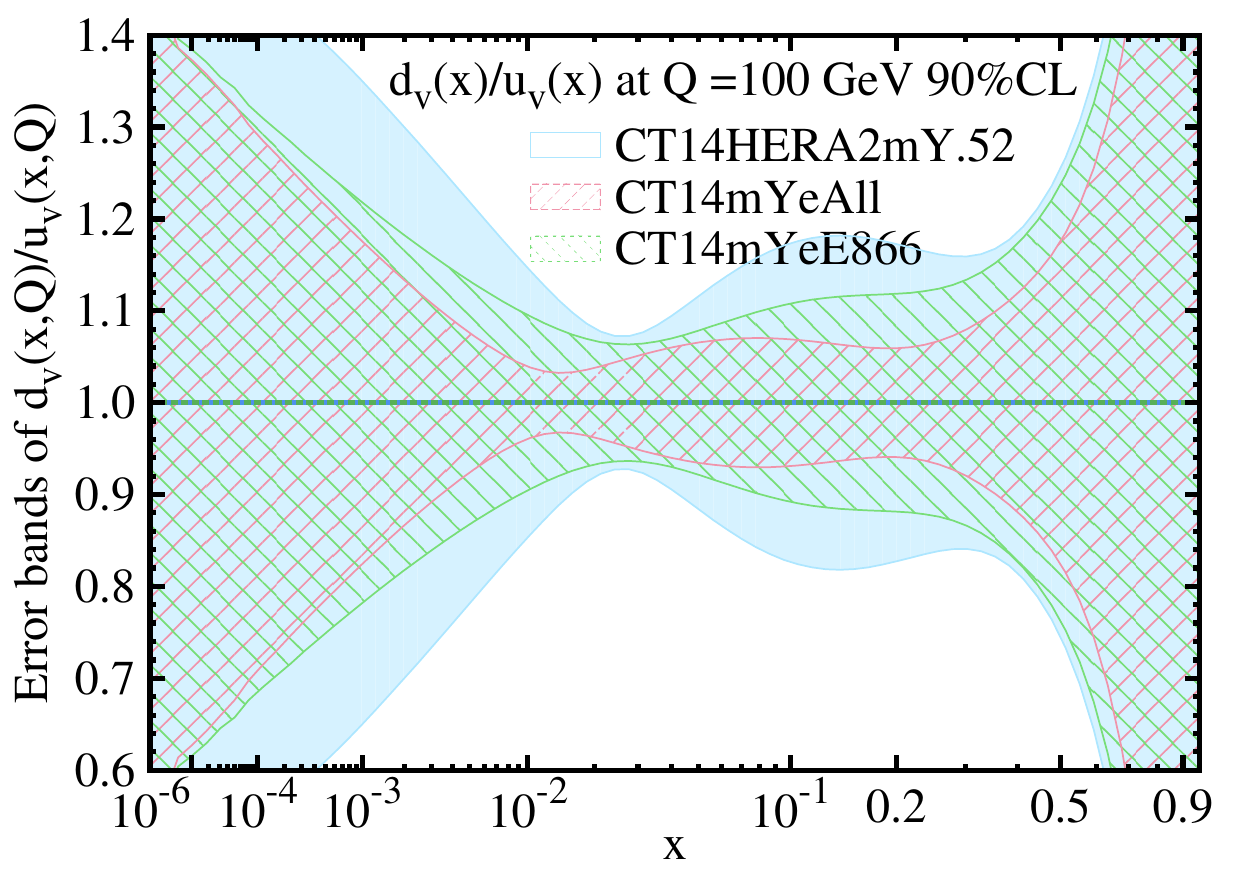}
	\caption{Updated PDF ratios $\bar d/\bar u$ and $d_v/u_v$ with their error bands, at $Q=100~\rm{GeV}$, when using \texttt{ePump} to add E866 data to CT14HERA2mY. Left panel: the PDF ratios for \texttt{ePump}-updated CT14mYeE866 and CT14mYeAll over the best-fit of the base CT14HERA2mY. Right side: the error bands relative to their own best-fit.}
	\label{Fig:DY203dbub}
\end{figure}

Another important Drell-Yan data set is the E866 data~\cite{Towell:2001nh} which measures the ratio of Drell-Yan production in proton-deuteron and proton-hydrogen collisions, $\sigma(pd)/2 \sigma(pp)$. These data impose important constraints on the PDF ratios $\bar d/\bar u$ and $d_v/u_v$ at larger values of $x$.
Using \texttt{ePump} to update CT14HERA2mY PDFs by taking the E866 data as ``new'' data, we obtained the CT14mYeE866 PDFs, which are compared to the CT14HERA2 PDFs in  Fig.~\ref{Fig:DY203dbub}.
This figure shows that at $x=0.02\sim 0.2$, the anti-quark PDF ratios, $\bar d/\bar u$, and the valence quark ratios, $d_v/u_v$, and their error bands are greatly constrained and are very close to the CT14mYeAll fit.
Comparing Figs.~\ref{DY asy ubar dbar}, \ref{DY asy du and dbarubar} and \ref{Fig:DY203dbub}, we conclude that both the lepton charge asymmetry data (from the Tevatron and the LHC) and the E866 data are needed to closely reproduce the CT14HERA2 fit result for the $\bar u$, $\bar d$, $\bar d/\bar u$ and $d_v/u_v$ PDFs.

\subsection{The impact of DIS data in the CT14HERA2 fit}\label{section: DIS }

\begin{figure}[h]
\includegraphics[width=0.45\textwidth]{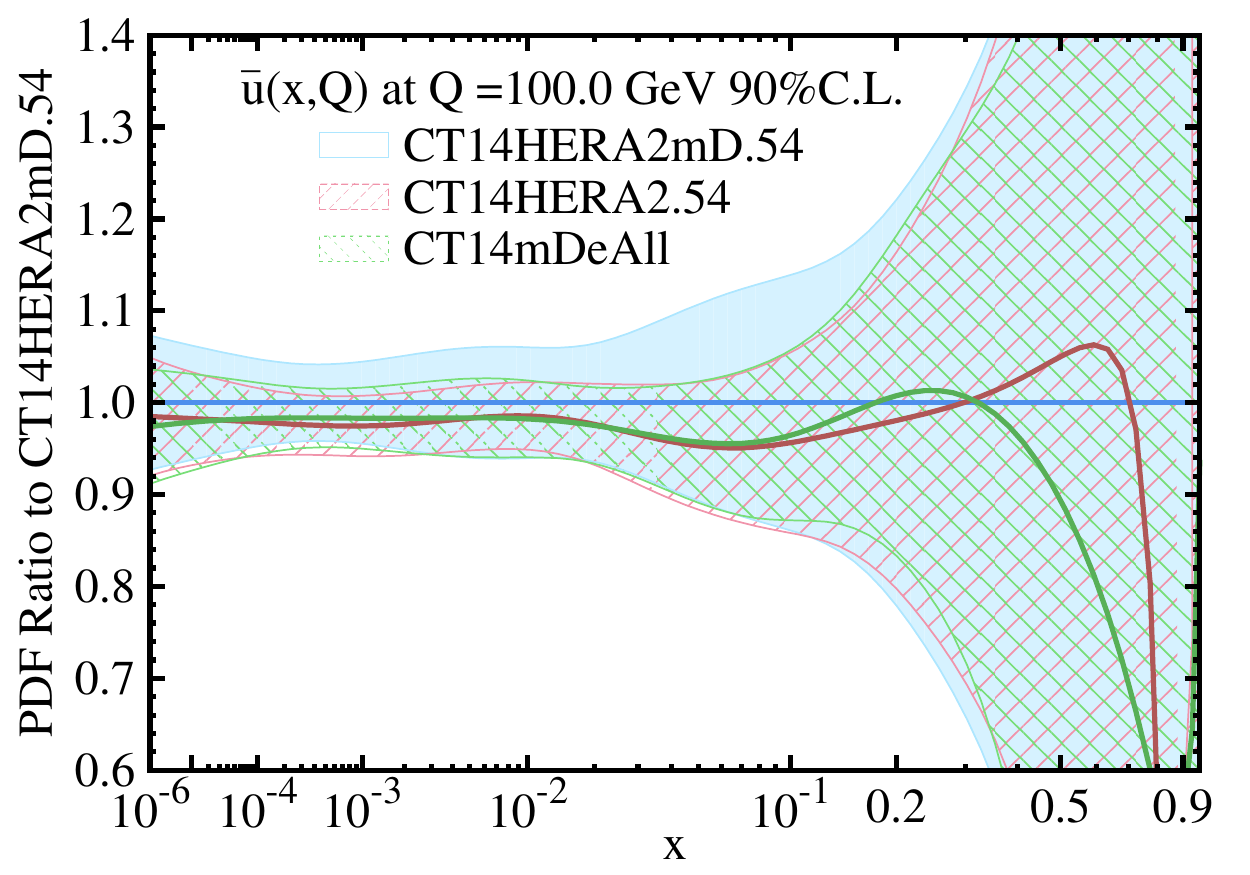}
\includegraphics[width=0.45\textwidth]{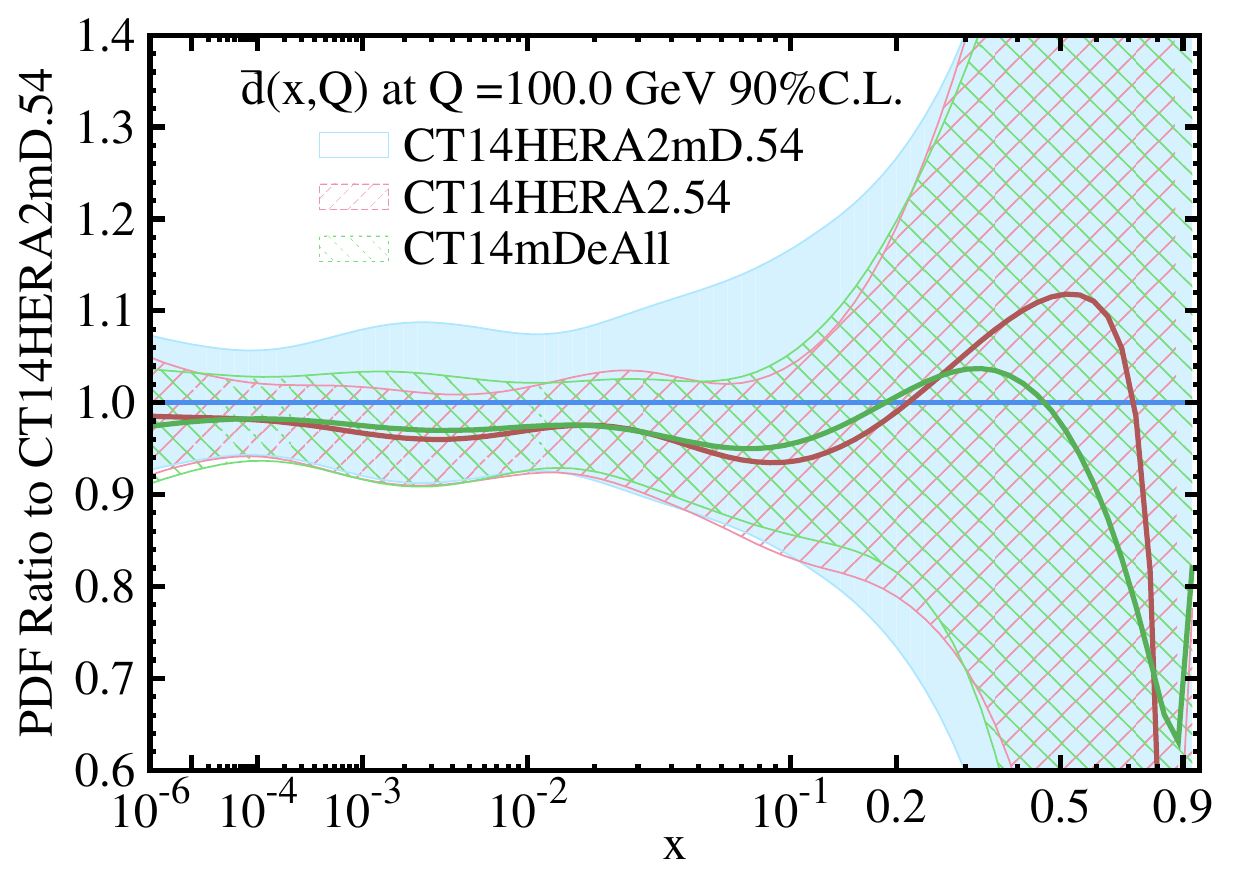}
\caption{Same as Fig.~\ref{Fig:Dud}, but for
	$\bar u$ and $\bar d$ quark PDFs.}
\label{Fig:Dudbar}
\end{figure}

\begin{figure}[h]
\includegraphics[width=0.45\textwidth]{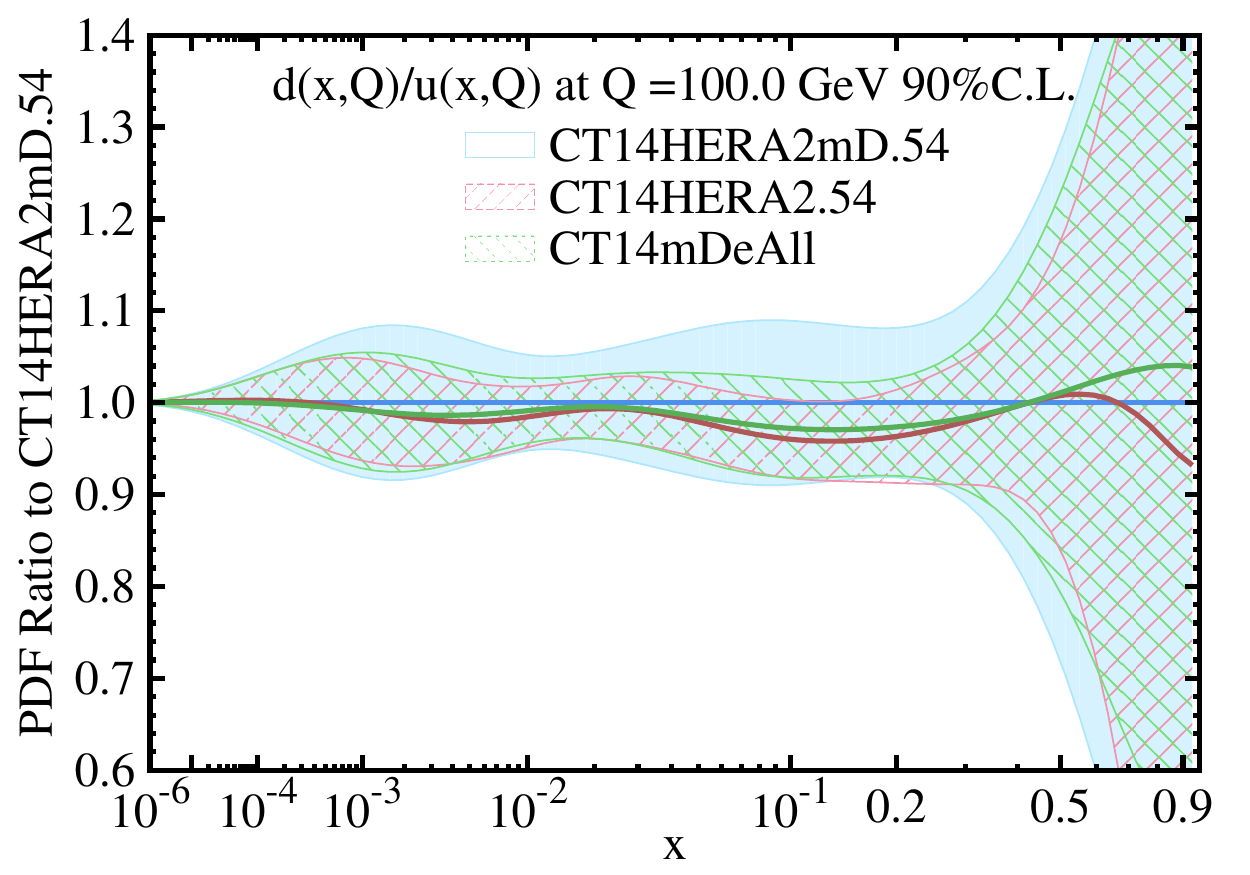}
\includegraphics[width=0.45\textwidth]{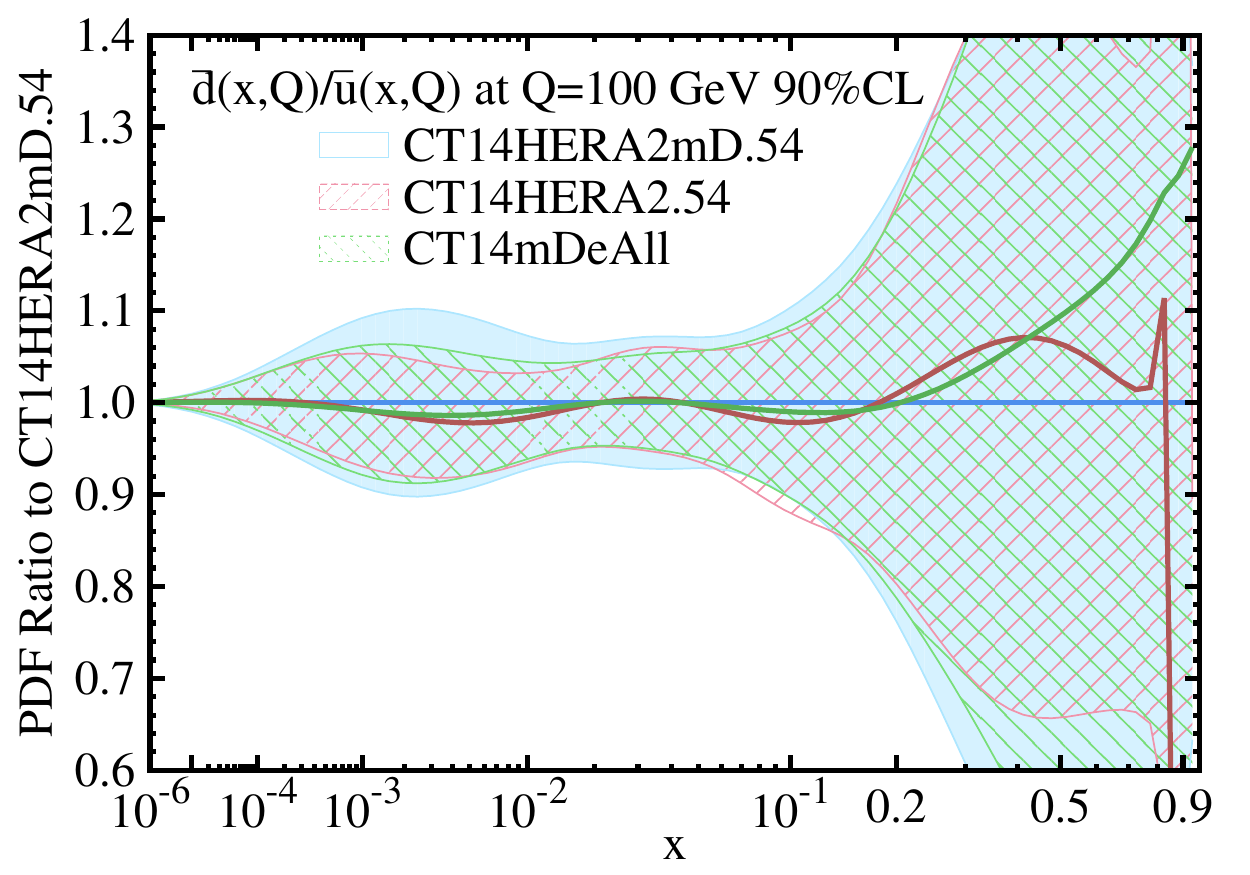}
\caption{
Same as Fig.~\ref{Fig:Dud}, but for
$d/u$ and $\bar d/\bar u$ PDF ratios.}
\label{Fig:D du dbub}
\end{figure}

\begin{figure}[h]
\includegraphics[width=0.45\textwidth]{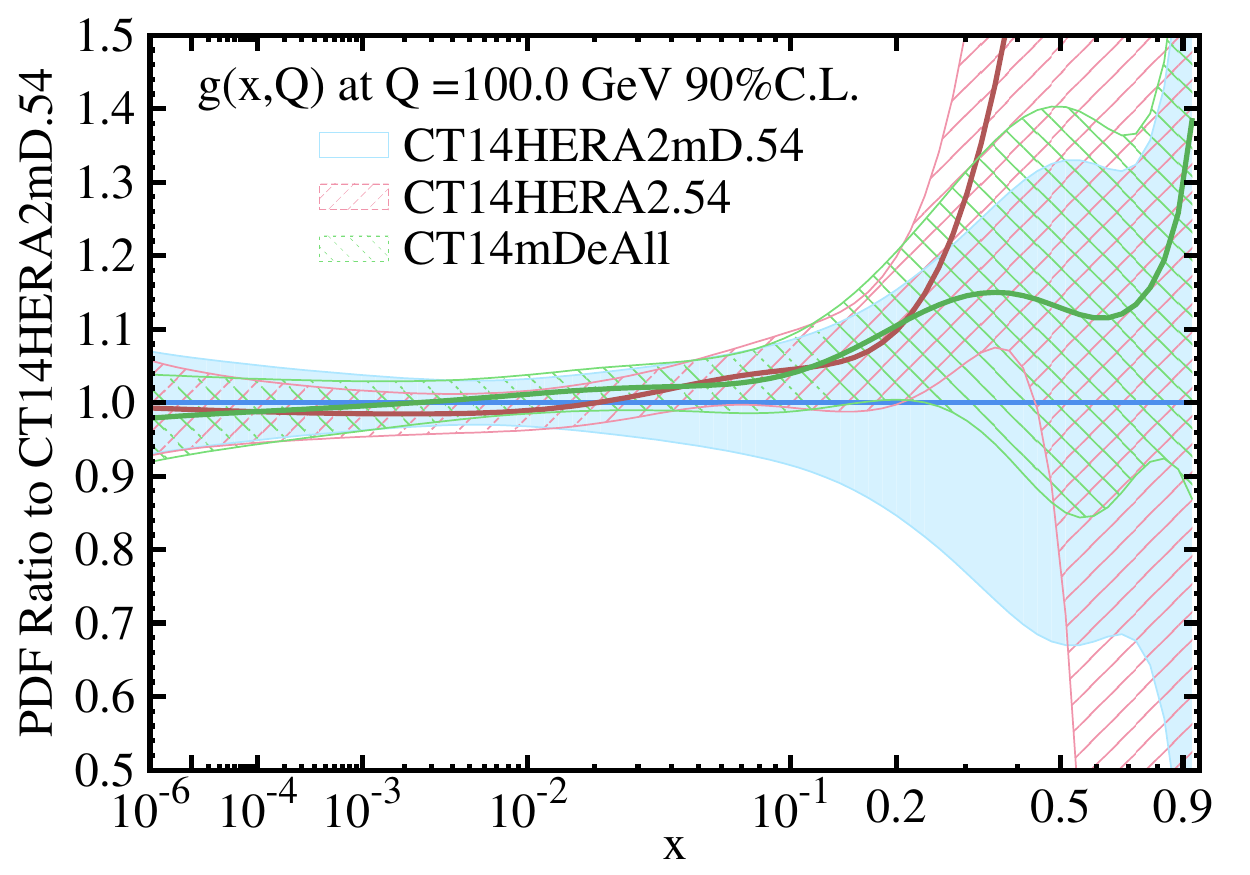}
\includegraphics[width=0.45\textwidth]{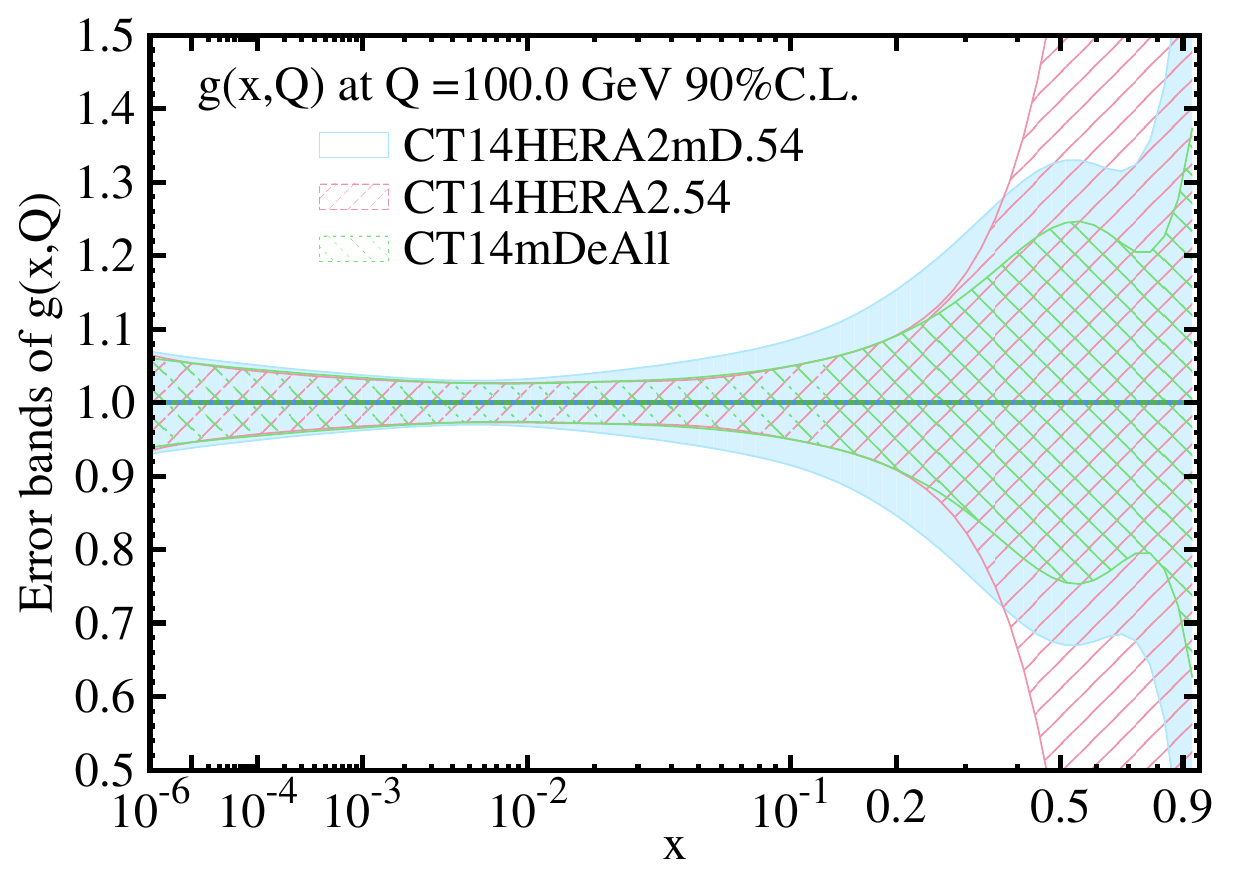}
\caption{Same as Fig.~\ref{Fig:Dud}, but for
	$g$-PDF.  }
\label{Fig:Dgc}
\end{figure}

\begin{figure}[h]
	\includegraphics[width=0.45\textwidth]{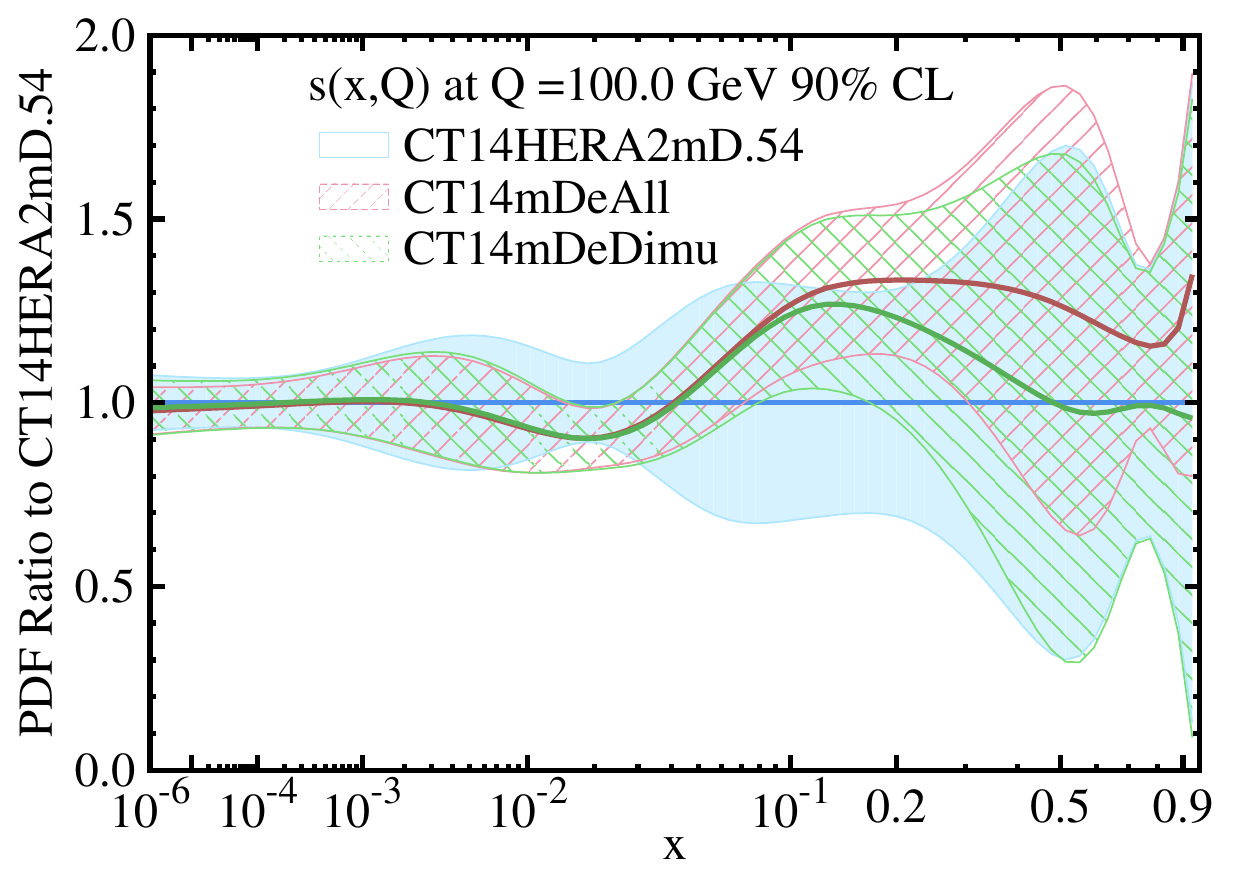}
	\includegraphics[width=0.45\textwidth]{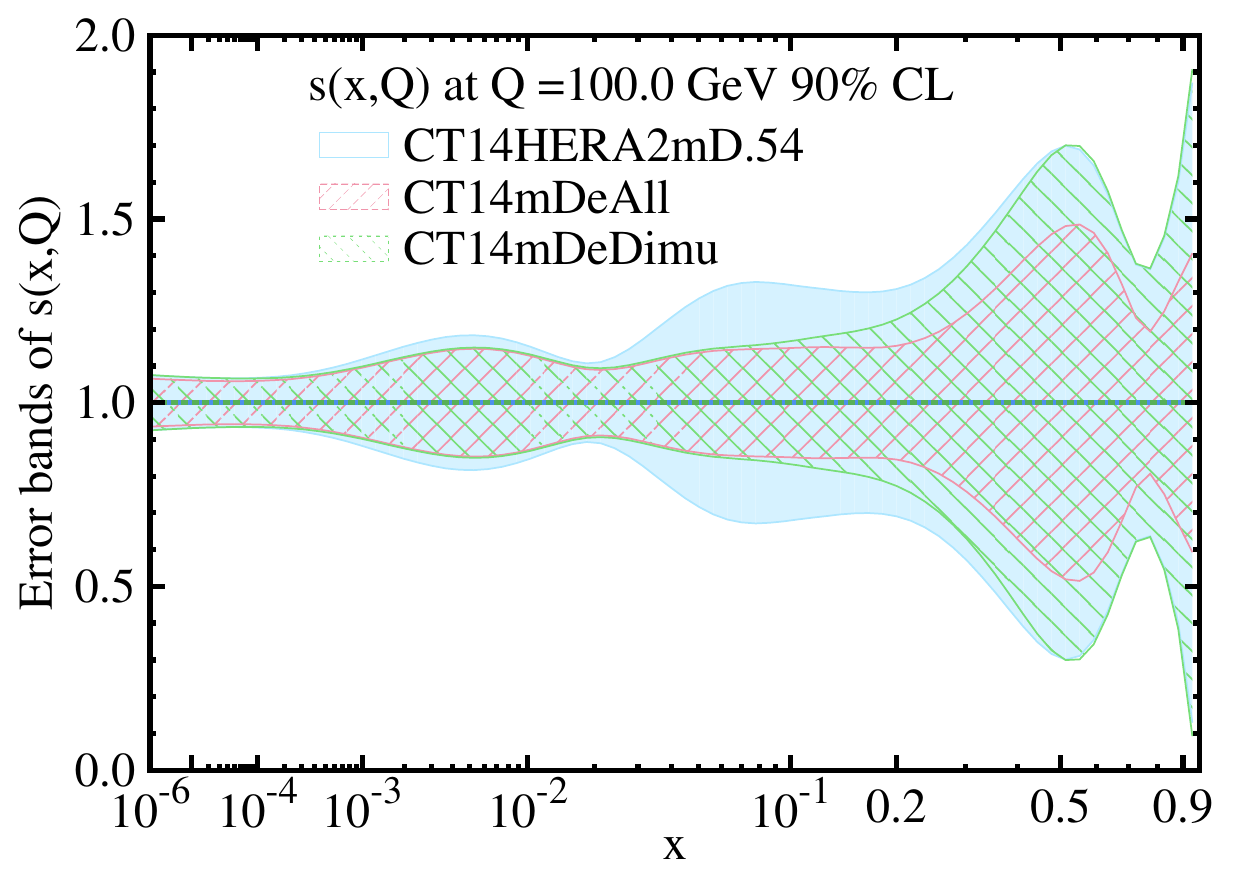}
	\caption{Comparison of \texttt{ePump}-updated $s$-PDF, at $Q=100~\rm{GeV}$.  CT14mDeDimu is obtained by adding only the DIS charged current dimuon data (NuTeV~\cite{Mason:2006qa}, and CCFR~\cite{Goncharov:2001qe}) to CT14HERA2mD with \texttt{ePump}. }
	\label{Fig:mDmmus}
\end{figure}

The DIS data sets provide important information on the $u$ and $d$ PDFs, even when the HERA run I+II combined data have already been included, as shown in Fig.~\ref{Fig:Dud}. In contrast to the Drell-Yan case, the DIS data pulls the $u$ and $d$ PDFs in the same direction. This feature also holds for the $\bar u$ and $\bar d$ PDFs, as shown in Fig.~\ref{Fig:Dudbar}.
The implication is that precision DIS data are more sensitive to the sum (or, rather, the weighted sum) than the difference or ratio between the $d$ and $u$ PDFs (or $\bar d$ and $\bar u$ PDFs).
This is because most of the precision DIS data measured the $F_2$ structure function. Only a few experiments provided precision measurements of the $F_3$ structure function, which probes the difference between $u$ and $d$ or $\bar u$ and $\bar d$.
In Fig.~\ref{Fig:D du dbub}, we show the impact of DIS data (excluding HERA I+II) on $d/u$ and $\bar d/\bar u$ in the CT14HERA2 fit, which is seen to be relatively small.

The impact of DIS data on the $g$-PDF is shown in Fig.~\ref{Fig:Dgc}. It is interesting to note that the DIS data prefer a harder gluon in the $x > 0.2$ region. This is opposite to the effect of the jet data, which prefer a softer gluon at $x>0.2$, cf. Fig.~\ref{Fig:Jg}.
The tension between these two kinds of data on the  $g$-PDF can be seen by noting that the error band of the true global fit CT14HERA2 is wider than that obtained by the \texttt{ePump} updating, cf. Fig.~\ref{Fig:Dgc}.

\begin{table}[tb]
\begin{tabular}{|l|lr|l|}
\hline
\textbf{ID}  & \textbf{Experimental data set}  & & Most prominent effects \tabularnewline
\hline
\hline
101 & BCDMS $F_{2}^{p}$ & \cite{Benvenuti:1989rh}                &  Reduce $g$ and $u_v$ uncertainties at $x: 0.01 \sim 0.7$ \tabularnewline
\hline
102 & BCDMS $F_{2}^{d}$ & \cite{Benvenuti:1989fm}                &  Reduce $g$ uncertainty at $x: 0.01 \sim 0.7$ and $u_v,~d_v$ at $x \sim 0.2$\tabularnewline
\hline
104 & NMC $F_{2}^{d}/F_{2}^{p}$ & \cite{Arneodo:1996qe}          & Reduce $d/u$, $\bar d/\bar u$ and $d_v/u_v$ uncertainties for all $x$  \tabularnewline
\hline
108 & CDHSW $F_{2}^{p}$ & \cite{Berge:1989hr}                    & Almost all the PDFs \tabularnewline
\hline
109 & CDHSW $F_{3}^{p}$ & \cite{Berge:1989hr}                    & $u_v$ and $d_v$ central fits for all $x$ \tabularnewline
\hline
110 & CCFR $F_{2}^{p}$ & \cite{Yang:2000ju}                      & Almost all the PDFs \tabularnewline
\hline
111 & CCFR $xF_{3}^{p}$ & \cite{Seligman:1997mc}                 & $u_v$ and $d_v$ central fits for all $x$ \tabularnewline
\hline
124 & NuTeV $\nu\mu\mu$ SIDIS & \cite{Mason:2006qa}              & $s$-PDF at $x: 0.01\sim 0.4$ \tabularnewline
\hline
125 & NuTeV $\bar\nu \mu\mu$ SIDIS & \cite{Mason:2006qa}         & $s$-PDF at $x: 0.01\sim 0.4$  \tabularnewline
\hline
126 & CCFR $\nu\mu\mu$ SIDIS & \cite{Goncharov:2001qe}           & Reduce $s$-PDF uncertainty at $x \sim 0.1$  \tabularnewline
\hline
127& CCFR  $\bar\nu \mu\mu$ SIDIS & \cite{Goncharov:2001qe}      & Reduce $s$-PDF uncertainty slightly at $x \sim 0.1$  \tabularnewline
\hline
145 & H1 $\sigma_{r}^{b}$ & \cite{Aktas:2004az}                  & Not much effect  \tabularnewline
\hline
147 & Combined HERA charm production & \cite{Abramowicz:1900rp}  & Not much effect   \tabularnewline
\hline
169 & H1 $F_{L}$ & \cite{Collaboration:2010ry}                   & Not much effect  \tabularnewline
\hline
\end{tabular}
\caption{Impact of individual DIS data sets in CT14HERA2 on the PDFs. For each data set, only its most prominent effects are listed. The base PDF set used for this study is CT14HERA2mD. Therefore, the effects refer to the ``net" impact when each individual data set is added, one at a time, to CT14HERA2mD.}
\label{tab:EXP_1}
\end{table}

\begin{table}[tb]
\begin{tabular}{|l|lr|l|}
\hline
\textbf{ID}  & \textbf{Experimental data set} &  & Most prominent effects  \tabularnewline
\hline
\hline
201 & E605 Drell-Yan process & \cite{Moreno:1990sf}                                    & Pull down $\bar u$ and $\bar d$ PDFs at $x\gtrsim 0.1$ \tabularnewline
\hline
203 & E866 Drell-Yan process, $\sigma_{pd}/(2\sigma_{pp})$ & \cite{Towell:2001nh}                               & $\bar d/\bar u$ and $d_v/u_v$ at $x: 0.01\sim 0.2$ \tabularnewline
\hline
204 & E866 Drell-Yan process, $Q^3 d^2\sigma_{pp}/(dQ dx_F)$ & \cite{Webb:2003ps}                                      & $\bar u$ at $x\gtrsim 0.04$ and $u_v$ at $x: 10^{-3}\sim0.4$  \tabularnewline
\hline
225 & CDF Run-1 electron  $A_{ch}$ & \cite{Abe:1996us}                                &  Reduce errors of $d/u$, $\bar d/\bar u$  at $x\sim 0.1$ and $d_v/u_v$ at all x \tabularnewline
\hline
227 & CDF Run-2 electron $A_{ch}$  & \cite{Acosta:2005ud}                               & Reduce $d_v/u_v$ by a little at $x\lesssim0.3$ \tabularnewline
\hline
234 & D\O~ Run-2 muon $A_{ch}$  & \cite{Abazov:2007pm}                                 & Reduce $d/u$, $\bar d/\bar u$, and $d_v/u_v$ \tabularnewline
&&& uncertainties at $x: 10^{-3}\sim 0.3$ \tabularnewline
\hline
240 & LHCb 7 TeV $35\mbox{ pb}^{-1}$ $W/Z$ $d\sigma/dy_{\ell}$ & \cite{Aaij:2012vn}    & Not much effect \tabularnewline
\hline
241 & LHCb 7 TeV $35\mbox{ pb}^{-1}$ $A_{ch}$ & \cite{Aaij:2012vn}  & Reduce $d/u$, $\bar d/\bar u$, $d_v/u_v$ uncertainties  \tabularnewline
       & & & slightly at $x: 10^{-4}\sim 10^{-2}$ \tabularnewline
\hline
260 & D\O~ Run-2 $Z$ rapidity & \cite{Abazov:2006gs}                                   &  Not much effect  \tabularnewline
\hline
261 & CDF Run-2 $Z$ rapidity & \cite{Aaltonen:2010zza}                                 &  Not much effect  \tabularnewline
\hline
266 & CMS 7 TeV $4.7\mbox{ fb}^{-1}$, muon $A_{ch}$  GeV& \cite{Chatrchyan:2013mza}         & Almost all the quark PDFs  \tabularnewline
\hline
267 & CMS 7 TeV $840\mbox{ pb}^{-1}$, electron $A_{ch}$  GeV & \cite{Chatrchyan:2012xt}     & Almost all the quark PDFs \tabularnewline
\hline
268 & ATLAS 7 TeV $35\mbox{ pb}^{-1}$ $W/Z$ cross sec., $A_{ch}$ & \cite{Aad:2011dm}   & Almost all the quark PDFs  \tabularnewline
\hline
281 & D\O~ Run-2 $9.7 \mbox{ fb}^{-1}$ electron $A_{ch}$  & \cite{D0:2014kma}               & Reduce $d/u$, $\bar d/\bar u$ and $d_v/u_v$ \tabularnewline
&&& uncertainties at $x: 10^{-4}\sim 0.3$ \tabularnewline
\hline
\end{tabular}
\caption{Same as Table~\ref{tab:EXP_1}, showing experimental data sets on Drell-Yan processes. The base PDF set for this study is CT14HERA2mY.  Therefore, the effects refer to the ``net" impact when each individual data set is added, one at a time, to CT14HERA2mY.}
\label{tab:EXP_2}
\end{table}

\begin{table}[tb]
\begin{tabular}{|l|lr|l|}
\hline
\textbf{ID}  & \textbf{Experimental data set} &  & Most prominent effects  \tabularnewline
\hline
\hline
504 & CDF Run-2 inclusive jet production & \cite{Aaltonen:2008eq}                      & $g$-PDF at $x: 0.02\sim 0.5$ \tabularnewline
\hline
514 & D\O~ Run-2 inclusive jet production & \cite{Abazov:2008ae}                       &  $g$-PDF at $x: 0.02\sim 0.5$ \tabularnewline
\hline
535 & ATLAS 7 TeV $35\mbox{ pb}^{-1}$ incl. jet production & \cite{Aad:2011fc}         & $g$-PDF at $x: 0.02\sim 0.5$  \tabularnewline
\hline
538 & CMS 7 TeV $5\mbox{ fb}^{-1}$ incl. jet production  & \cite{Chatrchyan:2012bja}   & $g$-PDF at $x: 0.02\sim 0.5$ \tabularnewline
\hline
\end{tabular}
\caption{Same as Table~\ref{tab:EXP_1}, showing experimental data sets on inclusive jet production. The base PDF set for this study is CT14HERA2mJ.  Therefore, the effects refer to the ``net" impact when each individual data set is added, one at a time, to CT14HERA2mJ.}
\label{tab:EXP_3}
\end{table}

\begin{figure}[h]
	\includegraphics[width=0.45\textwidth]{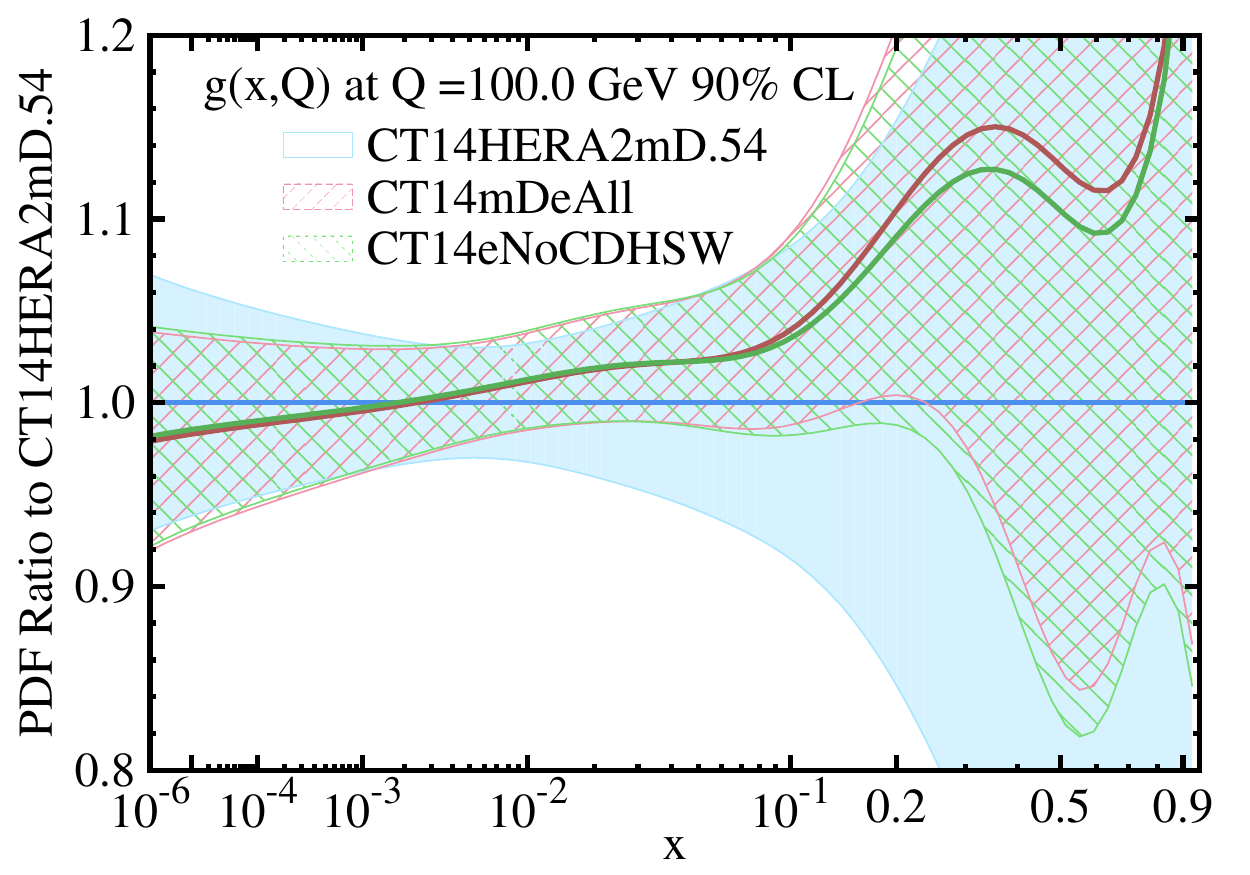}
	\includegraphics[width=0.45\textwidth]{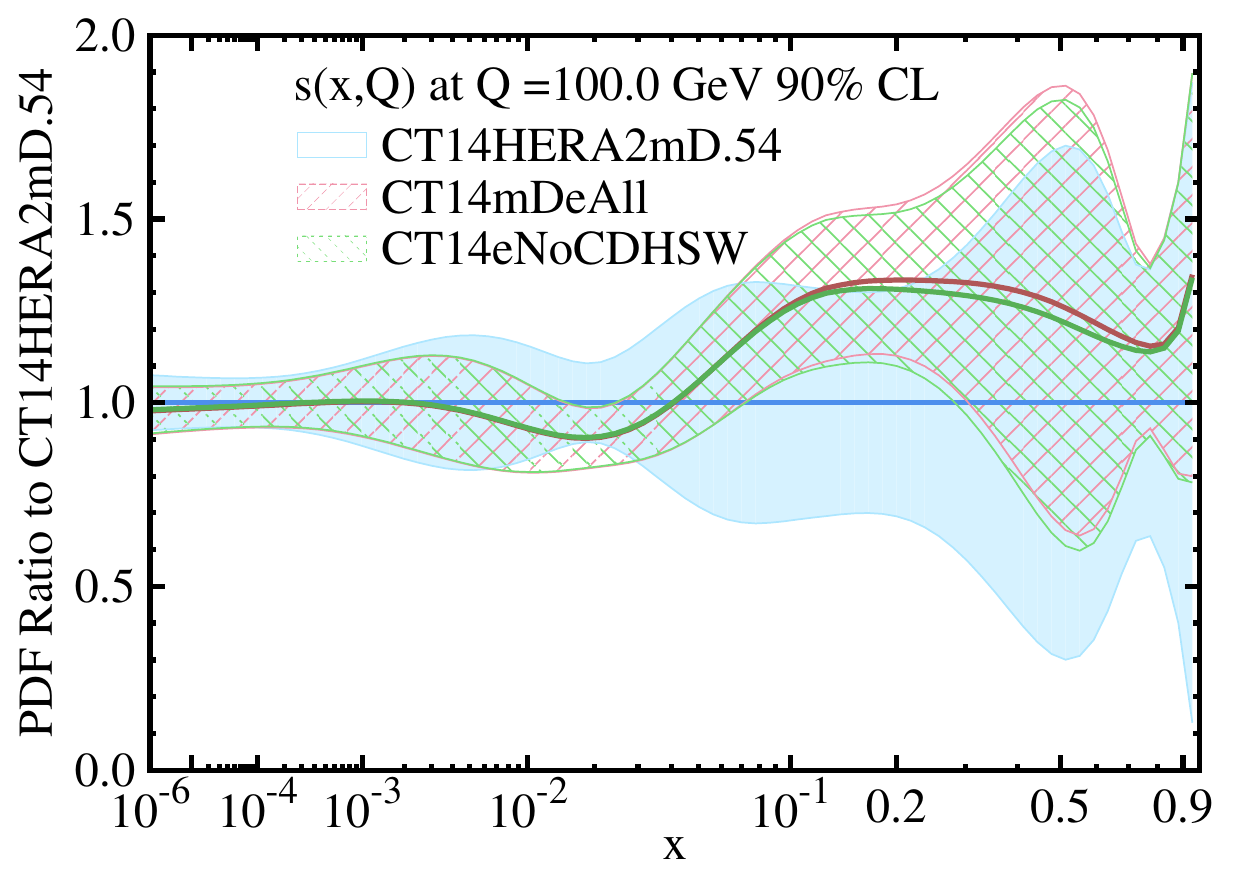}\\
	\includegraphics[width=0.45\textwidth]{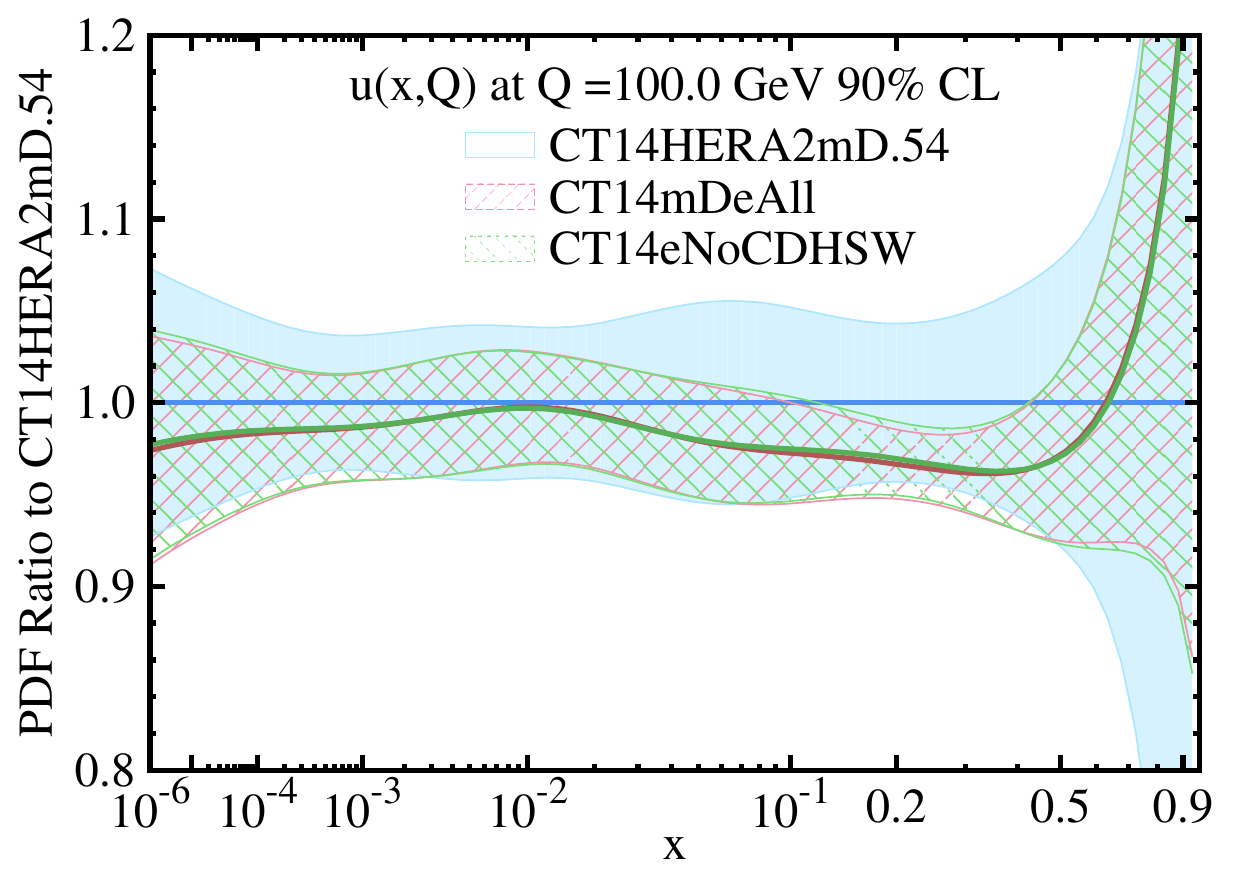}
	\includegraphics[width=0.45\textwidth]{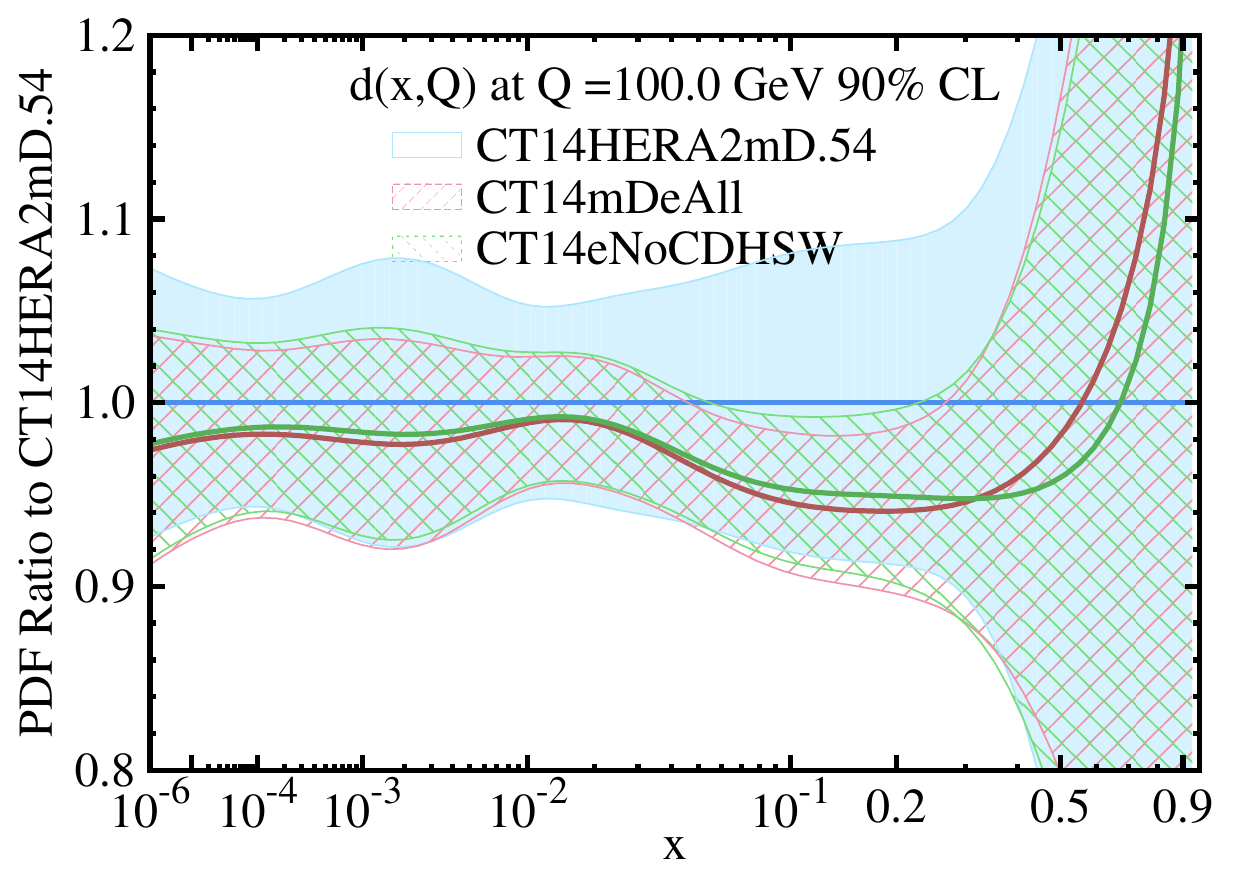}\\
	\includegraphics[width=0.45\textwidth]{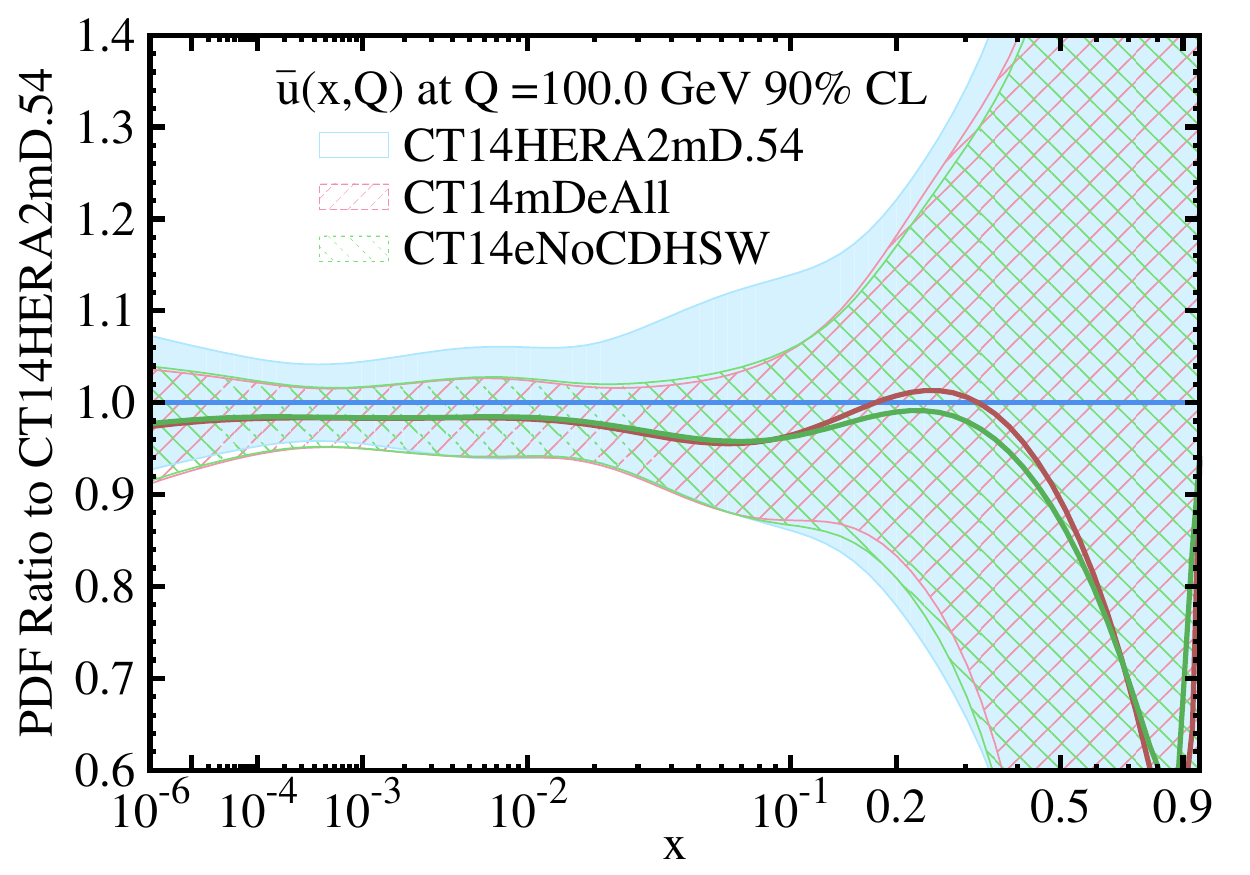}
	\includegraphics[width=0.45\textwidth]{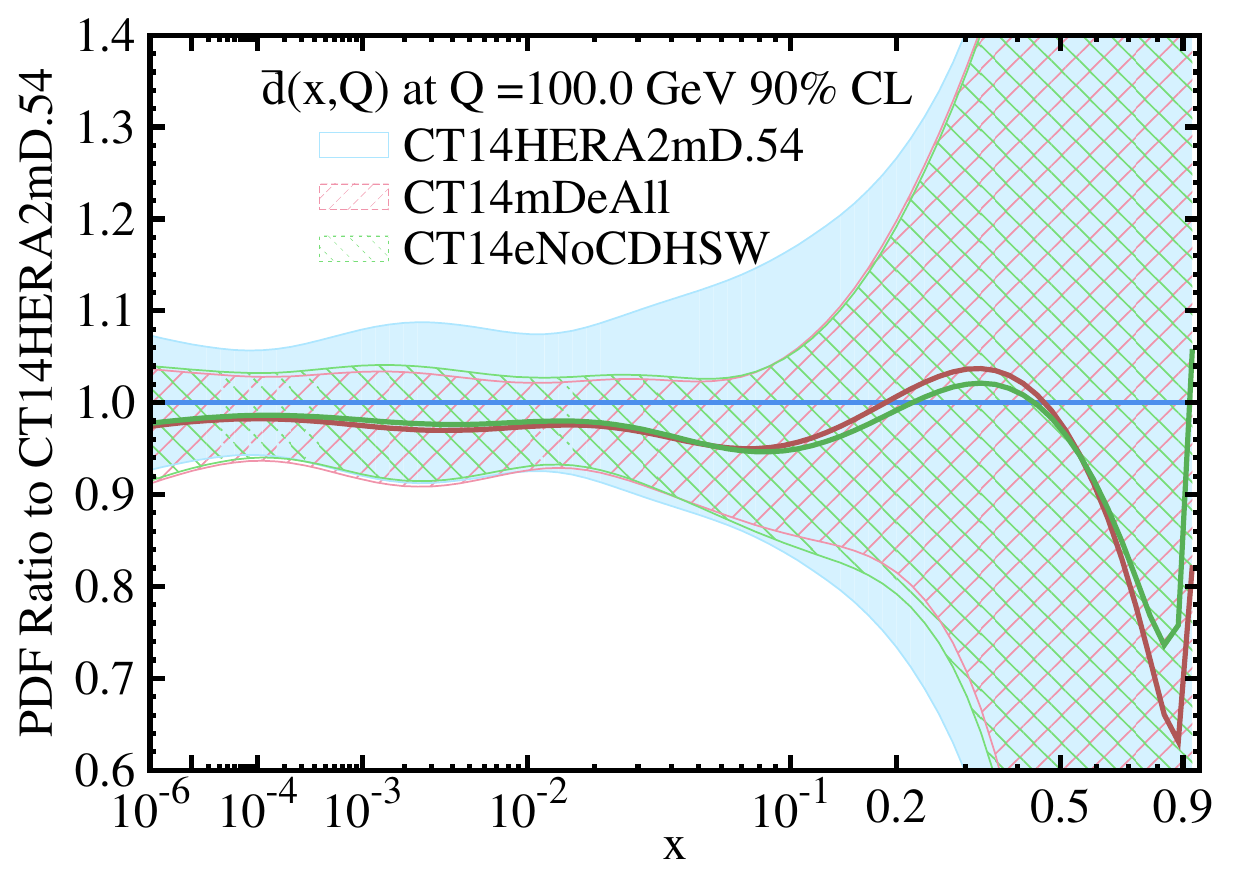}
	\caption{
		Comparison of \texttt{ePump}-updated PDFs, at $Q=100~\rm{GeV}$. CT14eNoCDHSW is obtained by taking all but the CDHSW DIS data as ``new'' data, and CT14HERA2mD as the base.
}
	\label{Fig:NoCDHSW}
\end{figure}

DIS data are also expected to constrain the strange quark ($s$) PDF.
For most of the DIS data, however, the contribution to the $s$-PDF is much smaller than to the $u$ and $d$ PDFs, except for the DIS charged current dimuon production experiments~\cite{Mason:2006qa}\cite{Goncharov:2001qe}, where the strange quark gives the dominant contribution.
Thus, it is interesting to see how much the dimuon data alone can constrain the $s$-PDF among all the DIS data. This can be quickly investigated with \texttt{ePump}.
In Fig.~\ref{Fig:mDmmus}, we compare the updated $s$-PDF, by including only these dimuon data, to the one with the full DIS data included.
Although dimuon data sets are not the only ones responsible for determining the $s$-PDF, they do fully constrain the $s$-PDF over a wide $x$ region, from $10^{-4} $ to $10^{-1}$, for both the central PDF and its error band.
Furthermore we will see that the effects mostly come from NuTeV dimuon data~\cite{Mason:2006qa}, and only a little from CCFR dimuon data~\cite{Goncharov:2001qe}.

As before, we can use the \texttt{ePump} updating code to investigate the effect from individual DIS data set on the CT14HERA2 PDFs. We find the bulk of the DIS data contributions comes from
CCFR $F_{2}^{p}$~\cite{Yang:2000ju},  $xF_{3}^{p}$~\cite{Seligman:1997mc},
CDHSW $F_{2}^{p}$ and $F_{3}^{p}$~\cite{Berge:1989hr},
NuTeV $\bar\nu \mu\mu$ SIDIS, and
NuTeV $\nu\mu\mu$ SIDIS~\cite{Mason:2006qa}
data sets.
As discussed above, NuTeV dimuon data
are almost solely responsible for constraints on the $s$-PDF, with
the stronger constraint coming from NuTeV $\bar\nu \mu\mu$ SIDIS data.
CCFR and CDHSW $F_{2}^{p}$ data produce the biggest changes to $u$ and $d$, and together with CCFR and CDHSW $F_{3}^{p}$, they give the biggest change to $\bar u$ and $\bar d$.
However, the error band on $d/u$ is not reduced until we also add in NMC $F_{2}^{d}/F_{2}^{p}$ data~\cite{Arneodo:1996qe}.
Although the other DIS data do not have as large of an impact as the above mentioned 7 data sets, they are still responsible for some ``fine structure'' of the PDFs.
For example, BCDMS $F_2^p$~\cite{Benvenuti:1989rh} and  $F_2^d$~\cite{Benvenuti:1989fm} data measured the structure functions $F_2$ of protons and deuterons, and cover the large $x$ region $x\lesssim 0.8$.
Hence, these two data sets constrain $u_v$ and $d_v$ quarks and $g$-PDFs in the large $x$ region.
Due to limited space, we shall not show all the corresponding plots in this paper, but instead will post them on the website of the \texttt{ePump} project~\cite{ePumpwebsite}.
For completeness, we summarize our findings in
Tables~\ref{tab:EXP_1}, \ref{tab:EXP_2} and \ref{tab:EXP_3}, where we  list the most prominent effects of each data set in CT14HERA2.
We note that such a study by \texttt{ePump} must start from a base set of global-fit PDFs, and the effects of the data listed in the tables refer to their ``net" impact when added, one at a time, to the particular base PDF set. For example, for the DIS data in Table~\ref{tab:EXP_1}, the base PDF set is CT14HERA2mD, which includes the Drell-Yan and jet data and the HERA Run I+II data~\cite{Abramowicz:2015mha}. Thus it may happen that the effects of some DIS data are similar to those of the HERA Run I+II data, with the result
that an individual data set, such as H1 $\sigma_{r}^{b}$ data~\cite{Aktas:2004az}, appears to have little or no impact to the updated central set PDFs.

Finally, before leaving this section, we
would like to investigate the impact of the CDHSW $F_2$ and $F_3$ data in the CT14HERA2 fit. It has long been argued that these data sets were not analyzed properly and therefore should not be used in a global fit~\cite{Barone:1999yv}.
Therefore, one may wonder how the CT14HERA2 PDFs would change if we exclude these two CDHSW data sets from the original CT14HERA2 fit.
Instead of redoing the whole global fit, we use \texttt{ePump} to quickly answer this question.
The procedure is to use \texttt{ePump} to update the CT14HERA2mD PDFs by taking all but the CDHSW DIS data as ``new''data. The resulting PDFs, CT14eNoCDHSW, are shown in
Fig.~\ref{Fig:NoCDHSW} to be compared with
CT14mDeAll, in which all DIS data were included.
From these figures,  we immediately note that they do not differ very much.
This is because the CCFR $F_2$ and $F_3$ data have similar effects to the CDHSW data on CT14HERA2 PDFs.
Furthermore, from Fig.~\ref{Fig:NoCDHSW} we can extract the consequences of removing the CDHSW data from the CT14HERA2 fit. Removing the CDHSW data leads to slightly softer $g$, $s$, $\bar u$, $\bar d$ PDFs at $x > 0.1$, and harder $d$ and $u$ PDFs at $x > 0.03$. Recall that the Tevatron and LHC jet data also prefer a softer $g$-PDF at $x > 0.2$, cf. Fig.~\ref{Jg1}. Thus, removing the CDHSW data is more consistent with jet data.

\section{Using \texttt{ePump} to study the impact of new data}\label{section:new data}

In the previous sections, we have validated \texttt{ePump} against the CT14HERA2 global fit by updating the PDFs with some subset of the CT14HERA2 data sets.
We have also used \texttt{ePump} to investigate the impact of individual data sets on the CT14HERA2 PDFs.
In this section, we will use \texttt{ePump} to study the potential impact of some new LHC data on improving the CT14HERA2 PDFs. An example was already given in
\cite{Schmidt:2018hvu}, where we analyzed the impact of the CMS inclusive jet production data at $\sqrt S = 8~{\rm TeV}$~\cite{Khachatryan:2016mlc}. Here, we consider two more examples of the new LHC data: the LHC 8 TeV $t \bar t$ differential cross section data, and the ATLAS 7 TeV $W^\pm$ and $Z$ data.

\subsection{$t\bar t$ data at the LHC}

\begin{table}[h]
\begin{center}
\begin{tabular}{| c | c | c | c |}
\hline
ID & data & $d^0$ of CT14HERA2 & $d^0$ of CT14HERA2mJ \\
\hline
561 & CMS 8 TeV Normalized $d\sigma/\sigma dp^t_T$ & 0.14 & 0.27 \\
\hline
562 & CMS 8 TeV Normalized $d\sigma/\sigma dy_t$ & 0.06 & 0.23 \\
\hline
563 & CMS 8 TeV Normalized $d\sigma/\sigma dm_{t\bar t}$ & 0.17 & 0.32 \\
\hline
564 & CMS 8 TeV Normalized $d\sigma/\sigma dy_{\ttb}$ & 0.25 & 0.64 \\
\hline
565 & ATLAS 8 TeV Absolute $d\sigma/dp^t_T$ & 0.01 & 0.02 \\
\hline
566 & ATLAS 8 TeV Absolute $d\sigma/d|y_t|$ & 0.09 & 0.31 \\
\hline
567 & ATLAS 8 TeV Absolute $d\sigma/dm_{\ttb}$ & 0.03 & 0.01 \\
\hline
568 & ATLAS 8TeV Absolute $d\sigma/d|y_{\ttb}|$ & 0.17 & 0.47 \\
\hline
\end{tabular}
\caption{List of $\ttb$ data sets from CMS~\cite{Khachatryan:2015oqa} and ALTAS~\cite{Aad:2015mbv}. $d^0$ is the length of shift of the best-fit point in parameter space, as explained in the text. The third column is $d^0$ values for each $\ttb$ data set when added to CT14HERA2 with \texttt{ePump} and fourth column to CT14HERA2mJ.
}
\label{table:ttblist}
\end{center}
\end{table}

\begin{table}[h]
\begin{center}
\begin{tabular}{| c | c | c | c |}
\hline
ID & data & $\chi^2/N$ of CT14HERA2 & $\chi^2/N$ of CT14HERA2mJ \\
\hline
561 & CMS 8 TeV Normalized $d\sigma/\sigma dp^t_T$           & 4.35  & 5.68  \\
\hline
562 & CMS 8 TeV Normalized $d\sigma/\sigma dy_t$              & 2.63   & 3.06  \\
\hline
563 & CMS 8 TeV Normalized $d\sigma/\sigma dm_{t\bar t}$      & 6.07  & 7.71  \\
\hline
564 & CMS 8 TeV Normalized $d\sigma/\sigma dy_{\ttb}$         & 2.19  & 3.24  \\
\hline
565 & ATLAS 8 TeV Absolute $d\sigma/dp^t_T$                  & 0.49   & 0.52  \\
\hline
566 & ATLAS 8 TeV Absolute $d\sigma/d|y_t|$                   & 2.89   & 6.09 \\
\hline
567 & ATLAS 8 TeV Absolute $d\sigma/dm_{\ttb}$                & 1.19   & 1.00  \\
\hline
568 & ATLAS 8TeV Absolute $d\sigma/d|y_{\ttb}|$               & 5.09   & 9.10  \\
\hline
\end{tabular}
\caption{List of $\ttb$ data sets from CMS~\cite{Khachatryan:2015oqa} and ALTAS~\cite{Aad:2015mbv}. $\chi^2$ per data point $\chi^2/N$ is shown in the third column for CT14HERA2 and fourth column for CT14HERA2mJ. Both are obtained by adding the $t\bar t$ data, one at a time, using \texttt{ePump}.
}
\label{table:ttblist_chi2}
\end{center}
\end{table}

\begin{figure}[h]
\includegraphics[width=0.45\textwidth]{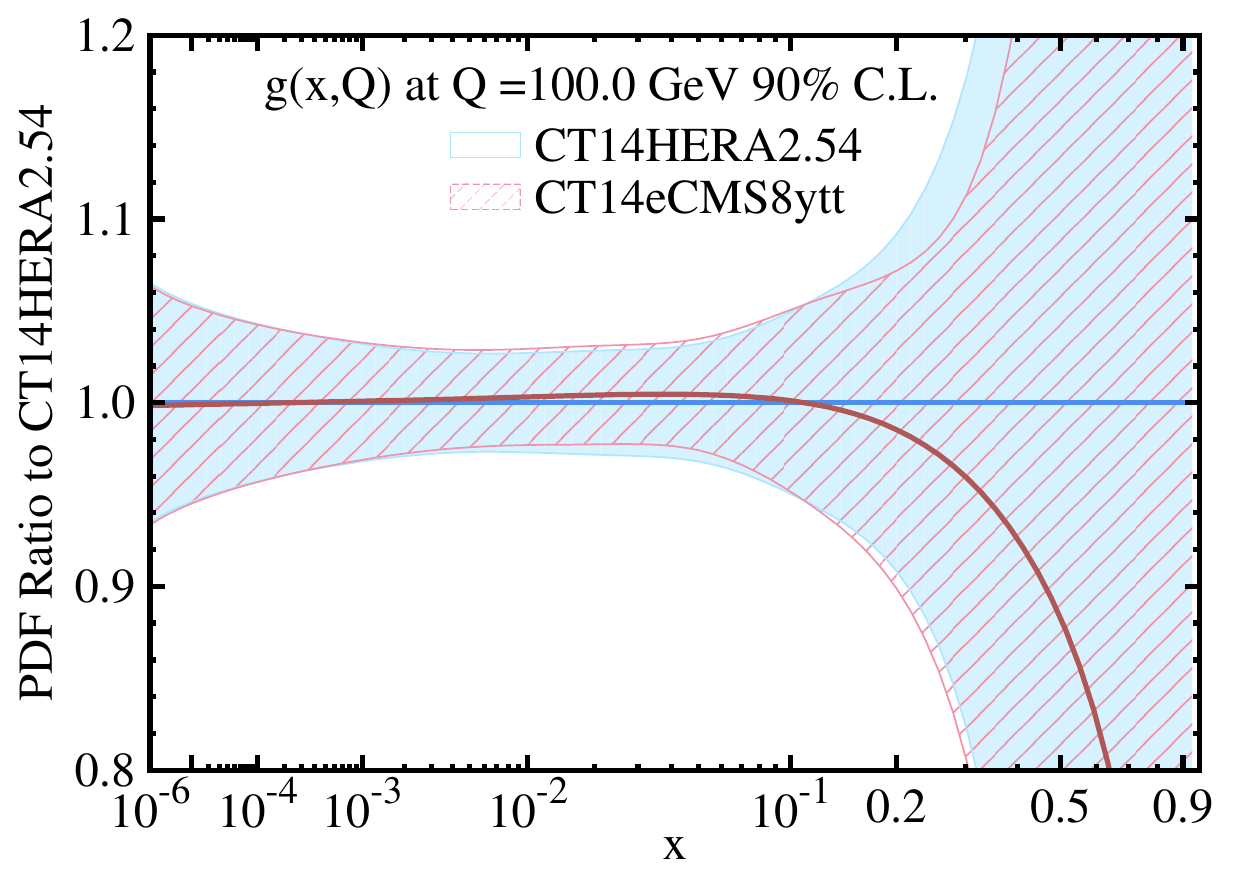}
\includegraphics[width=0.45\textwidth]{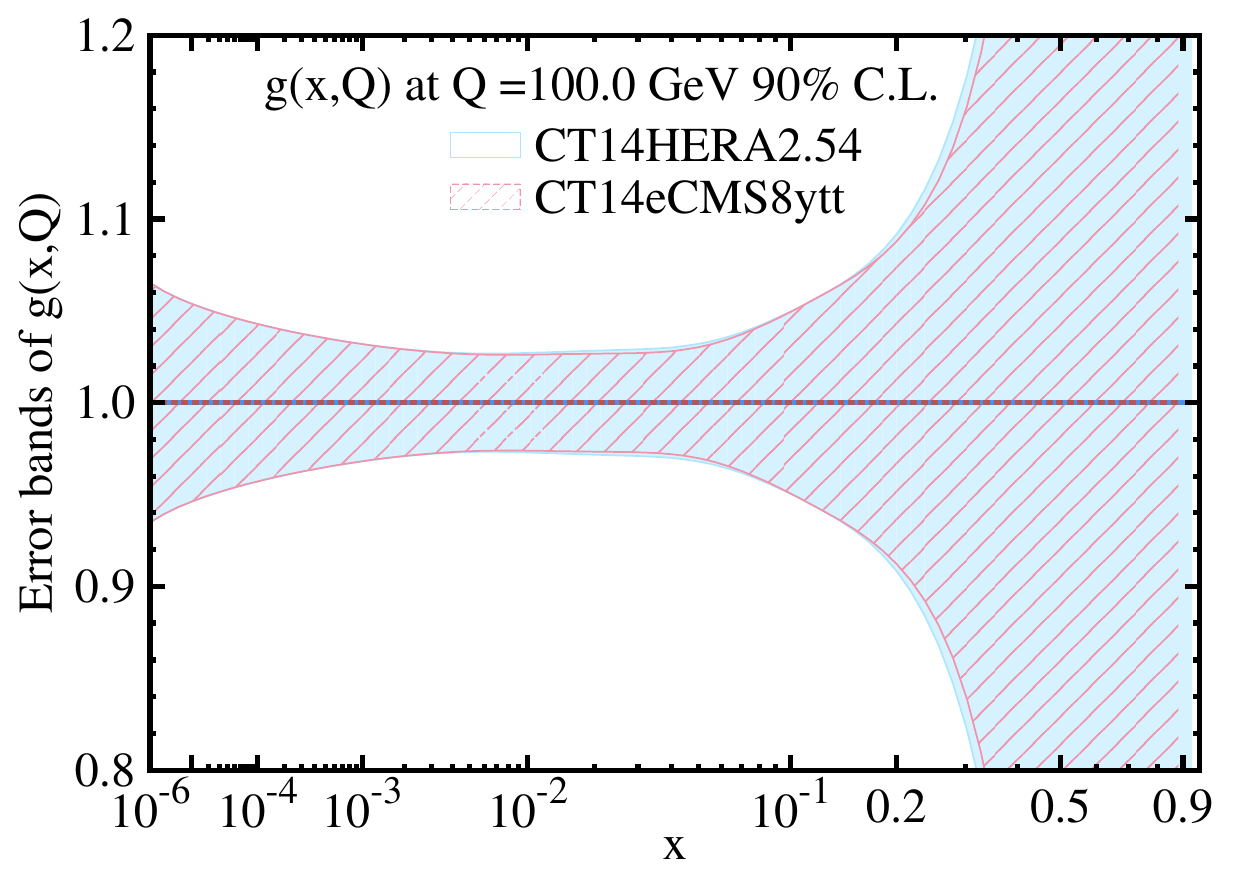}
\caption{\texttt{ePump}-updated $g$-PDF, when the CMS 8 TeV normalized $d\sigma/\sigma dy_{\ttb}$ data set is added to CT14HERA2. Left panel: PDF ratios to the CT14HERA2 best fit. Right panel: the error bands relative to their own best-fit.}
\label{Fig:CT14pttb564g}
\end{figure}

We shall consider eight $t\bar t$ data sets presented by the CMS~\cite{Khachatryan:2015oqa} and ALTAS~\cite{Aad:2015mbv}  collaborations, as listed in Table~\ref{table:ttblist}.
They are the absolute and normalized
one-dimensional differential cross sections of the transverse momentum ($p^t_T$) and rapidity ($y_t$) of top quark, and invariant mass ($m_{t\bar{t}}$) and rapidity ($y_{t\bar{t}}$) of $ t \bar t$ pair.
The dominant production of $t\bar t$ pairs at the LHC is through the gluon-gluon fusion process.   Thus, $t\bar t$ data can potentially constrain the $g$-PDF, especially at large values of $x$, due to the large $t\bar t$ invariant mass.
We also display in the third column of Table~\ref{table:ttblist} the measure $d^0$, introduced in Ref.~\cite{Schmidt:2018hvu}, which summarizes in a single value
the change in the best-fit PDFs after the new data has been added to the original CT14HERA2 fit.  To be precise, $d^0$ is the length of the shift of the best-fit point in parameter space, relative to the 90\% confidence level (C.L.) boundary of the original PDFs.  Thus, $d^0=1$ means that the new best-fit touches the 90\% C.L. boundary, while a value of $d^0\lesssim0.1$ implies a very small change to the best-fit PDFs.\footnote{One should note that $d^0$ only reflects the change in the best-fit PDFs, so that it is still possible for the new data to produce a significant reduction in the PDF error bands, even if $d^0$ is small.}
From Table~\ref{table:ttblist} we can see that most of the new $t \bar t$ data have a minimal effect on the best-fit CT14HERA2 PDFs, with the CMS 8 TeV normalized $d\sigma/\sigma dy_{\ttb}$ data set having the biggest impact.  The $g$-PDF, updated for this data set, is shown in Fig.~\ref{Fig:CT14pttb564g}. One can see that
the updated best-fit $g$-PDF slightly decreases at $x > 0.2$, with slightly reduced error band at $x\sim 0.3$, as compared to the CT14HERA2 PDFs.
Hence, this data set prefers a softer $g$-PDF in the large $x$ region.

\begin{figure}[h]
\includegraphics[width=0.45\textwidth]{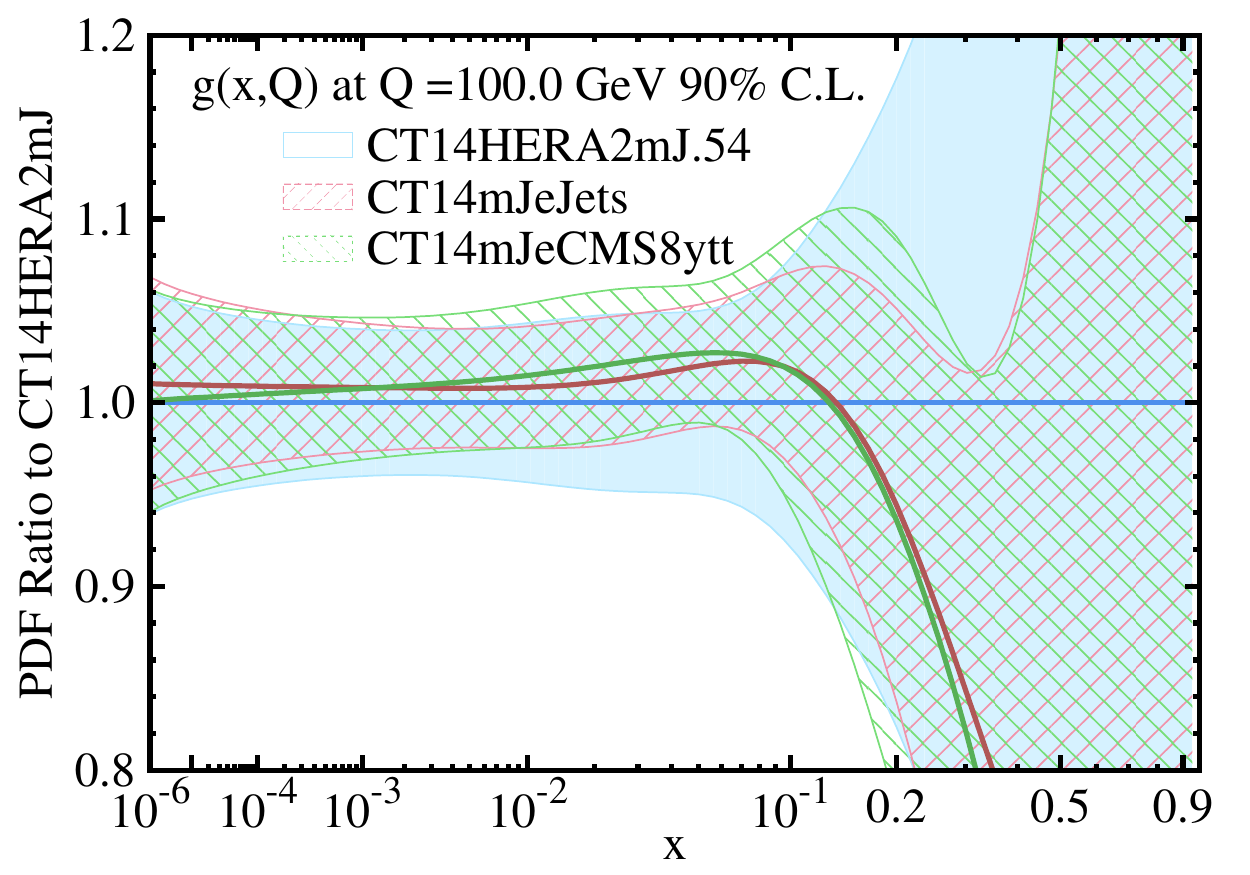}
\includegraphics[width=0.45\textwidth]{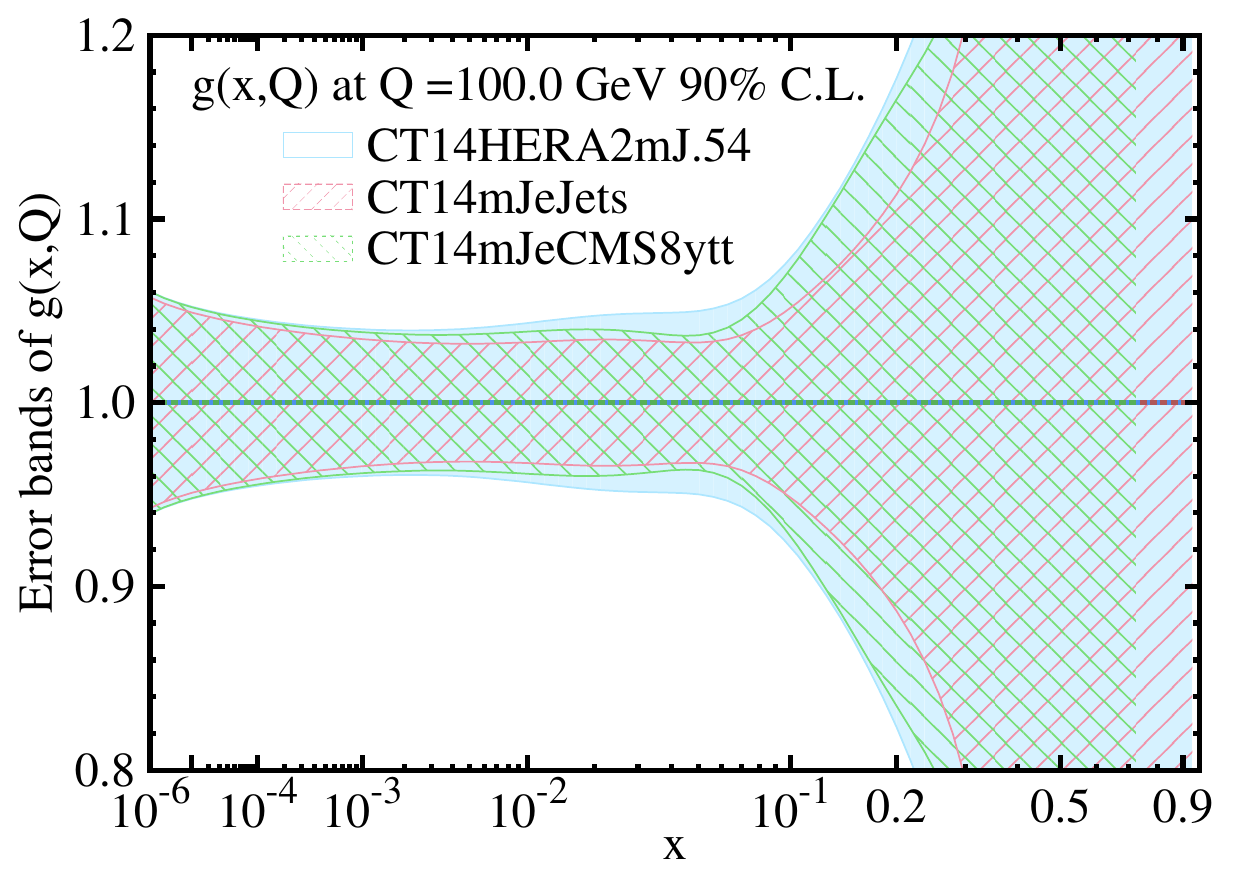}
\includegraphics[width=0.45\textwidth]{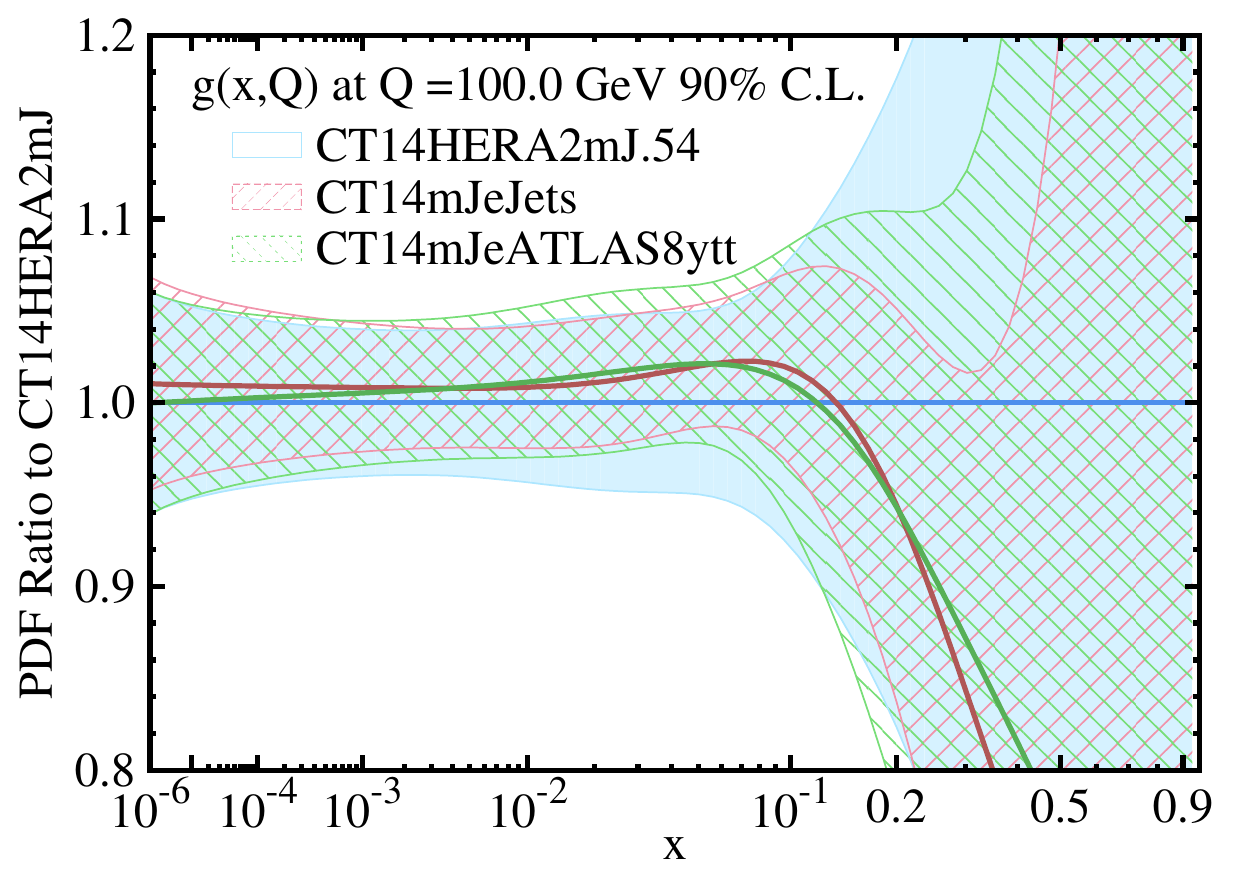}
\includegraphics[width=0.45\textwidth]{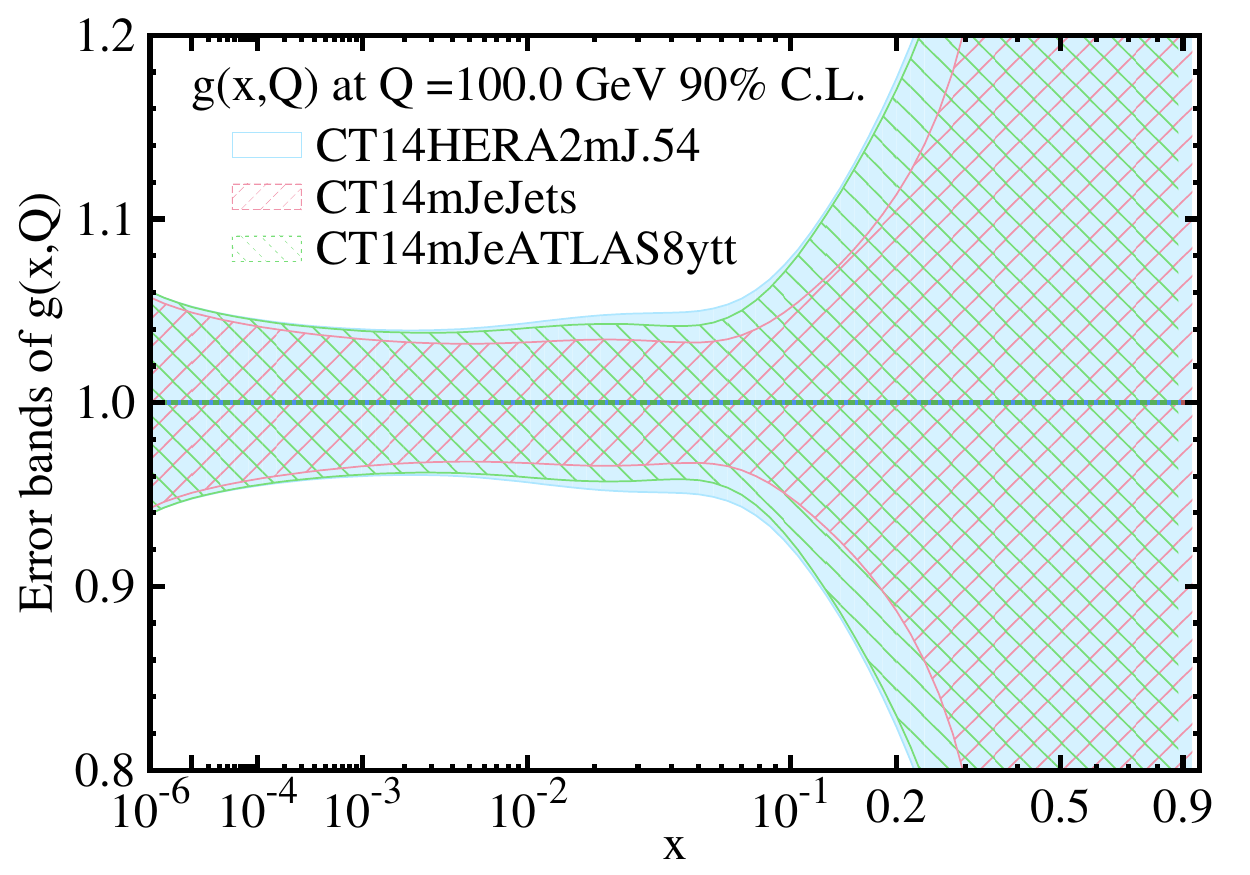}
\caption{Updated $g$-PDFs when CMS 8TeV normalized $d\sigma/\sigma dy_{\ttb}$ data or ATLAS 8TeV absolute $d\sigma/\sigma dy_{\ttb}$ data is added to CT14HERA2mJ. For comparison, we also display the updated $g$-PDF (CT14HERA2mJeJets), obtained when the jet data in CT14HERA2 are added to CT14HERA2mJ by \texttt{ePump}. Left panel: PDF ratios to the CT14HERA2mJ best fit. Right panel: the error bands relative to their own best-fit.}
\label{Fig: mjpttb564568g}
\end{figure}

Given the dependence of the $t\bar t$ distributions on the $g$-PDF, it is interesting to compare the impact of each of the new LHC $t\bar t$ data sets with that of the four jet data sets  already included in the CT14HERA2 fit.
To do this, we use  \texttt{ePump} to add each of the individual $t\bar t$ data sets as new data to update the CT14HERA2mJ PDFs.   Recall that the CT14HERA2mJ PDFs,
introduced in Sec.~\ref{subsection:mJ}, were obtained from a global fit to all of the CT14HERA2 data sets, except the four Tevatron and LHC jet data.
The resulting $d^0$ values, after updating from CT14HERA2mJ,  are listed in the fourth column of Table~\ref{table:ttblist}.
Comparing to the comparable values in the third column, obtained from updating CT14HERA2, we find that the new $t\bar t$ data sets have a much larger effect in the absence of the
jet data.  In this case,  the CMS 8 TeV normalized $d\sigma/\sigma dy_{\ttb}$
and ATLAS 8 TeV absolute $d\sigma/d|y_{\ttb}|$ data
have the largest impact, for which the updated $g$-PDFs are shown in Fig.~\ref{Fig: mjpttb564568g}. It can be seen that the $y_{t\bar t}$ distributions measured at both CMS and ATLAS have comparable effects, and they modify the $g$-PDF similarly to that of the jet data. But the $t\bar t$ data has less power to reduce the uncertainties of $g$-PDFs, especially in the $x$ range $0.1\sim 0.2$. This is consistent with our finding that in the presence of jet data, the new $t\bar t$ data sets have little effect on PDFs, because the $t\bar t$ data produces the same change on the central $g$-PDFs, but provides less constraining power on the error band. The reason for this can be traced to the simple fact that there are far fewer $t\bar t$ data points than jet data points, due to  a smaller production cross section. Thus, the statistical power of $t\bar t$ is smaller.

\begin{figure}[h]
\includegraphics[width=0.45\textwidth]{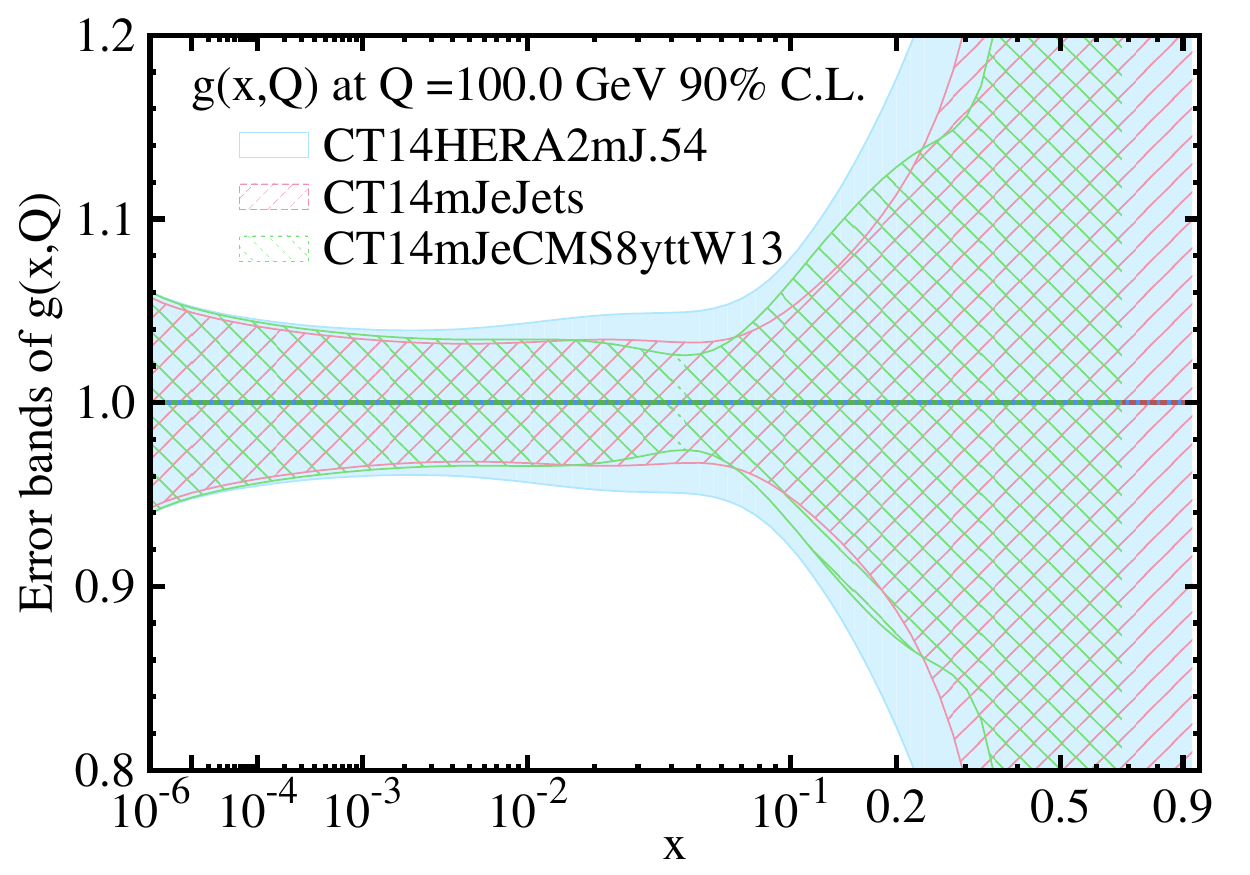}
\includegraphics[width=0.45\textwidth]{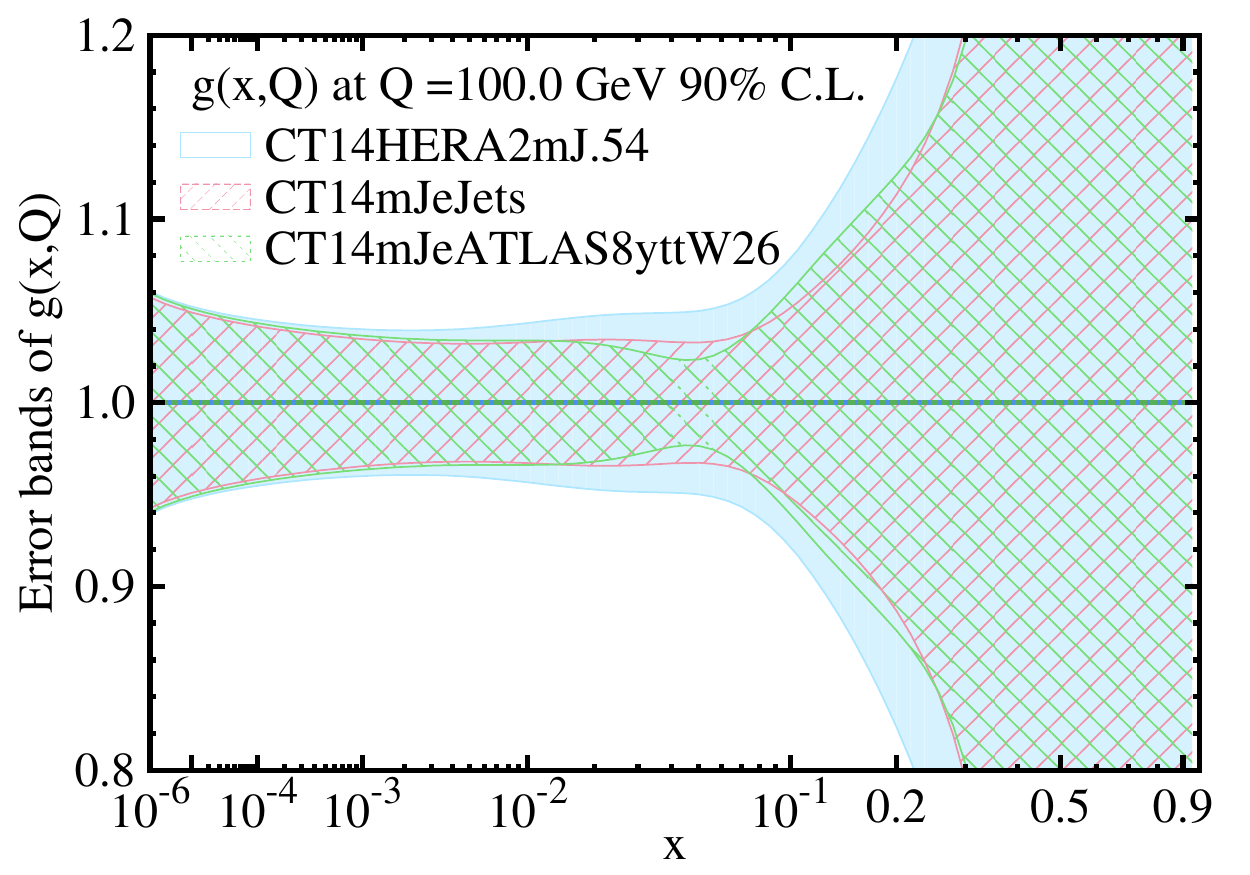}
\caption{Same as Fig.~\ref{Fig: mjpttb564568g}, with weights 13 and 26 for the CMS 8TeV normalized $d\sigma/\sigma dy_{\ttb}$ data and ATLAS 8TeV absolute $d\sigma/\sigma dy_{\ttb}$ data, respectively. Only the error bands are shown in this figure.}
\label{Fig: mjpttb564568wg}
\end{figure}

We can test this interpretation using \texttt{ePump}  by increasing the weight of the $t\bar t$ data in the \texttt{ePump} updating. A weight larger than 1 is equivalent to having more $\ttb$ data points with the same experimental uncertainties or, alternatively, to reducing the experiment uncertainties by a factor of the square root of the weight. Of course, increasing the weight is not exactly the same as  increasing the luminosity, since it does not change the central values of the data, which presumably have fluctuations described by the original experimental uncertainties. Nevertheless, one can get some estimate of the potential impact of the $\ttb$ data as the integrated luminosity is increased.  To compare with the effect of the jet data, we multiply the contribution of the new $t\bar t$ data set to $\chi^2$ by a weight equal to the ratio of the number of jet data points to the number of individual $\ttb$ data points. We have seen in Sec.~\ref{section:impact Jet} that the CMS 7 TeV jet data~\cite{Chatrchyan:2012bja}, with 133 data points, has the dominant effect among all the jet data in CT14HERA2, so we multiply by the weights $133/10=13$ for the CMS, and $133/5=26$ for the ATLAS $y_{t\bar t}$ distributions, respectively.
The $g$-PDFs, obtained by updating the CT14HERA2mJ fit with the weighted $y_{t\bar t}$ distributions using \texttt{ePump}, are shown in Fig.~\ref{Fig: mjpttb564568wg}.
The general shapes of the updated $g$-PDFs are similar to that obtained by including all four jet data in the CT14HERA2 fit.
However, the error band of the $g$-PDFs is not reduced as much as the CT14HERA2 fit for $x > 0.01$.
Hence, we conclude that the jet data will probably impose a stronger constraint on the $g$-PDF than the $t \bar t$ data, even with more integrated luminosity collected at a higher center-of-mass energy of the LHC.

\begin{figure}[h]
	\includegraphics[width=0.45\textwidth]{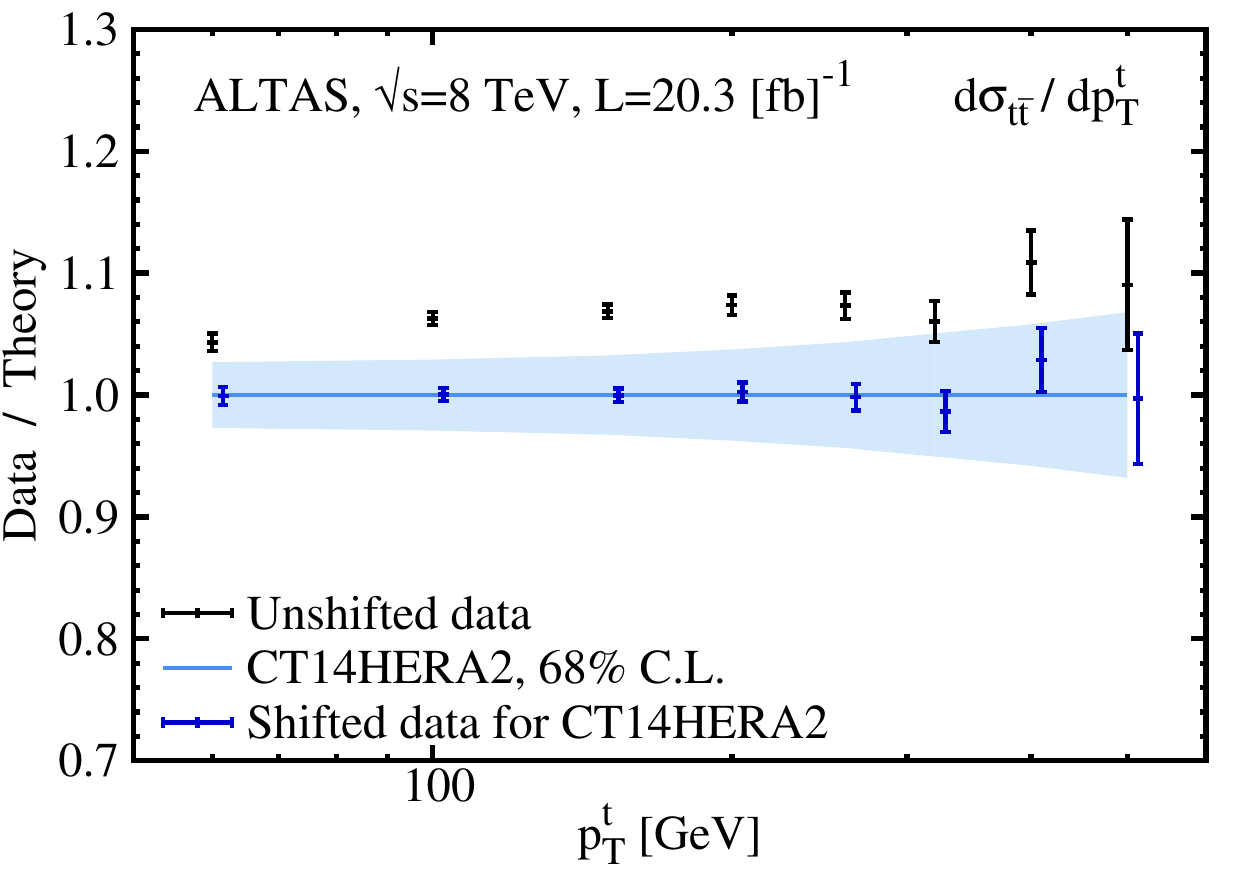}
	\includegraphics[width=0.45\textwidth]{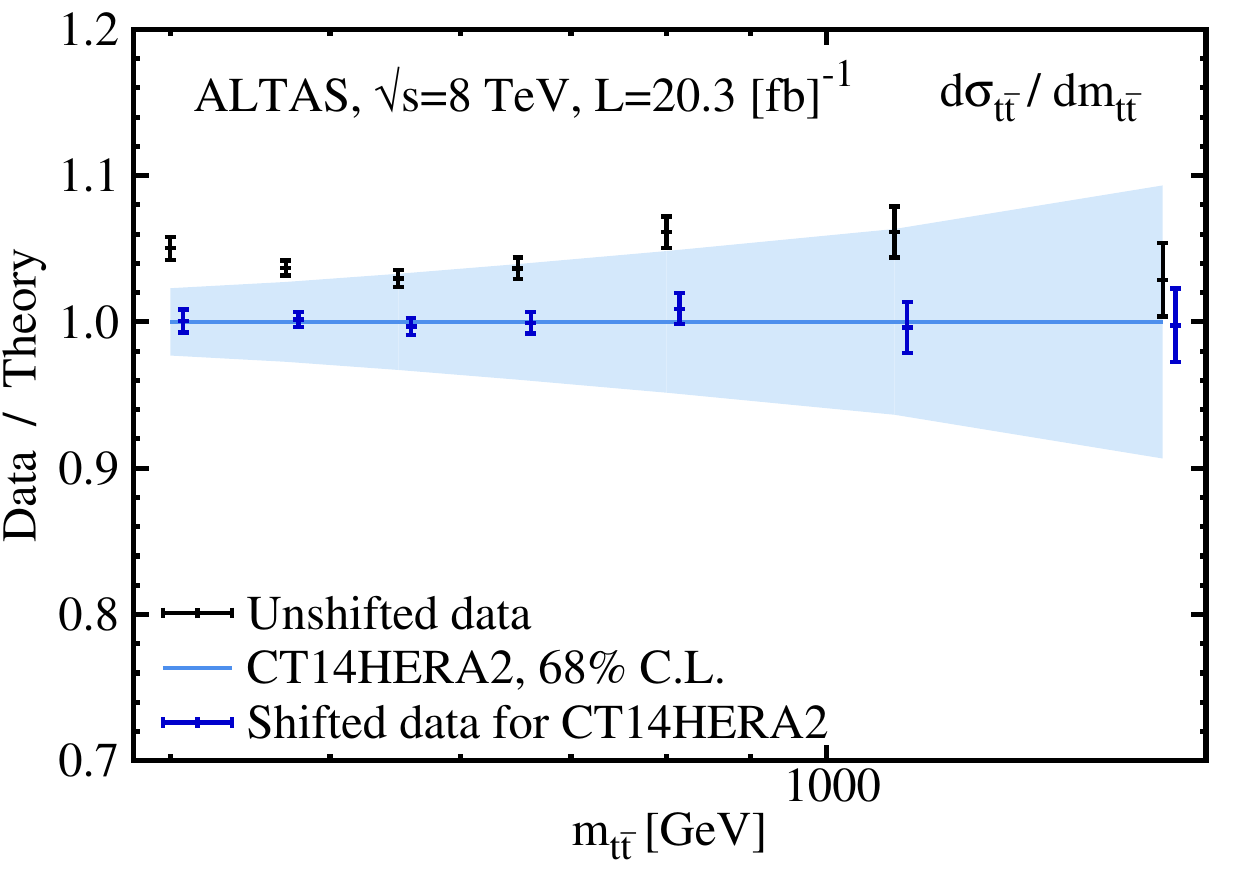}
	\caption{Data and theory comparison for ATLAS 8TeV $d\sigma/d p_T^t$ and $d\sigma/d m_{t\bar t}$ data. The uncertainties of theory and data are both at the 68\% C.L.
}
	\label{Fig: data theory comparison 565 567}
\end{figure}

Before leaving this section, we comment on the impact of the
ATLAS 8 TeV $d\sigma/dp_{T}^{t}$ and $d\sigma/dm_{\ttb}$ data.  Given the small values of $d^0$ in
 Table~\ref{table:ttblist}, we expect little change in the best-fit PDFs when only including these data to update either the CT14HERA2 or CT14HERA2mJ PDFs.
This can happen when the theory prediction is in good agreement with the data even before updating.
Fig.~\ref{Fig: data theory comparison 565 567} shows that this is indeed the case for CT14HERA2, and similar results were found for CT14HERA2mJ. This is also demonstrated by the small $\chi^2$ per data point for these two data sets in Table~\ref{table:ttblist_chi2}. Therefore, the best-fit $g$-PDF, updated by the $p_{T}^{t}$ or $m_{\ttb}$ distribution, does not need to move far from their original position to have a good fit to the data.
Another feature we observed from Table~\ref{table:ttblist_chi2} is that the impact of the $t\bar t$ data to $g$-PDF is consistent with the jet data included in the CT14HERA2 fit such that the $\chi^2/N$ values updated from CT14HERA2mJ are larger than those from CT14HERA2.

\subsection{ATLAS 7 TeV $WZ$ data}\label{sec: ATLAS7WZ}

\begin{figure}[h]
	\includegraphics[width=0.45\textwidth]{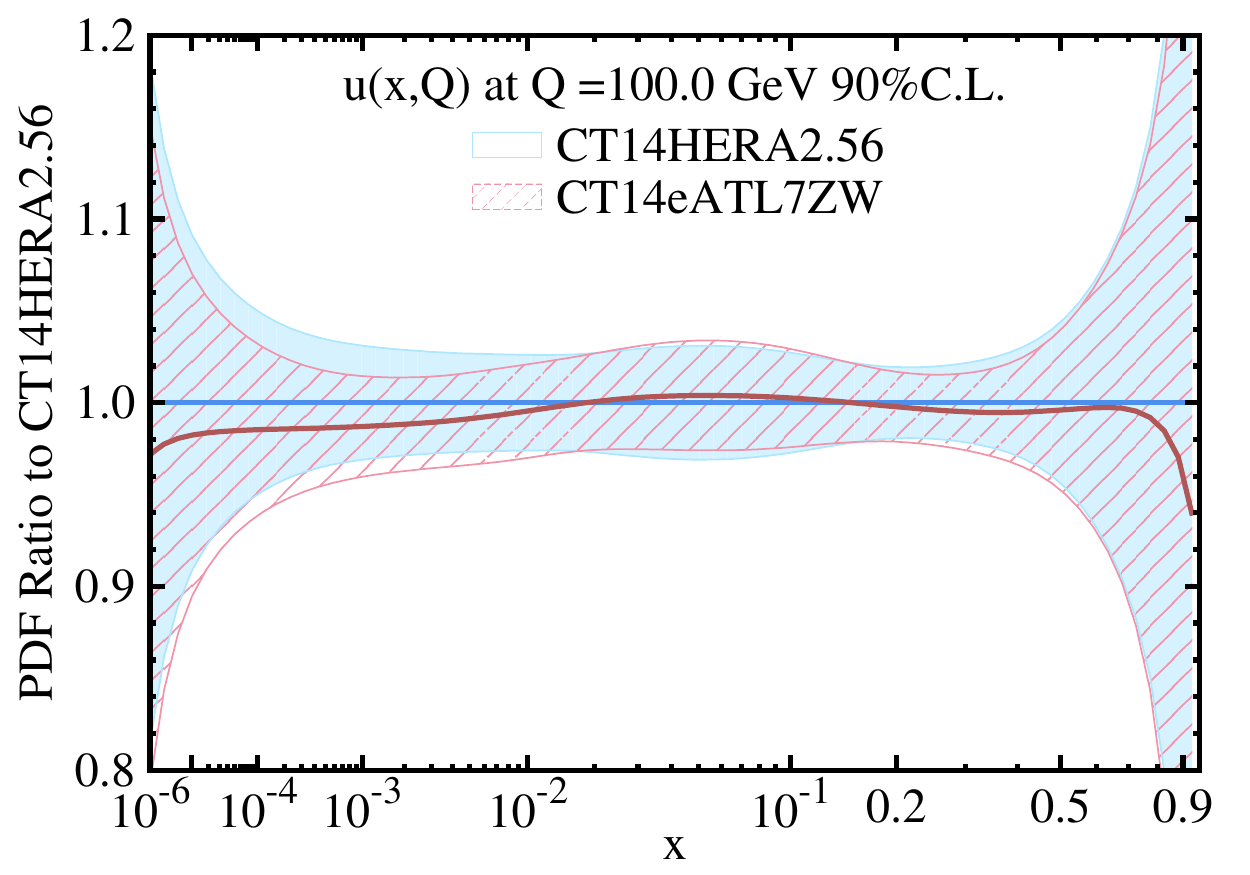}
	\includegraphics[width=0.45\textwidth]{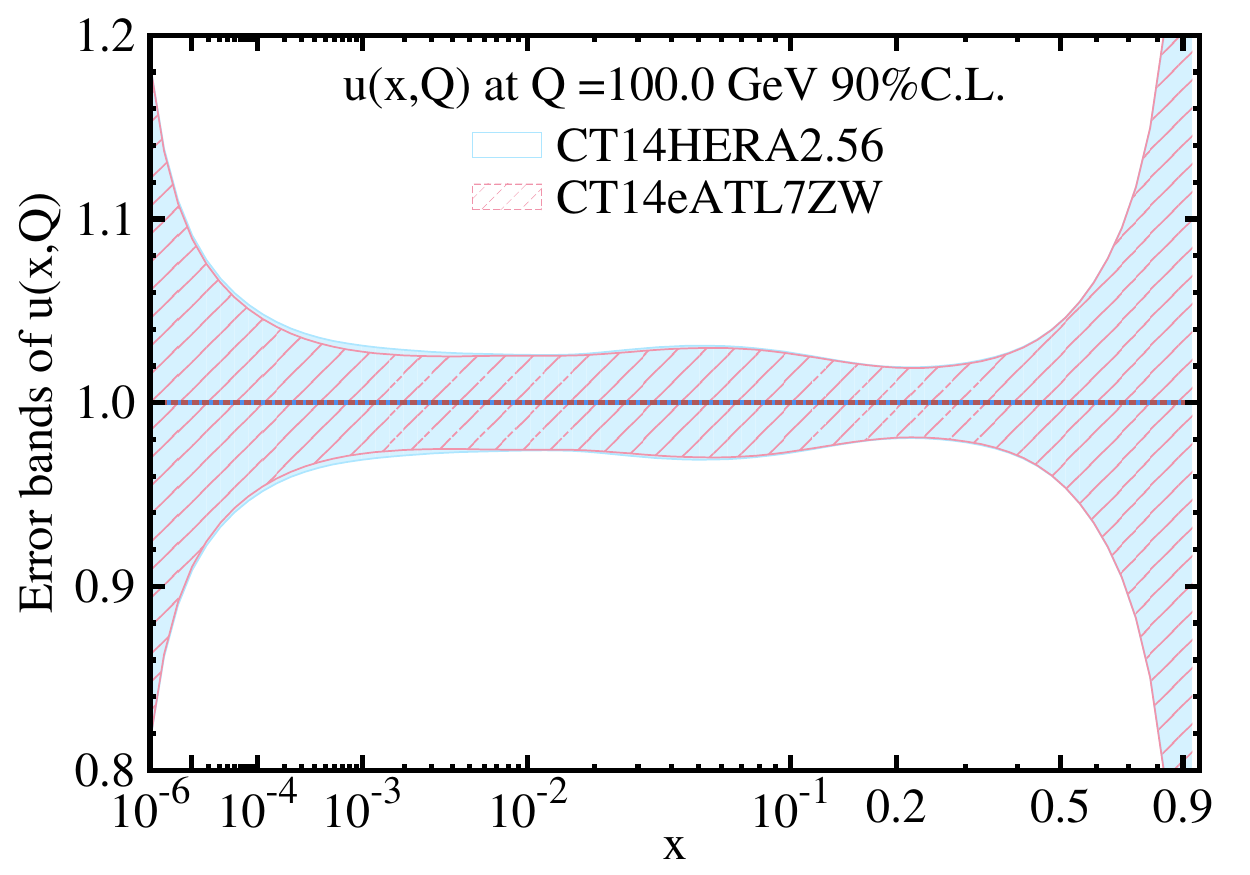}\\
	\includegraphics[width=0.45\textwidth]{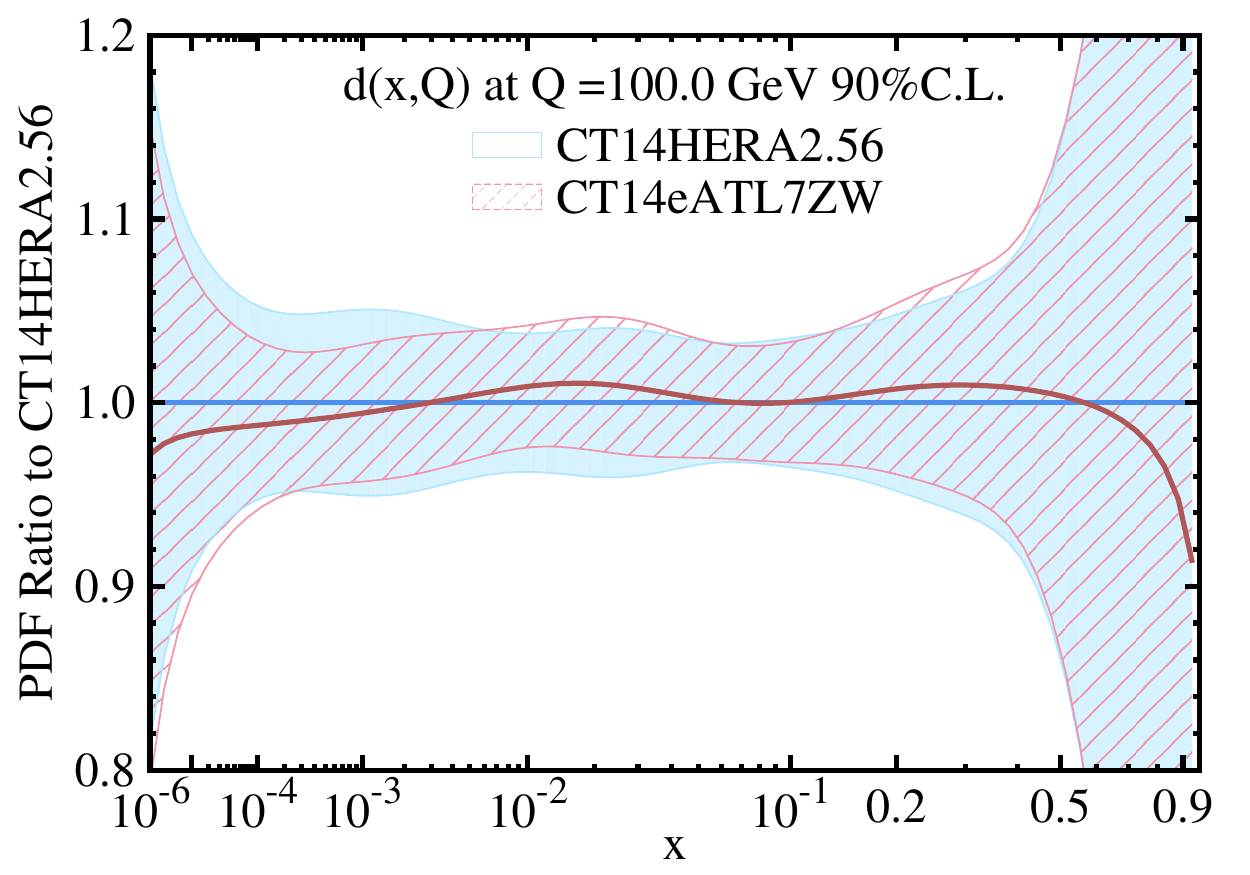}
	\includegraphics[width=0.45\textwidth]{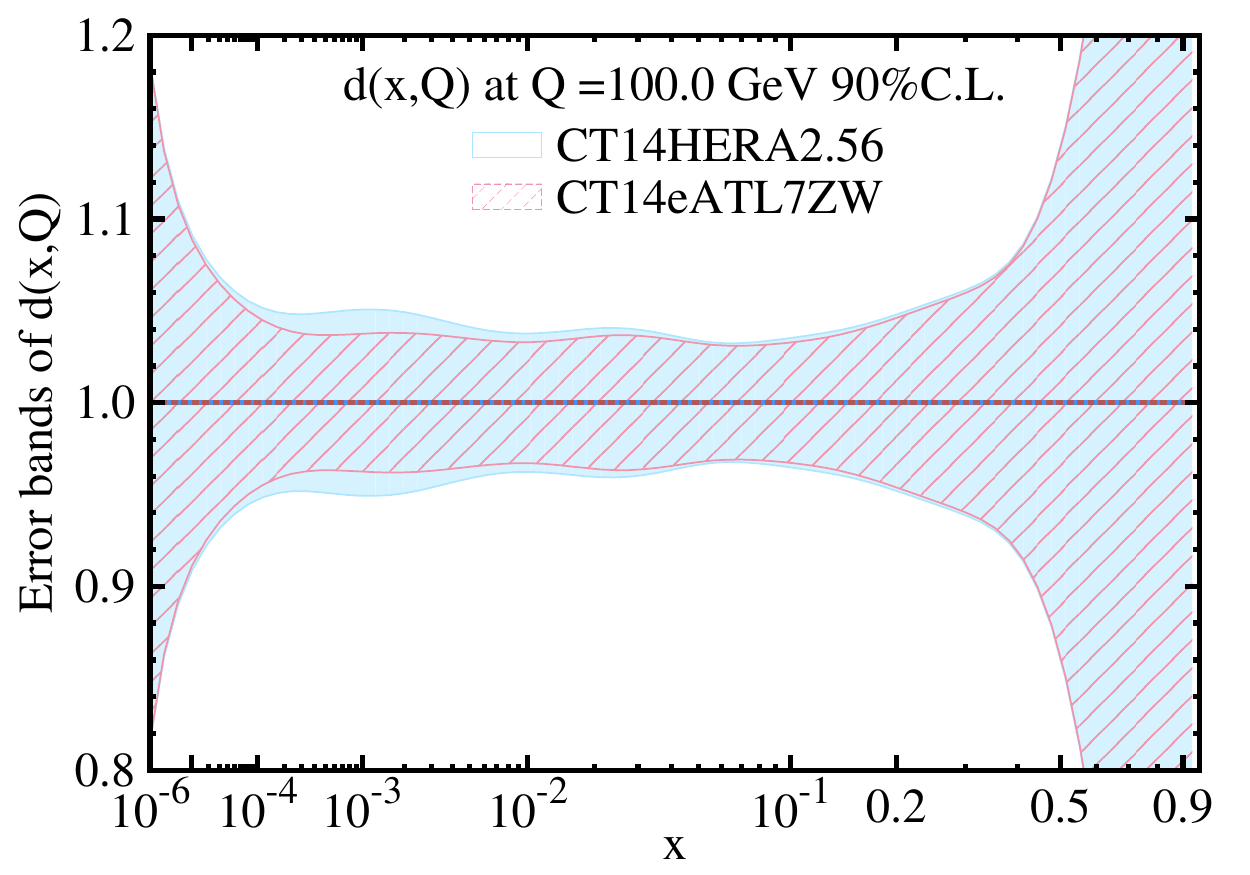}\\
	\includegraphics[width=0.45\textwidth]{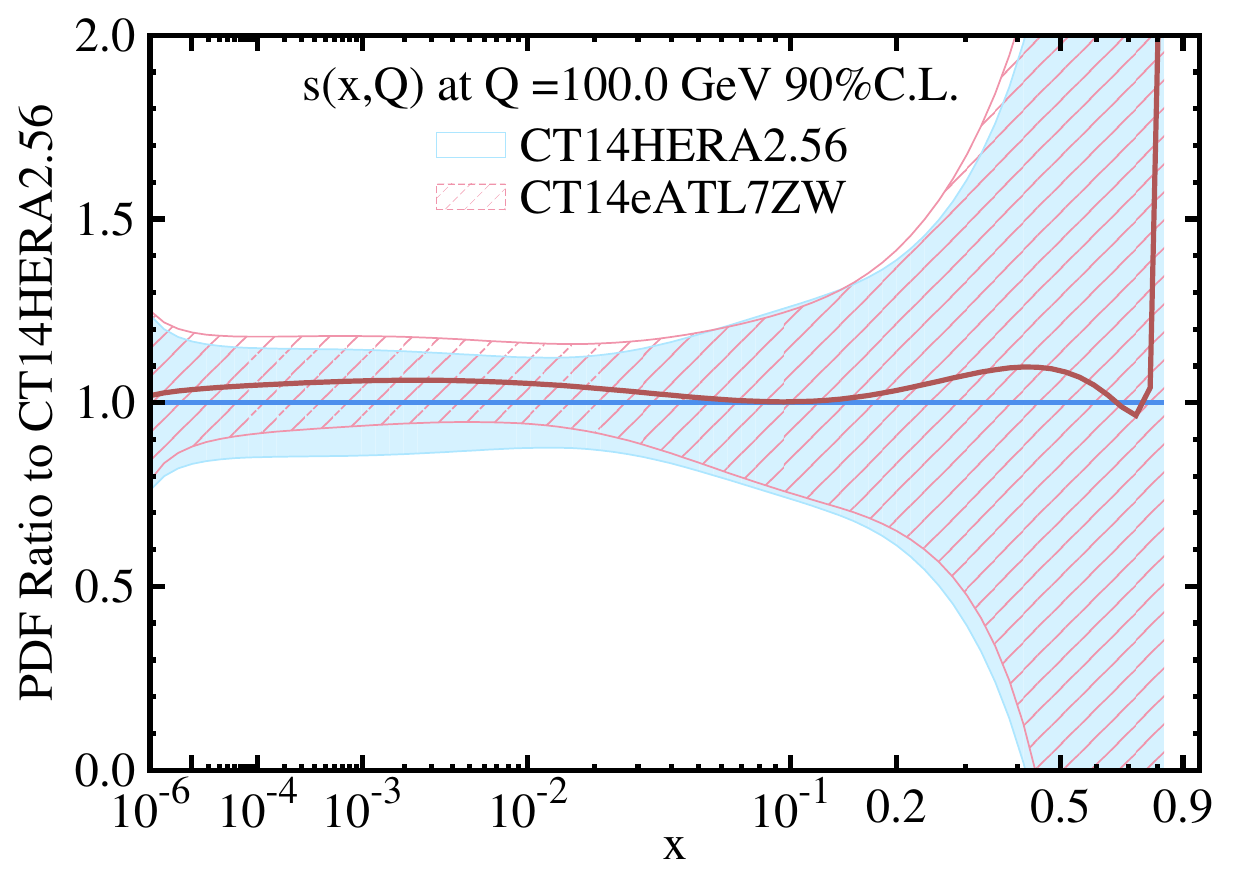}
	\includegraphics[width=0.45\textwidth]{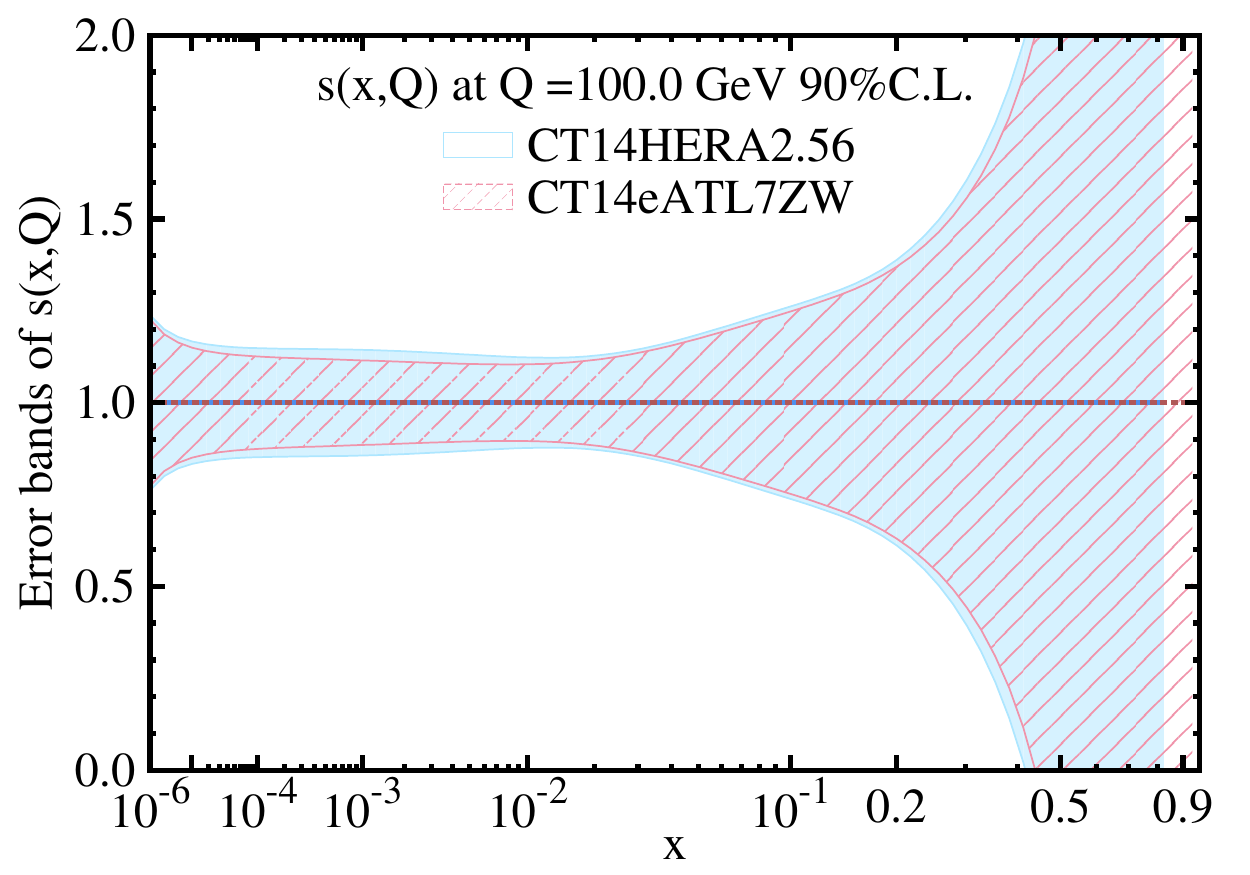}
	\caption{Updated $u$, $d$ and $s$-PDFs when the ATLAS 7 TeV $W$ and $Z$ data set is added to CT14HERA2. The suffix ``.56"
	indicates that here the error bands for CT14HERA2 are computed using 56 eigen-PDFs, including the two gluon extreme sets (labelled as the 55th and 56th eigen-sets). Left panel: PDF ratios to the CT14HERA2 best fit. Right panel: the error bands relative to their own best-fit.}
	\label{Fig: CT14p248}
\end{figure}

\begin{figure}[h]
   \includegraphics[width=0.45\textwidth]{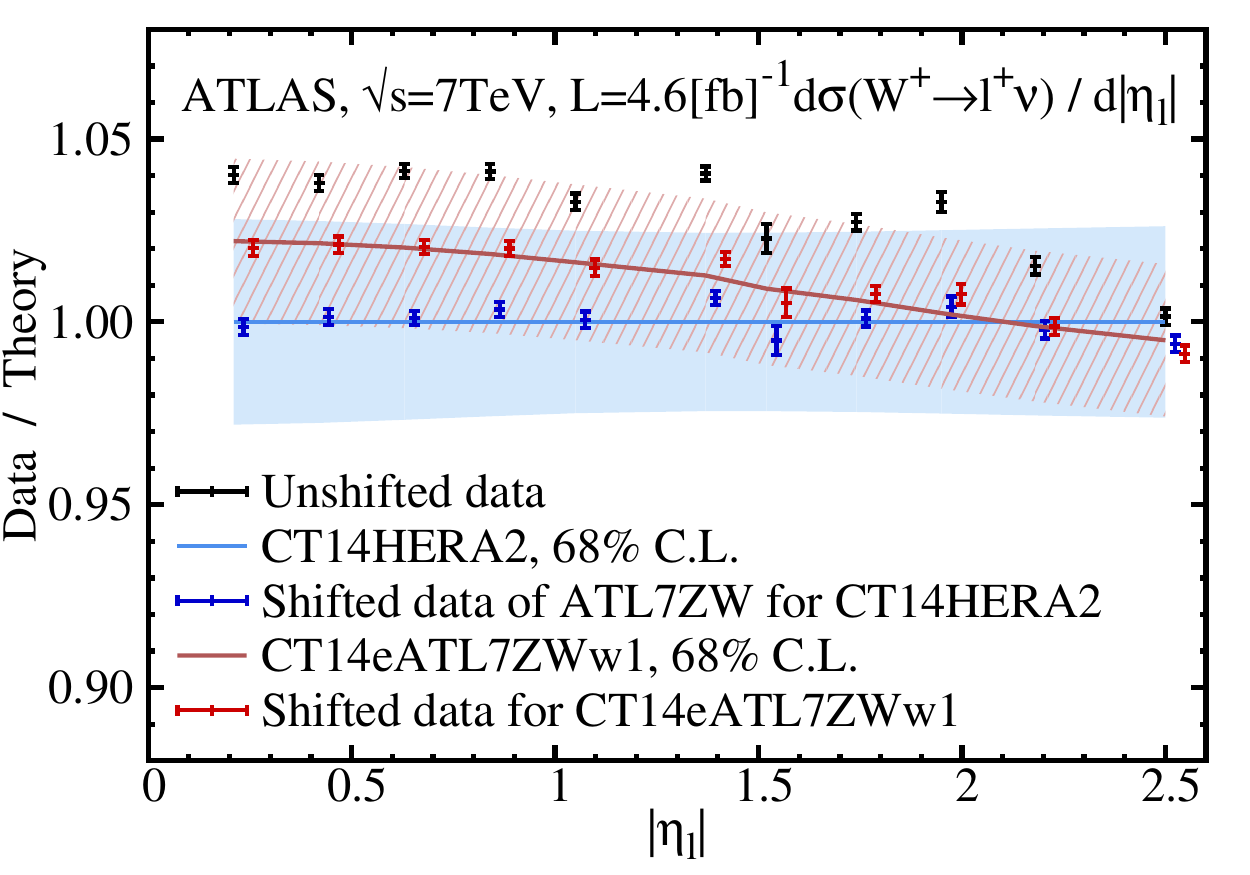}
   \includegraphics[width=0.45\textwidth]{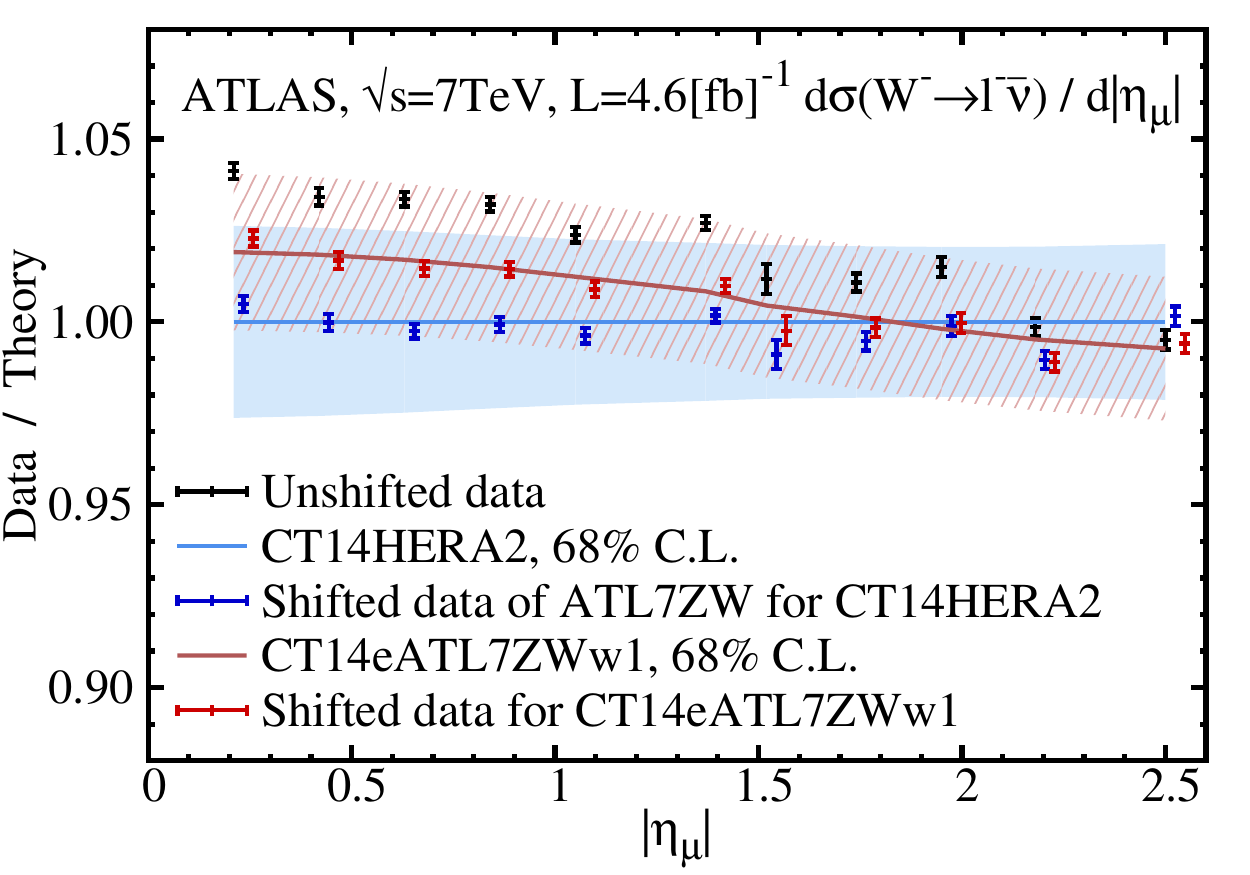}\\
   \includegraphics[width=0.45\textwidth]{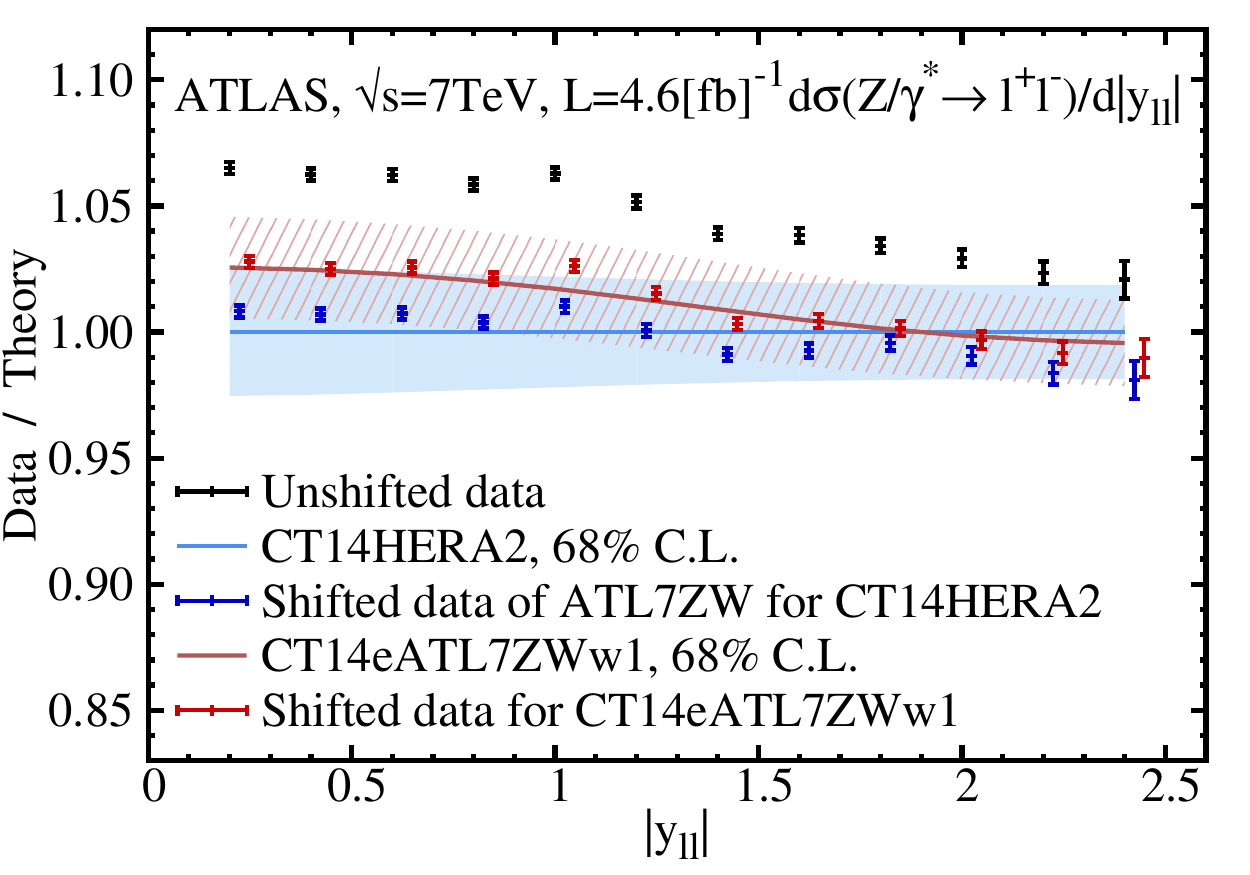}
   \caption{Comparison between ATLAS 7 TeV $W$ and $Z$ data and the theory predictions for each data point. The raw data are labelled as ``unshifted data''. The shifted data, after being corrected by the nuisance parameters of the correlated systematic errors, are compared to the theory predictions
      before  (labelled as CT14HERA2, in blue) and after the ATLAS 7 TeV $W$ and $Z$ data is included in the \texttt{ePump} updating with weight of 1 (labelled as CT14eATL7ZWw1, in red), respectively. The error bar on each data point includes both statistical and uncorrelated systematic errors, added in quadrature. The error band of theory prediction indicates the PDF induced error at the 68\% CL.}
\label{Fig: data theory comparison248}
\end{figure}

\begin{figure}[h]
  \includegraphics[width=0.32\textwidth]{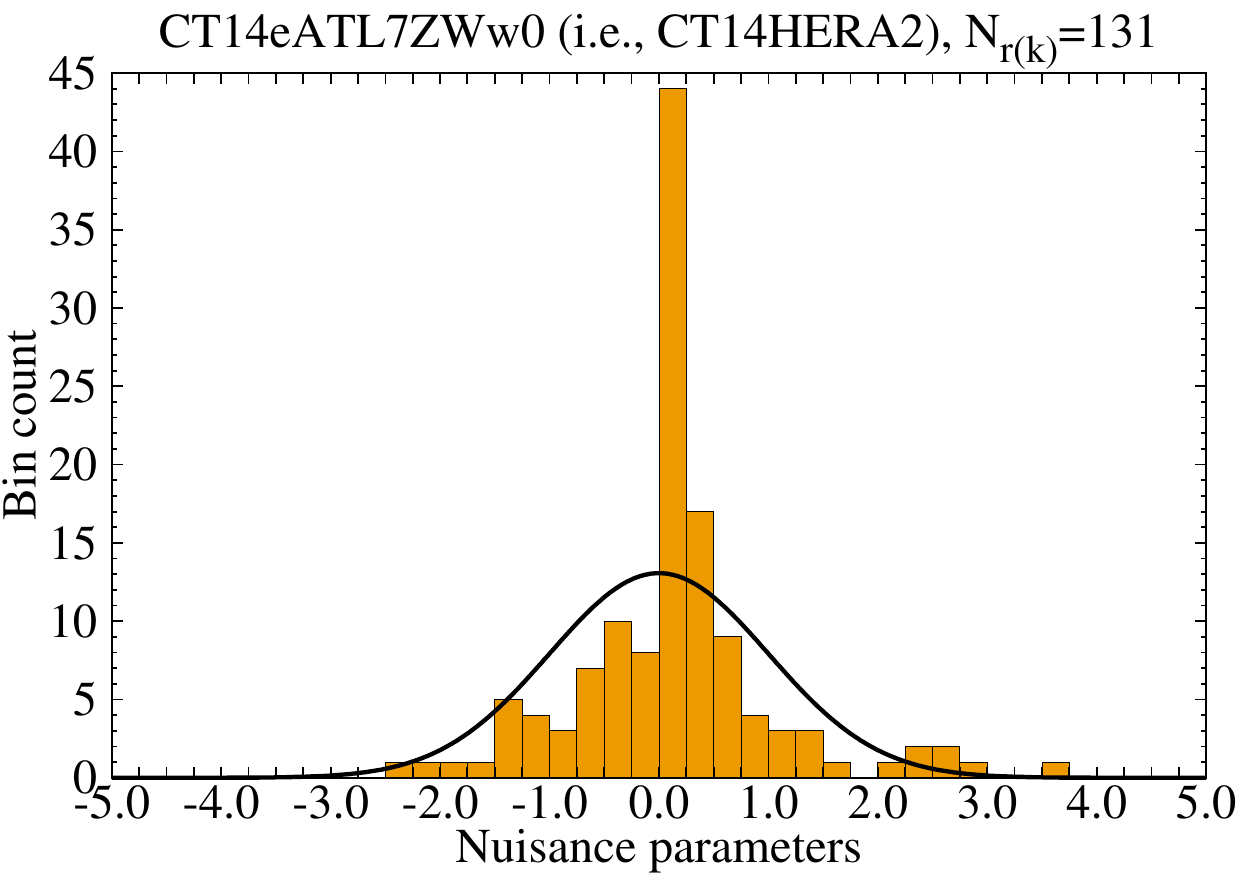}
  \includegraphics[width=0.32\textwidth]{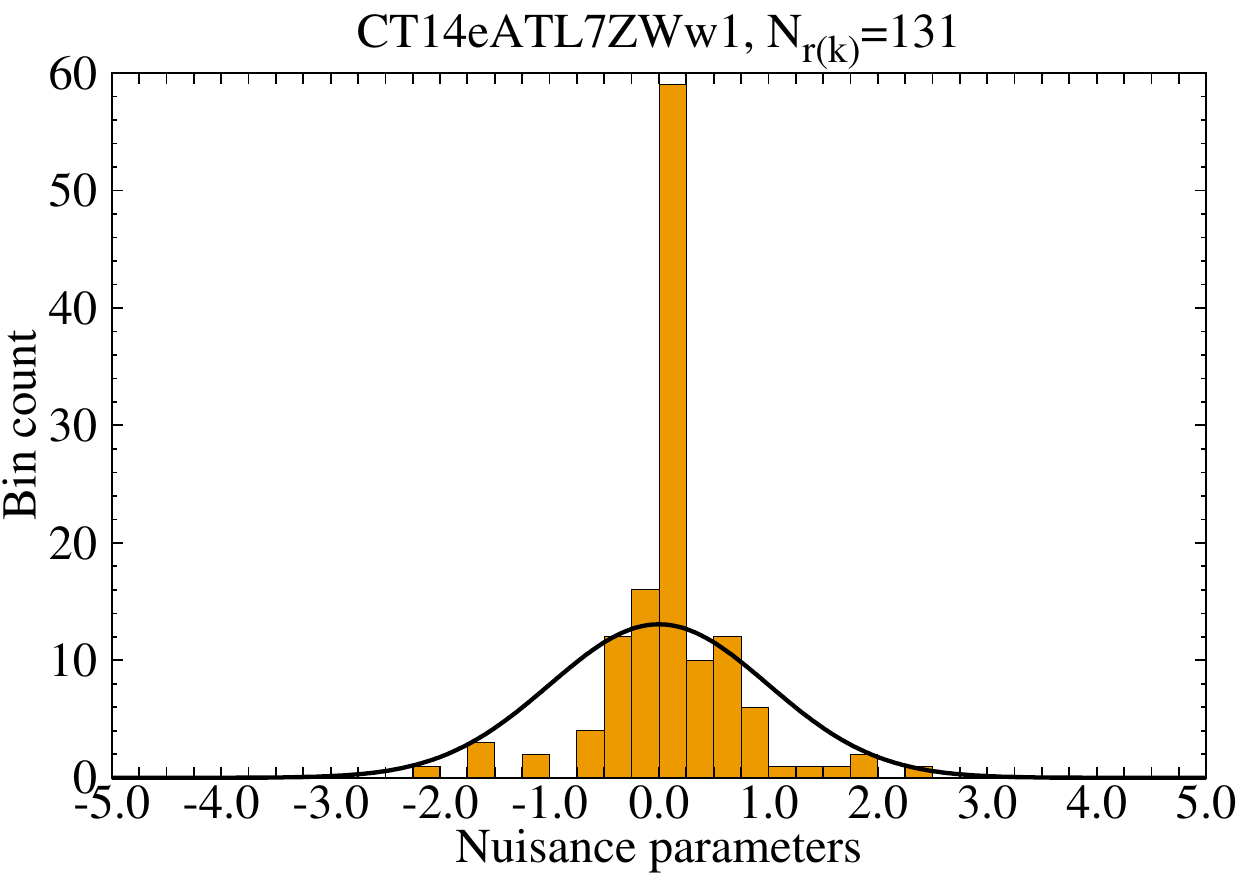}
  \includegraphics[width=0.32\textwidth]{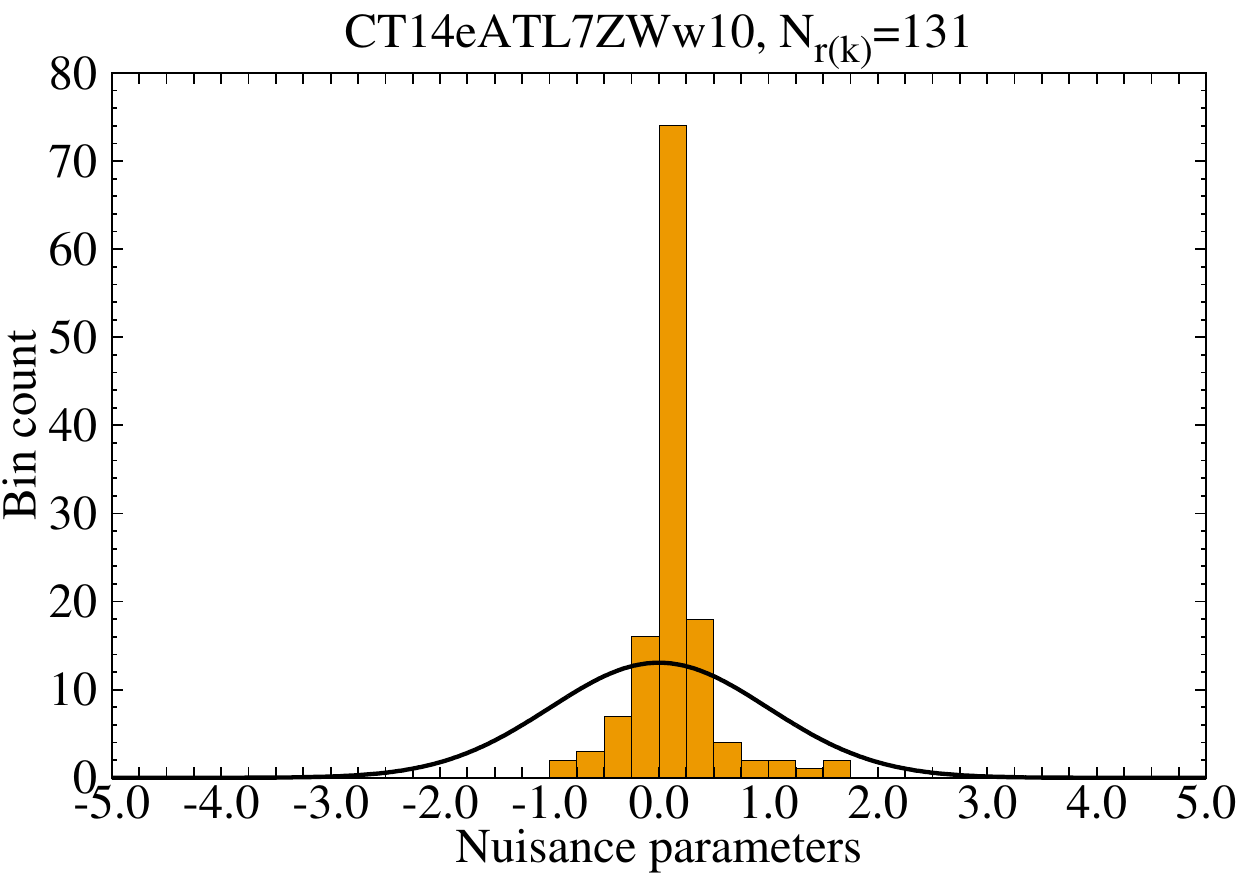}
  \caption{Distribution of the nuisance parameters for ATLAS 7 TeV $W$ and $Z$ data, after the \texttt{ePump} updating. {It shows, from left to right, the results of the \texttt{ePump} updated fits with the ATLAS 7 TeV $W$ and $Z$ data included with weight of 0} {(i.e., CT14HERA2,} { labelled as CT14eATL7ZWw0), 1 (labelled as CT14eATL7ZWw1), and 10 (labelled as CT14eATL7ZWw10), respectively. The solid curve is the standard normal distribution (with a mean value of 0 and standard deviation of 1) expected in the ideal case. }}
  \label{Fig: nuisanceDis248}
\end{figure}

After observing the constraints of new LHC jet data~\cite{Schmidt:2018hvu} and $\ttb$ data on the $g$-PDFs,
we would also like to see how new LHC Drell-Yan data could modify the quark PDFs. The low luminosity ($35\, {\rm pb}^{-1}$) ATLAS 7 TeV  $W^\pm$ and $Z$ cross section data~\cite{Aad:2011dm} were included in
the CT14HERA2 fit. Since then, ATLAS has published the more precise ATLAS 7 TeV $WZ$ data with an integrated luminosity of $4.6\, {\rm fb}^{-1}$~\cite{Aaboud:2016btc}.
Here, we will study the impact of this more precise data on further constraining the CT14HERA2 PDFs.
Strictly speaking, we should first remove the old ATLAS 7 TeV $WZ$ data from the CT14HERA2 global fit and then add
the new ATLAS 7 TeV $WZ$ data with \texttt{ePump}, so as not to double count the ATLAS 7 TeV $WZ$ data contributions.
However, since the two data sets are consistent and the new one has about 100 times the integrated luminosity as the old one, the impact of the double counting should be negligible.  Therefore, we shall simply add the new ATLAS 7 TeV $WZ$ data to update the CT14HERA2 PDFs using \texttt{ePump}.
Fig.~\ref{Fig: CT14p248} shows the updated PDFs. One can see that the new ATLAS 7 TeV $W$ and $Z$ data have a sizable impact on the quark PDFs and their uncertainties, particularly for $x$ ranging from $10^{-4}$ to a few times $10^{-2}$.  On the one hand, this is understandable, because these data are very precise, with uncertainties less than the percent level. On the other hand, such a large difference between  the updated PDFs and the original CT14HERA2 PDFs calls for further investigation.

We note that the new ATLAS 7 TeV $W$ and $Z$ data, to be denoted as ``ATL7ZW" data from now on, with a total of 34 data points,  cannot be fit well. Its $\chi^2$ per data point after the fit by \texttt{ePump} is found to be around 2.7, which is much larger than that  found for the full CT14HERA2 global fit (about 1.25) with a total of 3287 data points.
Let us consider the two measures, $d^0$ and $\tilde d^0$, introduced in Ref.~\cite{Schmidt:2018hvu} to assess the quality of the fit given by \texttt{ePump}.
Recall that these two measures are the length of the shift in the parameter space of the best-fit PDFs, relative to the original and to the updated 90\% C.L. boundaries, respectively.
The \texttt{ePump} output file gives $d^0=0.87$ after adding these data, which indicates that the shift of the best-fit parameters nearly touches the 90\% C.L. boundary of the original CT14HERA2 fit.  Furthermore, we find a value of $\tilde d^0=1.49$, which implies that the original best-fit parameter point falls outside of the 90\% C.L. region of the updated fit.
This latter
result, in particular, suggests that the points in the parameter space used to evaluate the new data may (though not necessarily) be outside the region of validity of the Hessian approximations.  Thus, in this case, the result from \texttt{ePump} should be taken with caution, and the shift in the best-fit PDFs found in a true global fit may likely be larger than that given by the \texttt{ePump} program.
With that said, we find from \texttt{ePump} updating that
adding the ATLAS 7 TeV $WZ$ data to CT14HERA2 fit would decrease the $u$ and $d$ quark PDFs and increase the $s$ quark PDF at $x=10^{-4} \sim 10^{-3}$, and
increase the $d$ PDF at $x$ around $10^{-2}$ and $0.3$. Also, the error band of the $d$ PDF is reduced significantly around $x=10^{-3}$, and the error band of the $s$-PDF
is reduced for nearly all values of $x$.

In addition to the single-value criteria, one can also compare the data and theory predictions point-by-point to reveal some more details about the quality of fit, as shown in  Fig.~\ref{Fig: data theory comparison248}.
First, we find that there is an overall shift for all the raw data points.  This means that the correlated systematic errors, weighted by their corresponding nuisance parameters, play an important role in the fitting.
Fig.~\ref{Fig: nuisanceDis248} shows the distributions of nuisance parameters, before and after updating with \texttt{ePump}. The solid curve in the figure shows a standard normal distribution with a mean value of 0 and standard deviation of 1. It shows that there are some large nuisance parameters before the updating. Given the large difference between data and theory for CT14HERA2 in Fig.~\ref{Fig: data theory comparison248}, we conclude that ATLAS 7 TeV $W$ and $Z$ data are not described well by the CT14HERA2 PDFs, so we expect a large impact of this data set to update the CT14HERA2 PDFs. Second, one can see that the  ATLAS 7 TeV $W$ and $Z$ data are more precise than the theory predictions, with PDF induced uncertainty included, and even after \texttt{ePump} updating the precision data still cannot be described well by the theory. This, together with the large contributions from the nuisance parameters, leads to the large $\chi^2$ for this data set.

\begin{figure}[h]
	\includegraphics[width=0.6\textwidth]{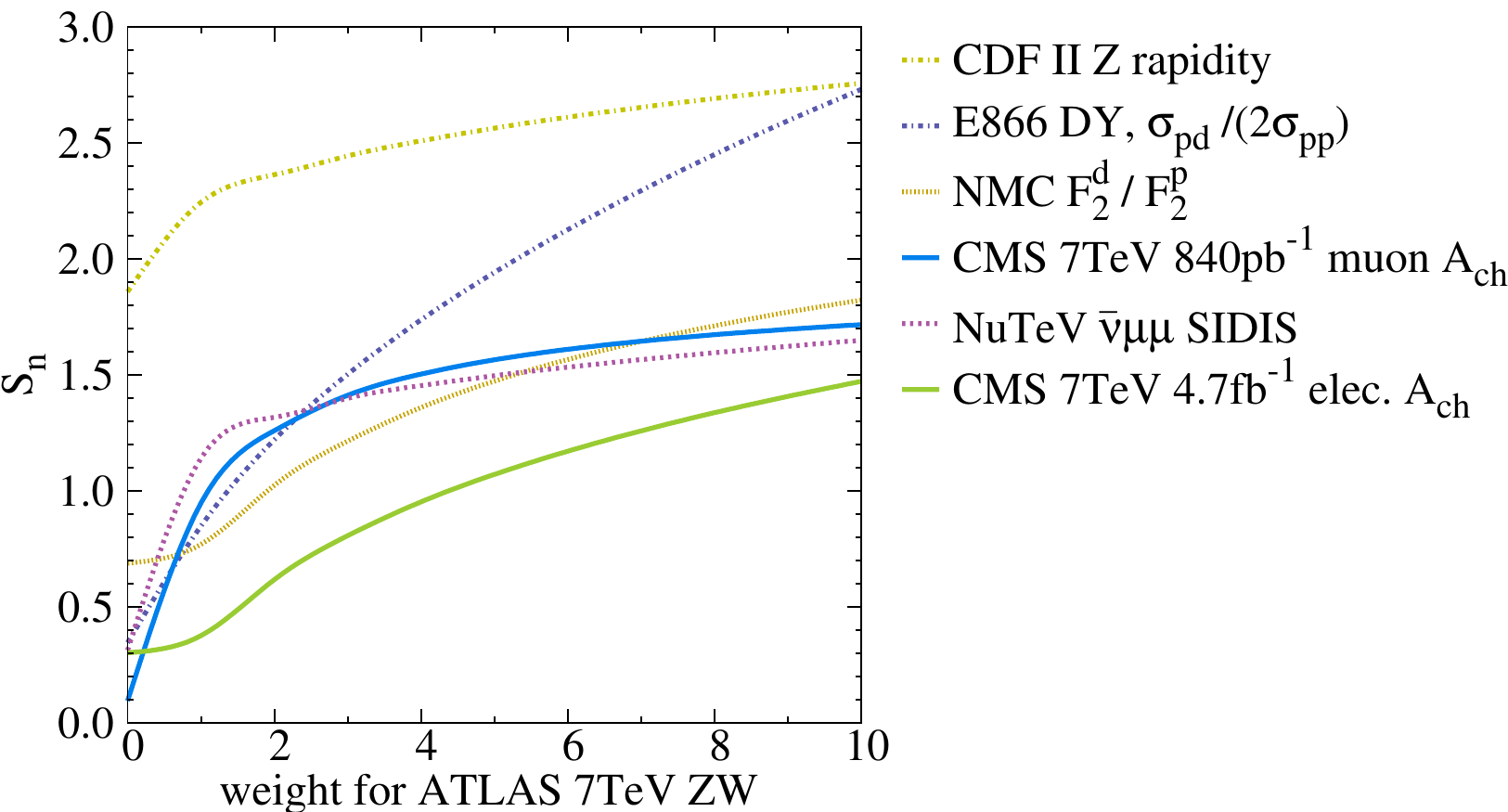}
	\caption{The change of spartyness $S_n$ for some data sets in CT14HERA2 as the weight of ATLAS 7 TeV $WZ$ data is increased from 0 to 10.  Only the data sets with a large  change in $S_n$ are shown.  Note that a weight of zero corresponds to CT14HERA2 fit.}
	\label{Fig: S248w}
\end{figure}

\begin{figure}[h]
	\includegraphics[width=0.45\textwidth]{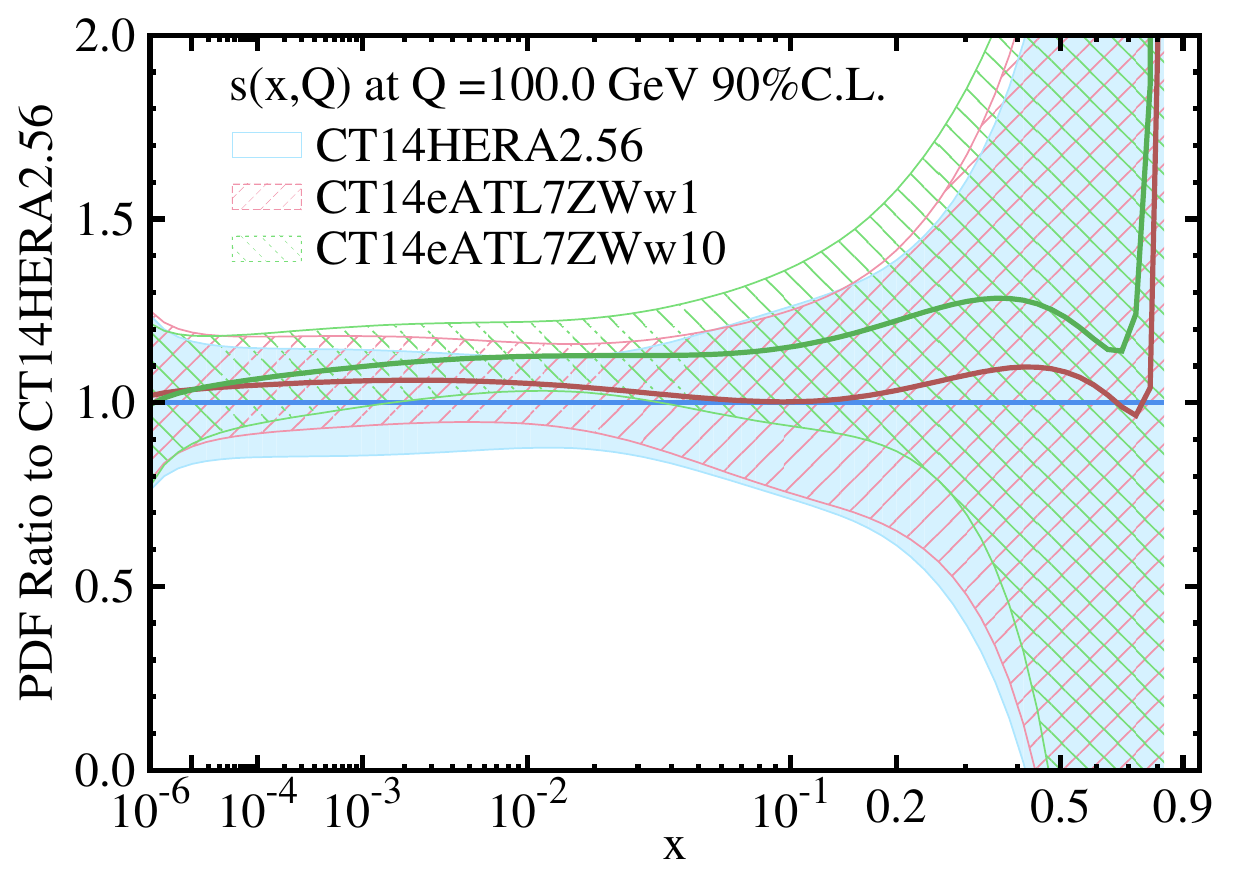}
	\includegraphics[width=0.45\textwidth]{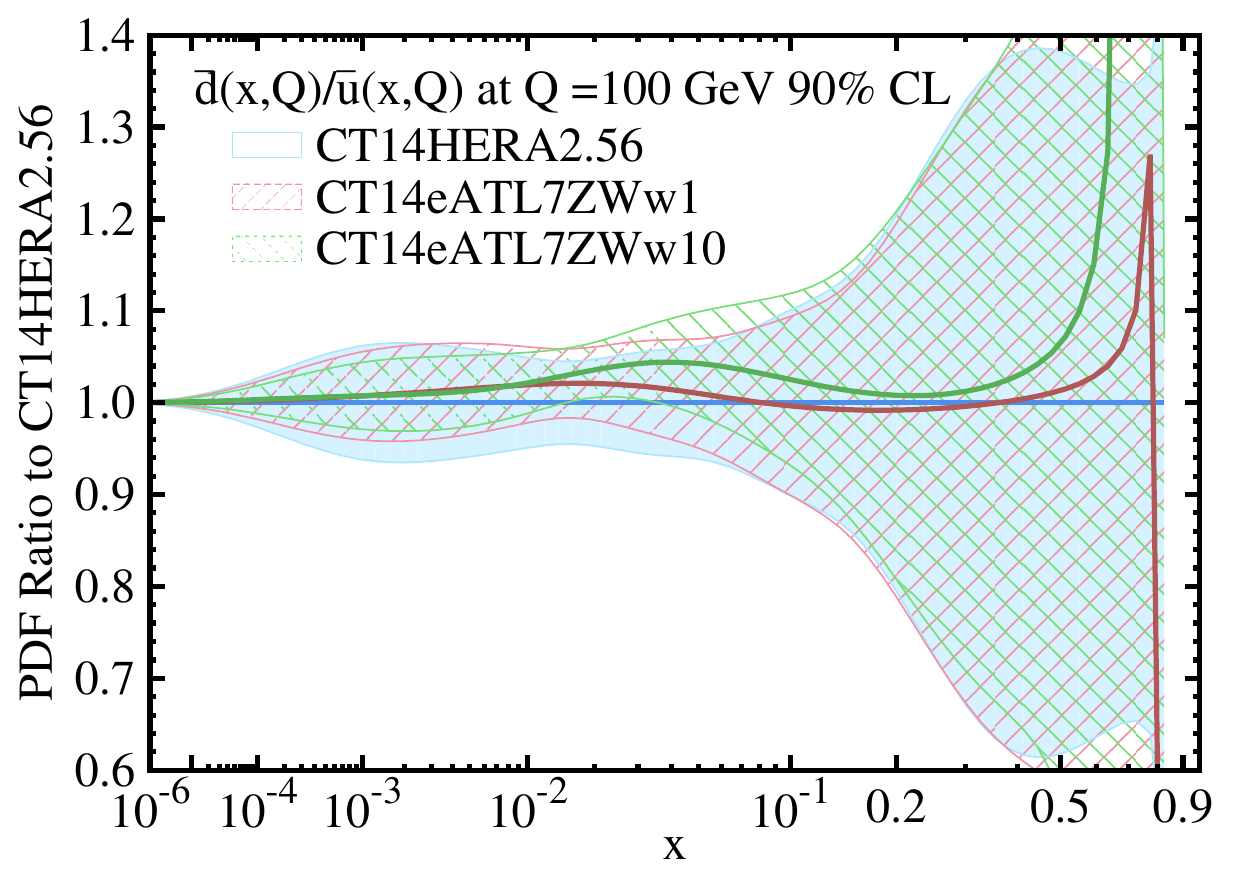}
	\caption{The updated $s$-PDF and the $\bar d/\bar u$ PDF ratio, at $Q=100~{\rm GeV}$, after ATLAS 7 TeV $WZ$ data are added to CT14HERA2 using \texttt{ePump}, with weights of 1 and 10. As in Fig.~\ref{Fig: CT14p248}, the CT14HERA2 error bands are computed using 56 eigen-PDFs.}
	\label{Fig: 248w10dbuba}
\end{figure}

\begin{figure}[h]
	\includegraphics[width=0.45\textwidth]{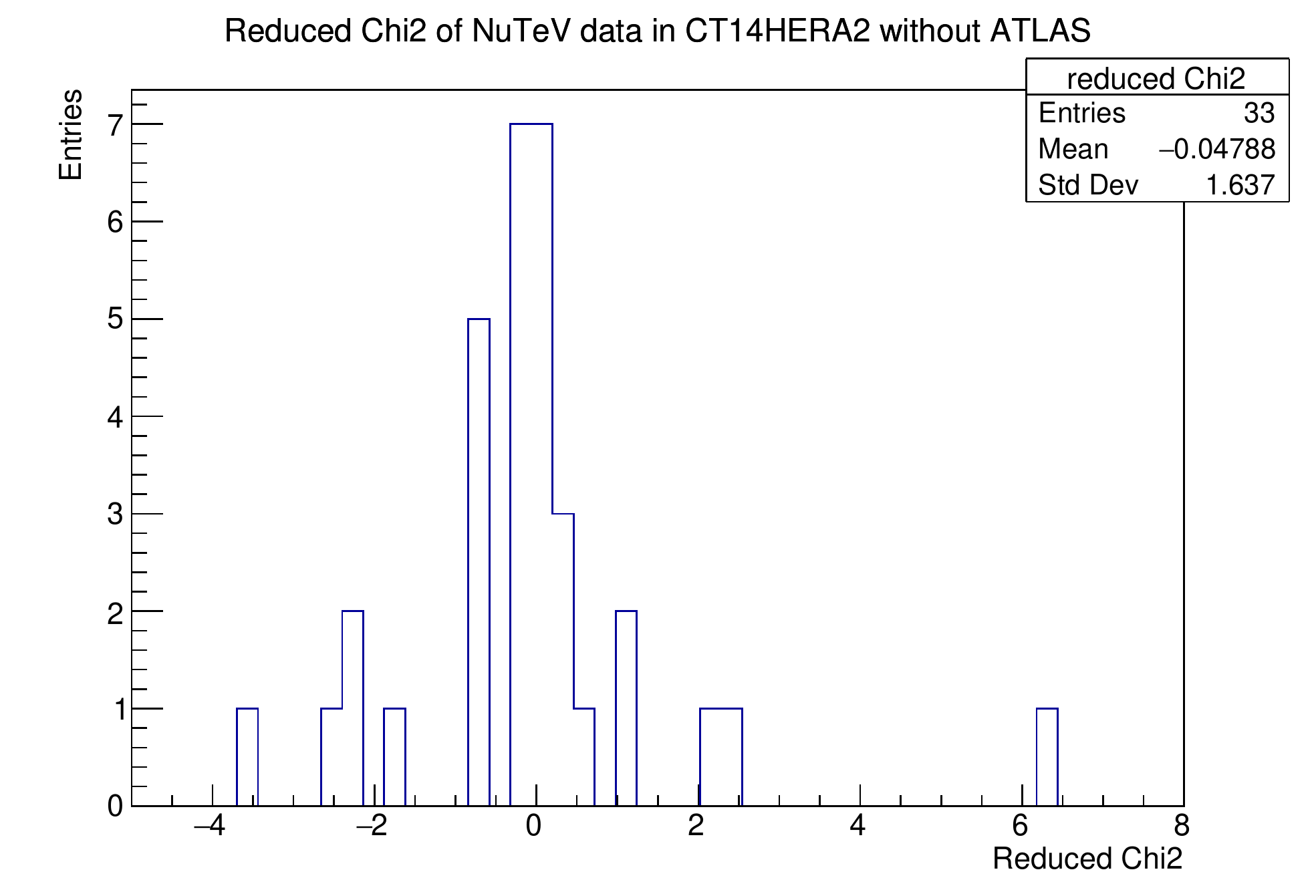}\\
	\includegraphics[width=0.45\textwidth]{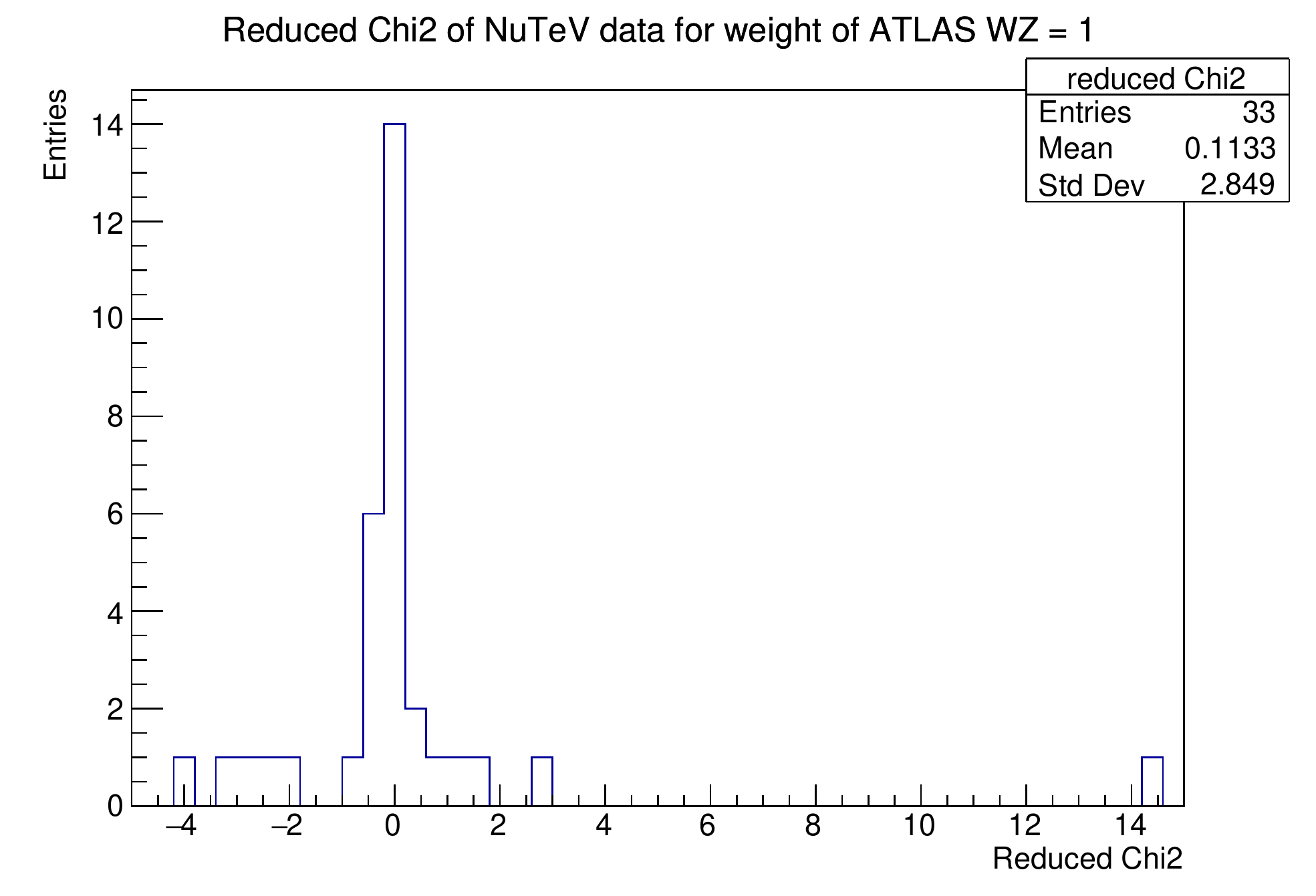}
	\includegraphics[width=0.45\textwidth]{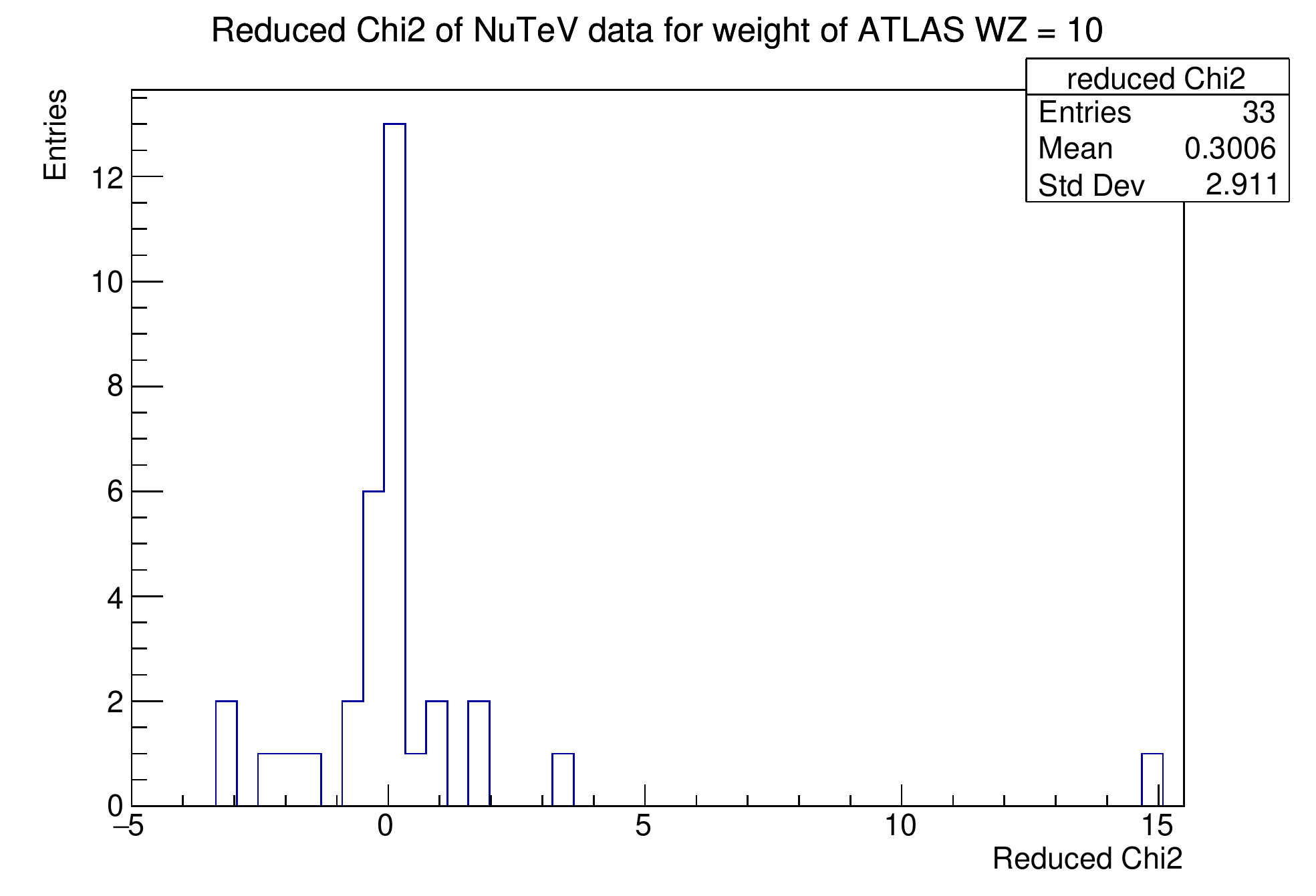}
	\caption{{The distribution of reduced $\chi^2$ for the NuTeV $\bar\nu\mu\mu$ data. The upper one is the reduced $\chi^2$ in CT14HERA2 before \texttt{ePump} updating. The lower two are the distribution of reduced $\chi^2$ when the ATLAS 7TeV $W$ and $Z$ data set is included using \texttt{ePump}, with weight equal to 1 and 10, respectively. We see that with the NuTeV $\bar\nu\mu\mu$ data included, one data point shows dramatically increasing reduced $\chi^2$. In Fig.~\ref{Fig: theory_data_comparison_125 for 248}, it is recognized to be the data point with $x=0.015$ and $y=0.776$. When the weight increases from 1 to 10, their features do not change largely.}}
	\label{Fig: 125chi2w}
\end{figure}

\begin{figure}[h]
	\includegraphics[width=1.0\textwidth]{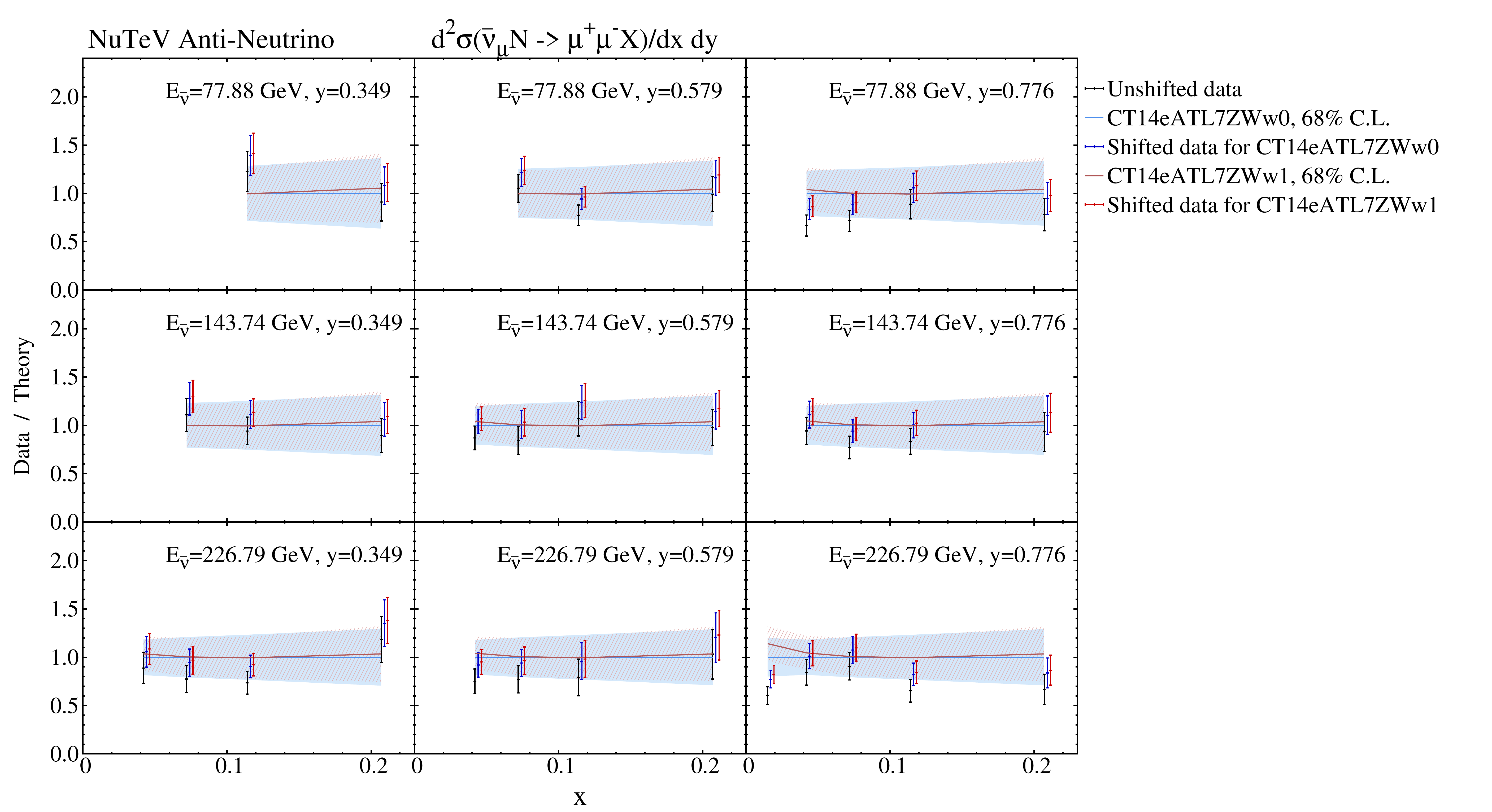}
	\caption{{Comparison between NuTeV $\bar\nu\mu\mu$ data and the theory predictions for each data point, before (labelled as CT14eATL7ZWw0) and after (labelled as CT14eATL7ZWw1) ATLAS 7 TeV $WZ$ data is included to update the CT14HERA2 PDFs. The unshifted data are shown at their original $x$ values and the shifted data are shown at slightly shifted $x$ values for clear comparison. From the bottom to the right figure we can identify the data point with $x=0.015$ and $y=0.776$ as the one possessing tension with the ATLAS 7 TeV $WZ$ data.}}
	\label{Fig: theory_data_comparison_125 for 248}
\end{figure}

Given the above discussion, we might expect some tension between the new ATLAS 7 TeV $W$ and $Z$ data and the old data sets included in the CT14HERA2 fit.
To examine this, we increase the weight of the ATLAS 7 TeV $WZ$ data while updating the CT14HERA2 PDFs using the \texttt{ePump} program.
We can simultaneously obtain the updated predictions for all of the other CT14HERA2 data sets, by including them in \texttt{ePump} as new data, but with zero weight.
In this fashion we can see how the fit to the original data sets change as the the new data is added, in order to investigate for possible tensions.
Increasing the weight of the ATLAS $WZ$ data forces \texttt{ePump} to fit this data better; however, if some of the original CT14HERA2 data sets have tension with the $WZ$ data,
they will be fitted worse as the weight of this $WZ$ data increases.
As discussed in Ref.~\cite{Dulat:2013hea}, the goodness of fit to individual data set can be quantified by the variable ``spartyness'' $S_n$, an equivalent Gaussian variable.
A well-fitted data set should have $S_n$ between $-1$ and 1. An $S_n$ smaller than $-1$ means the data set is fitted ``too'' well and an $S_n$ larger than 1 indicates poor fitting.

We find that most of the data sets in CT14HERA2 do not show appreciable tension with the ATLAS $WZ$ data. However, some data sets do exhibit tension, as shown in
Fig.~\ref{Fig: S248w}, which displays the change of spartyness $S_n$ for these affected data sets as the weight of the ATLAS $WZ$ data is increased from 0 to 10.
Some of these data sets were not well-fitted before (weight=0) and become worse as the weight is increased, e.g., the CDF Run-2 $Z$ rapidity data. Other of these data sets were well-fitted before, but become poorly fitted after the weight is increased, the most significant ones being the
NuTeV $\bar\nu \mu\mu$ SIDIS, the E866 $\sigma_{pd}/(2\sigma_{pp})$ data sets and the CMS 7 TeV $\mu$ and electron asymmetry data.

As discussed in Sec.~\ref{section: DIS }, the $s$-PDF is mainly constrained by the (anti-)neutrino DIS charged current di-muon data, cf. Fig.~\ref{Fig:mDmmus}, and the NuTeV $\bar\nu \mu\mu$ SIDIS data impose the strongest constraint on $s$-PDF among those four data sets.
Fig.~\ref{Fig: 248w10dbuba} shows the \texttt{ePump} updated $s$-PDF with weights of 1 and 10 on the ATLAS $WZ$ data.
We see that the ATLAS $WZ$ data prefer
larger values of the $s$-PDF, while the NuTeV data prefer smaller values at $x$ around a few times $10^{-2}$.
When the weight of the ATLAS 7 TeV $W$ and $Z$ data increases, the reduced $\chi^2$ of the NuTeV data point with $x=0.015$ and $y=0.776$ increases dramatically, as shown in Figs.~\ref{Fig: 125chi2w} and ~\ref{Fig: theory_data_comparison_125 for 248}, which indicates the tension between the ATLAS 7 TeV $WZ$ data and the NuTeV data.

\begin{figure}[h]
\includegraphics[width=0.50\textwidth]{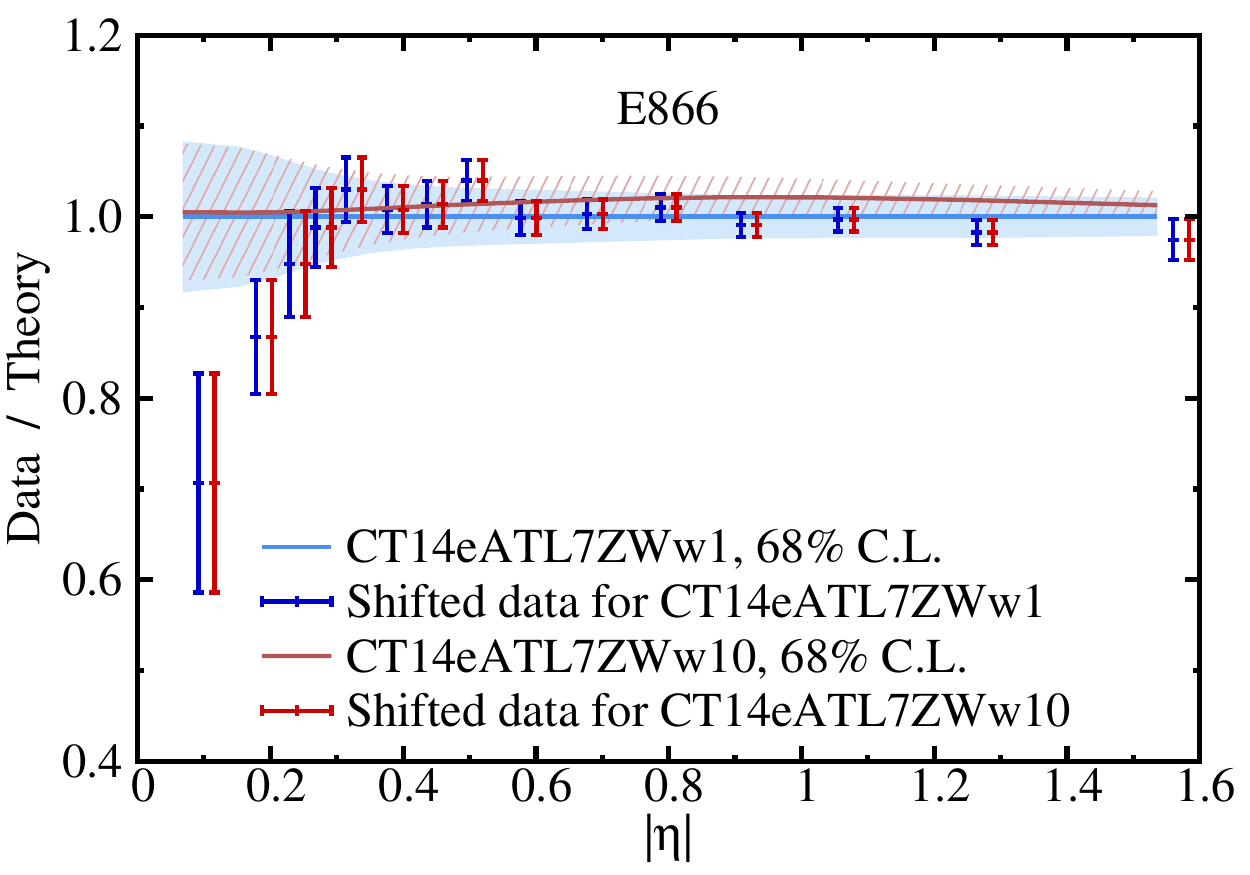}
\caption{Comparison between E866 data and the theory predictions for each data point, as the weight of ATLAS 7 TeV $WZ$ data is increased from 1 to 10.}
\label{Fig: theory_data_comparison_203 for 248}
\end{figure}

In Fig.~\ref{Fig: 248w10dbuba}, we also show the updated $\bar d/\bar u$ PDF ratio plot with weights of 1 and 10 on the ATLAS $WZ$ data.
We find that the ATLAS $WZ$ data prefer a larger value of $\bar d/\bar u$ ratio at $x$ around $10^{-3}$ to $10^{-1}$.
This is to be compared with what we concluded in Sec.~\ref{section:} that E866 $\sigma_{pd}/(2\sigma_{pp})$ data set is crucial for constraining $\bar d/\bar u$ and $d_v/u_v$  at $x$ around $10^{-2}$ to $0.2$, cf. Fig.~\ref{Fig:DY203dbub}. Therefore, increasing the weight of ATLAS 7 TeV $WZ$ data contradicts the fit of E866 data and leads to the tension. This can also be illustrated by the comparison between data and theory before and after ATLAS 7 TeV $WZ$ data set is included, see Fig.~\ref{Fig: theory_data_comparison_203 for 248}, where we find a deviation the theory predictions from the data for large rapidity when the weight of ATLAS 7 TeV $WZ$ data is increased from 1 to 10 .

\begin{figure}[h]
\includegraphics[width=0.45\textwidth]{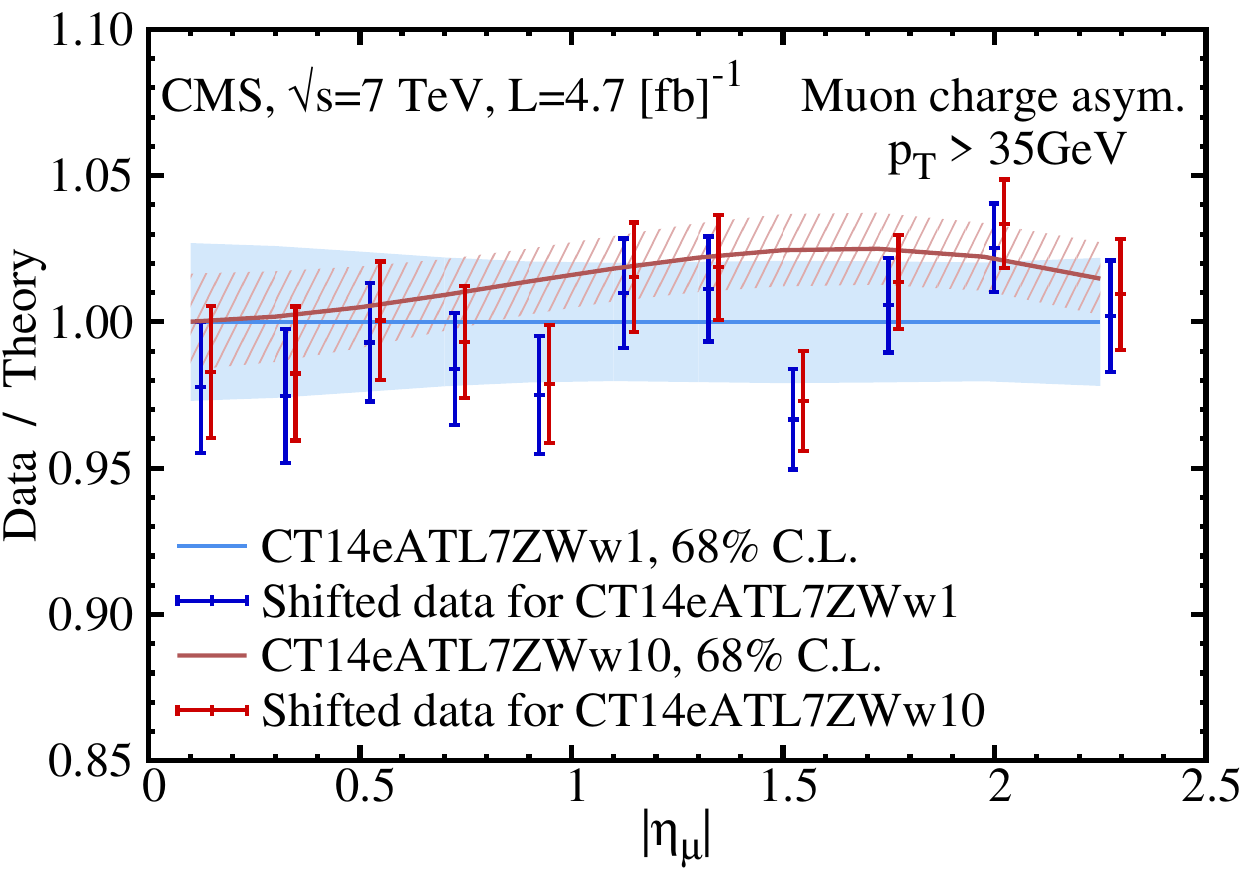}
\includegraphics[width=0.45\textwidth]{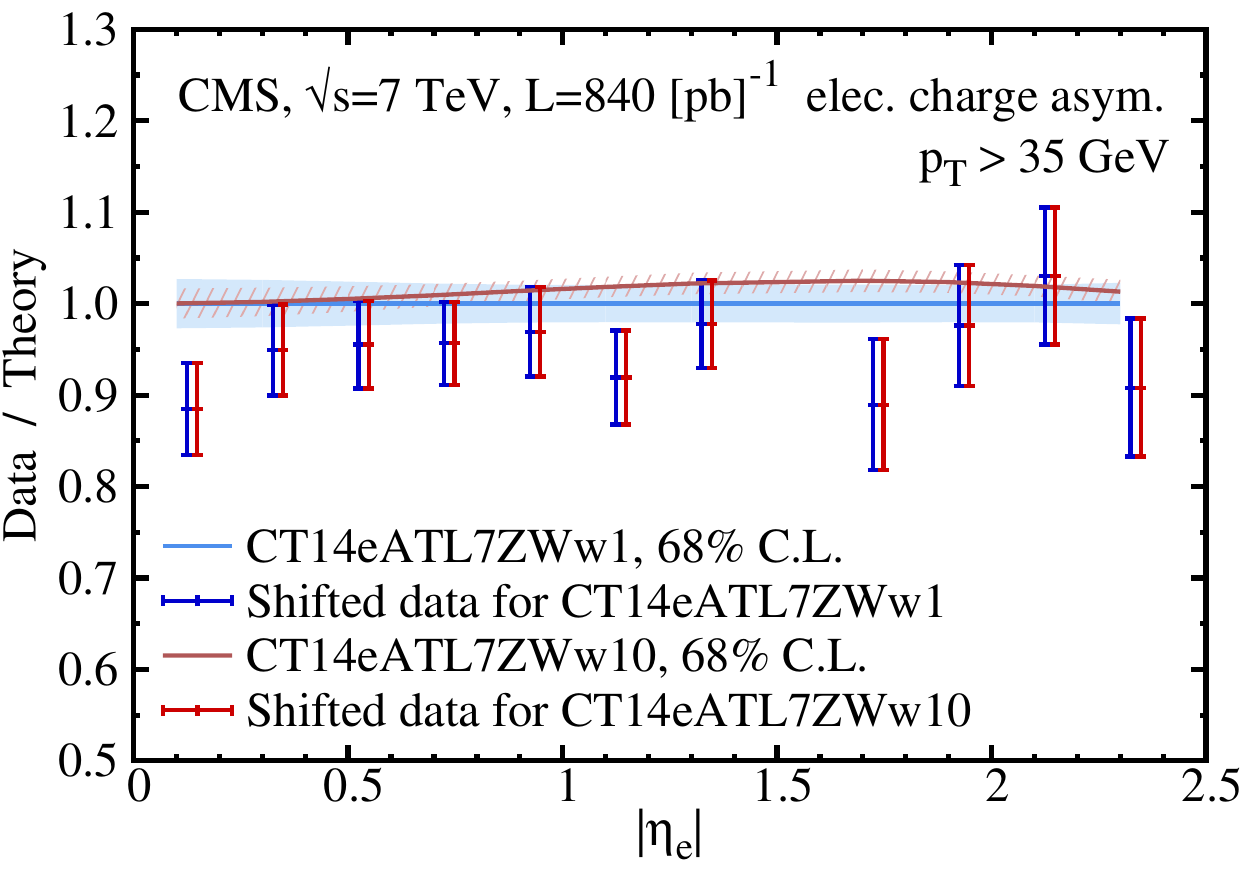}
\caption{Comparison between CMS 7 TeV $\mu$ and electron asymmetry data and the theory predictions for each data point, as the weight of ATLAS 7 TeV $WZ$ data is increased from 1 to 10.}
\label{Fig: theory_data_comparison_266_267 for 248}
\end{figure}

For the CMS asymmetry data, the tension with the ATLAS $WZ$ data can be demonstrated in the same way, by comparing theory with data for each data point, as the weight of ATLAS 7 TeV $WZ$ data is increased. In Figs.~\ref{Fig: theory_data_comparison_266_267 for 248}, the comparison are shown for CMS $\mu$ and electron asymmetry data. It is apparent that as the weight of ATLAS 7 TeV $WZ$ data is increased from 1 to 10, both the theory predictions of CMS $\mu$ and electron asymmetry have an overall upward shift compared to the data for almost all of the data points. Given the precision of the CMS data, this leads to a large $\chi^2$, which is reflected by the rapid increase of spartyness in Fig.~\ref{Fig: S248w}.

\section{Summary and Prospects}\label{section:summary}

A fast and efficient tool for estimating the impact of new data on the PDFs is essential in this high precision era of the LHC.  In this paper we have tested just such a tool, \texttt{ePump},
both as to its effectiveness and its validity.

We have validated \texttt{ePump} in three trials, where we started with a base PDF set obtained from a global fit with some subset of the CT14HERA2 data removed.  We used
\texttt{ePump} to update these base PDFs with the missing data sets, and then compared with the CT14HERA2 PDFs.  In all three trials (updating with DIS, Drell-Yan, or jet data sets)
the \texttt{ePump} results are very close to the CT14HERA2 global-fit results. This is important, because the goal is to have the best approximation to the full global fit as possible.
Of course, there are some differences, but they are either small compared to the error bands, or happen in very small or large $x$ regions, where the PDFs depend strongly on the parameterization forms.  Another case where the \texttt{ePump} approximations break down is when there are strong tensions between the new and old data.
As we have seen, the global fit may increase the error bands, but this will never happen with \texttt{ePump}. Again, we emphasize that \texttt{ePump} is not meant to replace the global fit. However, even in situations with tensions between new and old data, it still gives qualitatively correct results, and therefore provides a useful tool for judging the impact of new data.
In addition to updating PDFs, \texttt{ePump} can also update the observables at the same time without the need for recalculating.  An example of this use of \texttt{ePump} was given for the predictions of $\sigma(gg\to h)$.

A big advantage of \texttt{ePump} is that it can run very fast. This was exploited to study the impact of different data sets in the CT14HERA2 fit.  Summaries of the impact of each of the data sets were given in Tables.~\ref{tab:EXP_1}, \ref{tab:EXP_2}, and \ref{tab:EXP_3}.  The impact of each data set is strongest for some particular flavors and for its relevant region of $x$.  But it also depends on the precision of the data and its agreement with the current PDFs.  Therefore, even for two data sets that
that are sensitive to the same kinematic range
and flavor content, they do not necessarily have the same effect on the PDFs.   One remarkable thing we found is that among the 33 data sets in CT14HERA2, only 1 jet, 5 Drell-Yan and 8 DIS data sets\footnote{This includes HERA I+II data as well.}
have the dominant effects.   Just by including these data sets, we can reproduce the bulk part of the CT14HERA2 fit. The other data sets are only responsible for some fine structures of the PDFs.

It is incredible that we can fit thousands of data points with only 27 or 28 parameters. This triumph strongly shows the effectiveness of the QCD improved parton model. In such an era of precision, it has become an important and indispensable task to reduce the uncertainties of PDFs. So the natural question is: ``What kinds of observables can reduce the PDF uncertainties?'' The purpose of \texttt{ePump} is to help answer this question. The ``new" data to be investigated by \texttt{ePump} can be new experimental data, or it can be simulated pseudo-data, whose impact one might be interested to see. An example of this second scenario was presented in Ref.~\cite{Willis:2018yln}, where \texttt{ePump} was used to show that with increased precision and optimal choice of kinematic variables, the high-invariant-mass Drell-Yan processes can greatly reduce the PDF uncertainties. In this paper, we examined the
impact of the latest $\ttb$ data and $W$ and $Z$ data at the LHC on the CT14HERA2 PDFs.  We found that the $\ttb$ data has the potential to reduce $g$-PDF uncertainties given increased luminosity and that the high-precision $W$ and $Z$ data also can provide strong constraints on the quark PDFs. Of course, these results will be refined quantitatively by
a full global fit, but \texttt{ePump} can quickly assess the qualitative . Similar studies can also be done for other processes. We expect \texttt{ePump} to play an important role in the study of PDFs, and to assist in the understanding and reduction of theoretical errors in the current era of the high-luminosity LHC. The complete {\tt ePump} package, together with detailed instructions for installing and file formatting, and additional output files relevant to this study, can be found at the website \url{http://hep.pa.msu.edu/epump/}.

Before closing this section, we would like to make two additional remarks. First, we note that the comparison made in this paper between the ePump and global fit analyses hold at any given fixed order, either next-to leading order or next-to-next-to leading order (NNLO), of theory calculations. Second, we note that the default  xFitter profiling analysis~\cite{Camarda:2015zba} can be reproduced by \texttt{ePump}, but using a global tolerance set to 1. However, as discussed in the Appendix, setting tolerance to be 1 will greatly overestimate the impact of a given new data set when updating the existing PDFs in the CT PDF global analysis framework.

\begin{acknowledgments}
We thank our CTEQ-TEA colleagues for support and discussions.
This work was supported by the U.S. National Science Foundation under Grant No. PHY-1719914.
C.-P. Yuan is also grateful for the support from the Wu-Ki Tung endowed chair in particle physics.
The work of S.~Dulat was supported by the National Natural Science Foundation of China under the Grant No. 11965020.	
\end{acknowledgments}

\appendix
\section{The role of the tolerance $T$ in updating of PDFs}

When constructing Hessian eigenvector PDFs for error estimation, different PDF groups have made different choices for defining the PDF errors.
This includes the choice of 68\% versus 90\% confidence levels (C.L.), the inclusion of a global tolerance $T$, or the imposition of dynamical tolerances $T^\pm_i$ along different eigenvector directions.  This has caused some confusion when the eigenvector PDFs are then used in the updating of PDFs with new data.  We shall clarify this issue here in the context of  \texttt{ePump}.

The basic premise of the Hessian approximation is that, near its global minimum, $\chi^2$ can be written as a quadratic function of the PDF parameters  $\{z_i; i=1,N\}$. For CT14HERA2 fit, $N=28$ or 27, depending on whether the gluon extreme sets are included or not. By shifting, rotating, and rescaling the PDF parameters, we can write $\Delta\chi^2$ for the original data sets as
\beq
\Delta\chi^2_{\rm old}=T^2 \sum_{i=1}^N z_i^2\,.\label{app:chiold}
\eeq
The parameters $z_i$ have been chosen here so that $z_i=0$ (for all $i$) for the best-fit PDF set $f^0$, while $z_i=\pm\delta_{ij}$ for the $2N$ eigenvector PDFs $f^{\pm j}$.  In this way, each of the eigenvector PDF sets correspond to a $\Delta\chi^2_\mathrm{old}=T^2$, where $T$ is an overall global tolerance parameter.  If the various data sets were all internally consistent and satisfied Gaussian statistics, then one should use $T=1$ at the 68\% C.L. or $T=1.645$ at the 90\%
C.L.  However, due to inconsistencies between the various data sets, as well as uncertainties arising from the initial choice of PDF parametrization forms, the CTEQ-TEA group has historically chosen a larger value of $T=10$ at the 90\% C.L.

Another variation in defining the Hessian PDF errors is in the imposition of dynamical tolerances, which have been used in recent CT~\cite{Dulat:2015mca,Hou:2016nqm} and MMHT~\cite{Harland-Lang:2014zoa} PDF sets.  The idea here is that if the constraints on a given PDF eigenvector direction comes dominantly from a single data set (or several self-consistent sets), then the overall global tolerance produces too large of a PDF error.  By incorporating separate constraints from individual experiments one determines that a $\Delta\chi^2_\mathrm{old}=(T^\pm_i)^2<T^2$ in the particular eigenvector direction should correspond to the given C.L.  Keeping Eq.~(\ref{app:chiold}) unchanged, this implies that the Hessian eigenvector PDFs now must correspond to
$z_i=\pm(T^\pm_j/T)\delta_{ij}$. (It can easily be seen that in the presence of dynamical tolerances, the global parameter $T$ scales out of all calculated observables.)

We emphasize here that, although the different choices of tolerance parameters (including whether global or dynamical), as well as the choice of 68\% or 90\% C.L, give different results and/or interpretations for the PDF errors, they all give a self-consistent description of $\chi^2_\mathrm{old}$ around the global minimum.

The next step in updating the PDFs in the Hessian approach is to add the contribution of the new data set (or sets) to $\chi^2$, yielding
\beq
\chi^2_{\rm new}=T^2 \sum_{i=1}^N z_i^2+w\sum_{\alpha,\beta=1}^{N_X}\left(X_{\alpha}^E-X_{\alpha}(\mathbf{z})\right) C^{-1}_{\alpha\beta}\left(X_{\beta}^E-X_{\beta}(\mathbf{z})\right)\,,
\label{total Chi2}
\eeq
where $N_X$ is the number of new data points, $X_\alpha^E$ are the experimental data values, $X_\alpha({\bf z})$ are the theoretical predictions, and $C^{-1}_{\alpha\beta}$ is the experimental inverse covariance matrix.  We have also included a weight factor $w$ that may be assigned to the new set of data, which by default is set to be 1.  To linear order in the PDF parameters (assuming that a global tolerance is used), we can express the theoretical predictions as
\beq
X_{\alpha}(\mathbf{z})=X_{\alpha}(0)+\sum_{i=1}^N \left(\frac{X^{+i}_{\alpha}-X^{-i}_{\alpha}}{2}\right) z_i\,,\label{X derivative}
\eeq
where $X^{\pm i}_{\alpha}$ is the theoretical prediction of $X_{\alpha}$ calculated with the error PDFs $f^{\pm i}$.
We emphasize that Eq.~(\ref{X derivative}) is valid only when $z_i=\pm \delta_{ij}$ corresponds to the error PDFs $f^{\pm j}$.
At this stage, $\chi^2_\mathrm{new}$ can now be minimized to obtain the new value of the best-fit parameters, which can be used to obtain updated best-fit PDFs.
The generalization of Eq.~(\ref{X derivative}) for dynamical tolerances, as well as the extension of this equation to include diagonal quadratic terms
are given  in Ref.~\cite{Schmidt:2018hvu}, and are implemented in \texttt{ePump}. In addition, \texttt{ePump} produces an updated set of error PDFs, under the same tolerance and confidence level assumptions of the original error PDFs.

Although these results were given previously in Ref.~\cite{Schmidt:2018hvu}, our purpose for restating them here is to make it clear that the choice of the global tolerance $T$ or the use of dynamical tolerances in the updating of the PDFs should not be chosen freely, but rather is determined by the original Hessian error PDF sets.  In particular, if the error PDFs were determined using a global tolerance of $T=10$, then the use of Eq.~(\ref{X derivative}) is only consistent
if the value $T=10$ is used in Eq.~(\ref{total Chi2}).  It is straightforward to show that if the tolerance were set to $T=1$ in Eq.~(\ref{total Chi2}) while using such error PDFs, it is equivalent to weighting the new data by a factor $w=100$ in the update.

We exemplify this by performing the following exercise.
 As demonstrated in Section.~\ref{section: DIS }, the dimuon data are almost entirely responsible for constraining the $s$-PDF. Therefore, we remove the dimuon data from the CT14HERA2 data sets and perform a new global fit, named CT14HERA2mDimu. Then we use \texttt{ePump} to add back the dimuon data. If everything works perfectly, we should expect that the  \texttt{ePump}-updated $s$-PDF will agree with CT14HERA2. 
Also, based on the present discussion, we should perform the update using the dynamical tolerances, but we shall also try other choices for the tolerance to see the results.
 
\begin{figure}[h]
	\includegraphics[width=0.45\textwidth]{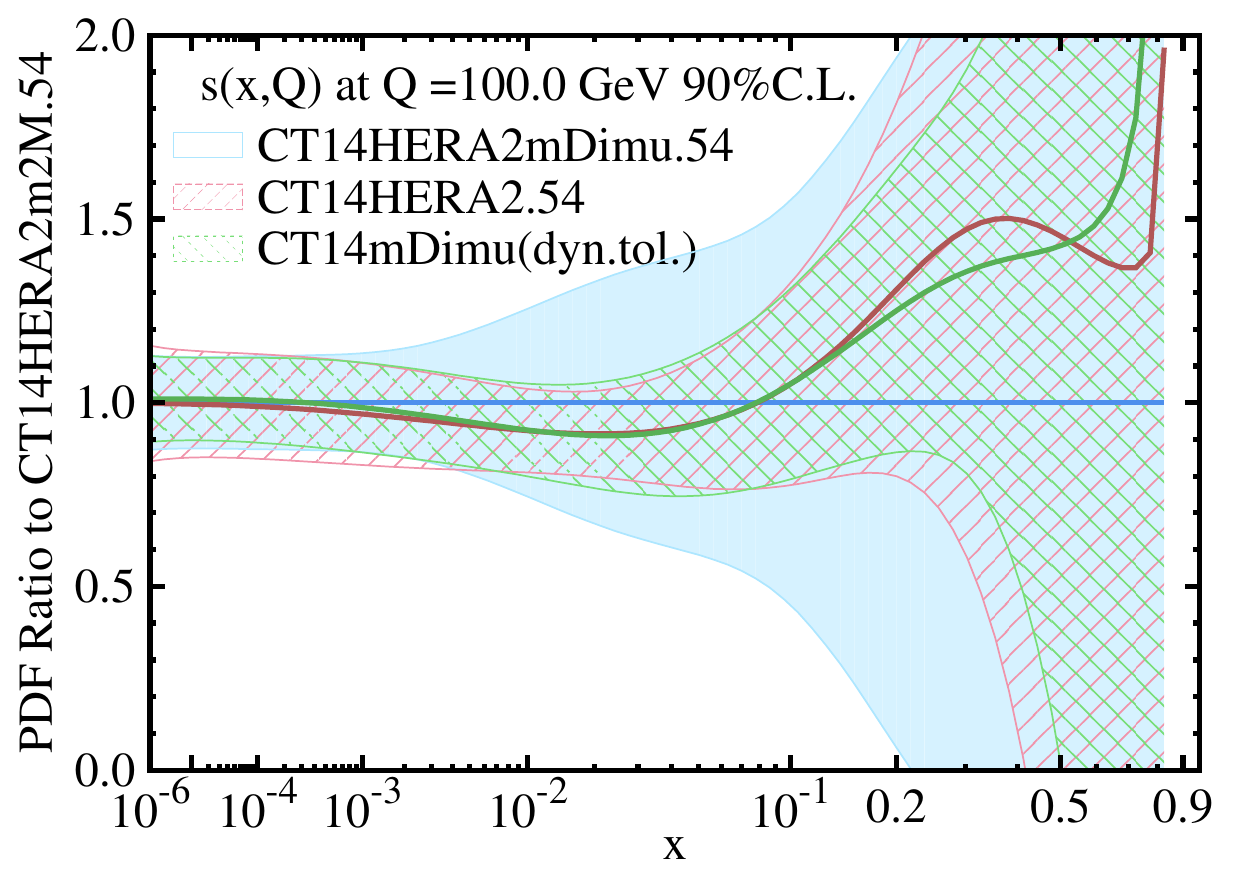}
	\includegraphics[width=0.45\textwidth]{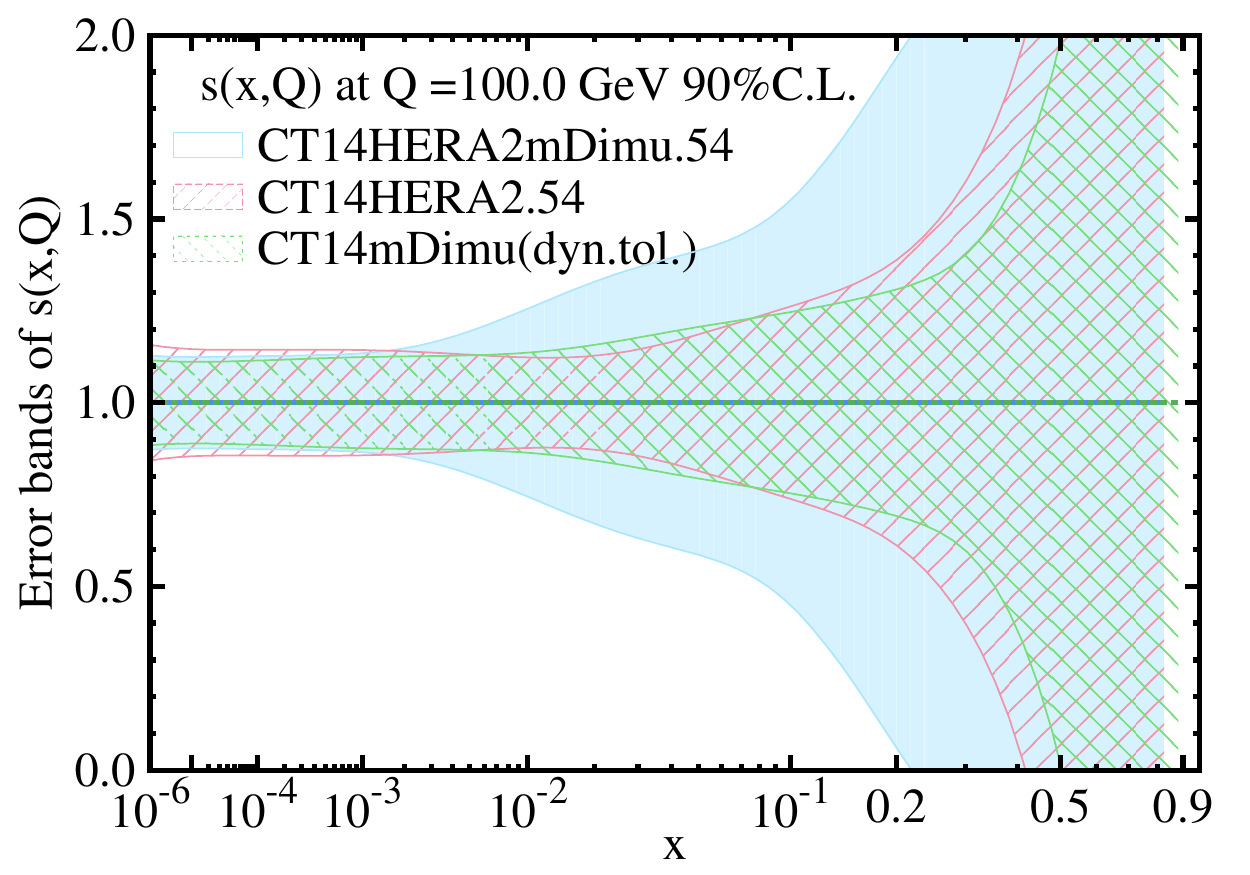}\\
	\includegraphics[width=0.45\textwidth]{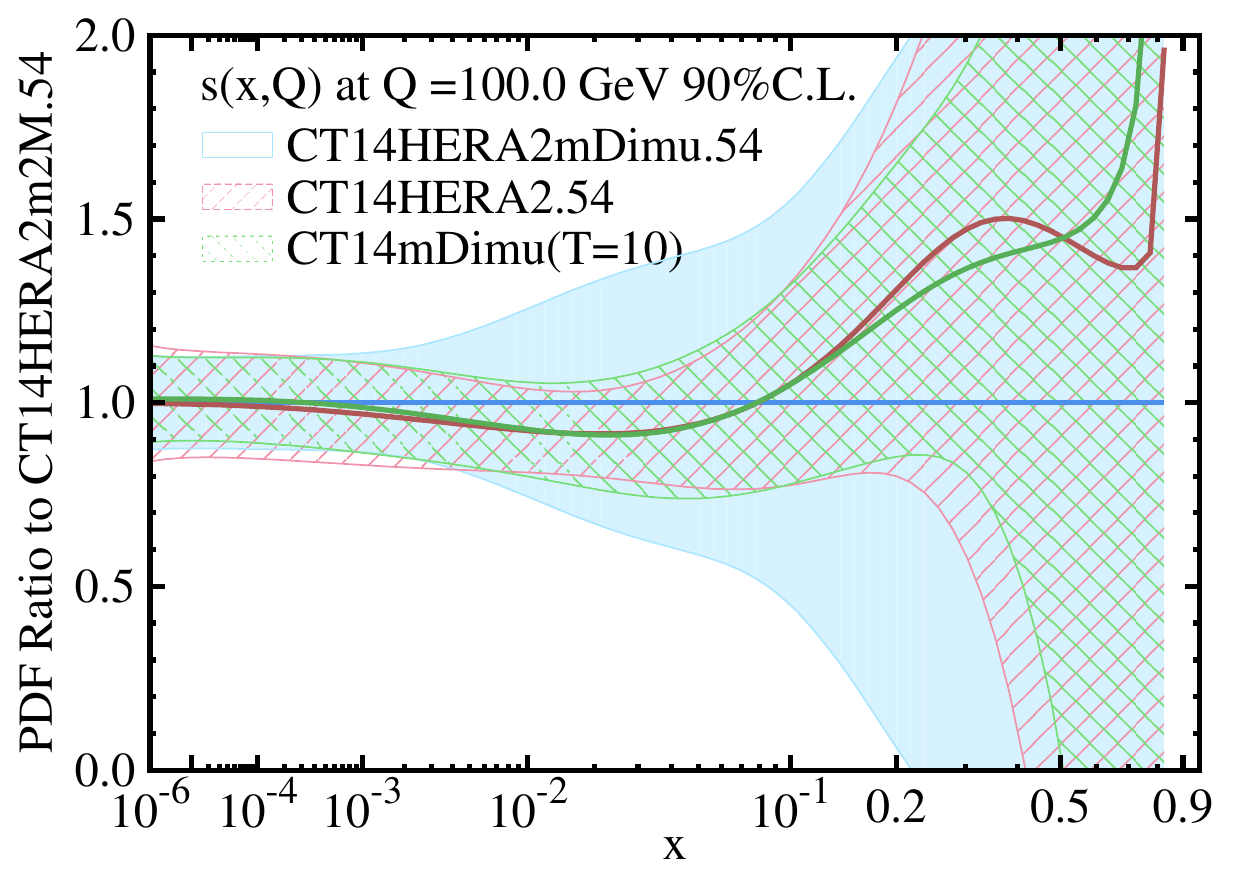}
	\includegraphics[width=0.45\textwidth]{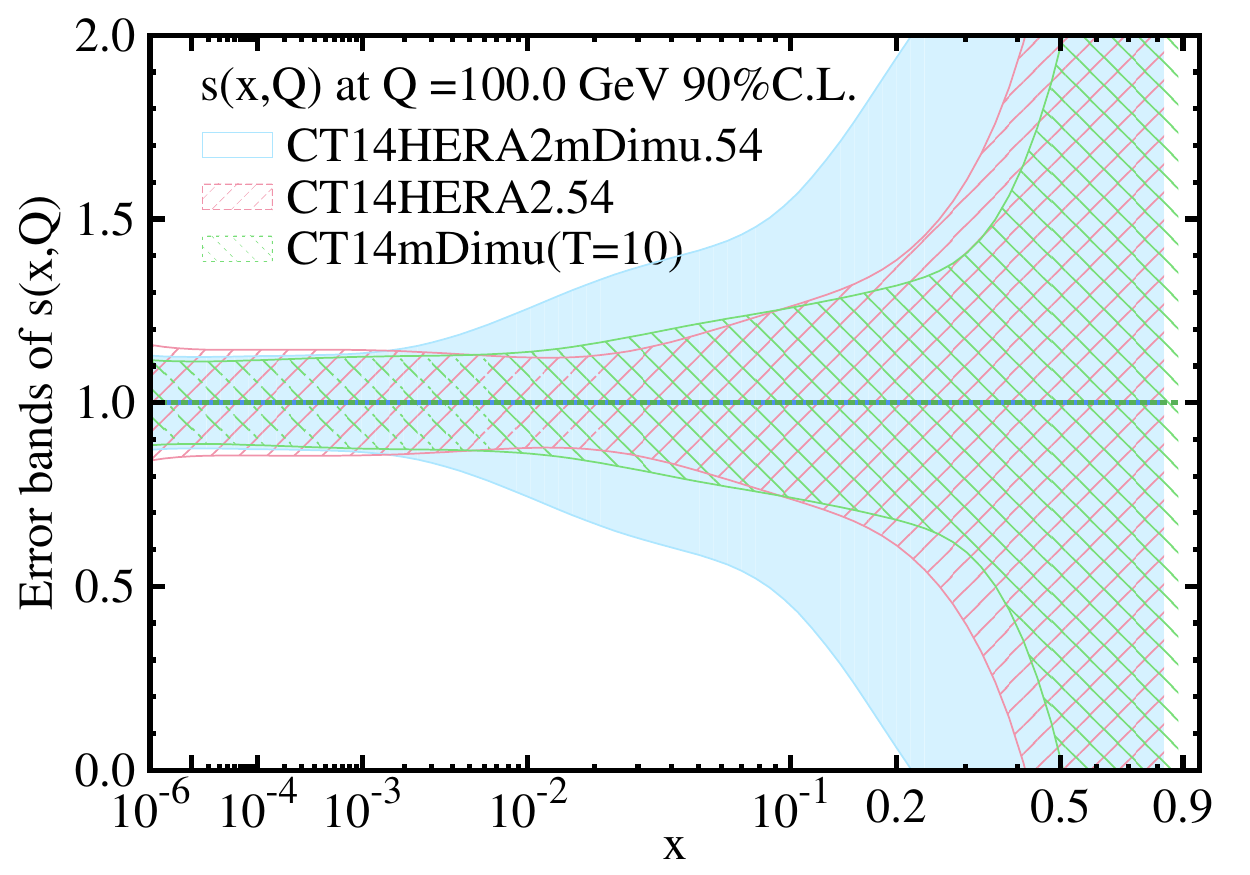}
	\caption{
		Comparison of \texttt{ePump}-updated $s$ PDFs with the global-fit result, at $Q=100~\rm{GeV}$. In the upper two plots, dynamical tolerance was used (labelled as CT14mDimu(dyn.tol.)). In the lower two plots, dynamical tolerance was turned off and $T^2=100$ was assigned (labeled as CT14mDimu(T=10)). These two results are very similar and both reproduce CT14HERA2 with good agreement. Left panel: the PDF ratios over the best-fit of the base CT14HERA2mD. Right panel: the error bands relative to their own best-fit.}
	\label{Fig:CT14m2DeDimuon}
\end{figure}

 We first compare the update using dynamical tolerances with that using a global tolerance of $T^2=100$. The results are shown in Fig.~\ref{Fig:CT14m2DeDimuon}. In this case we find that these two updates give very similar predictions and reproduce CT14HERA2 very well. 
On checking the dynamical tolerance values for CT14HERA2mDimu, we discovered that the eigenvector directions that are most sensitive
to the dimuon data have $(T^\pm_i)^2\simeq 90$, which explains why the update is not much different when using a global $T^2=100$.

\begin{figure}[h]
	\includegraphics[width=0.45\textwidth]{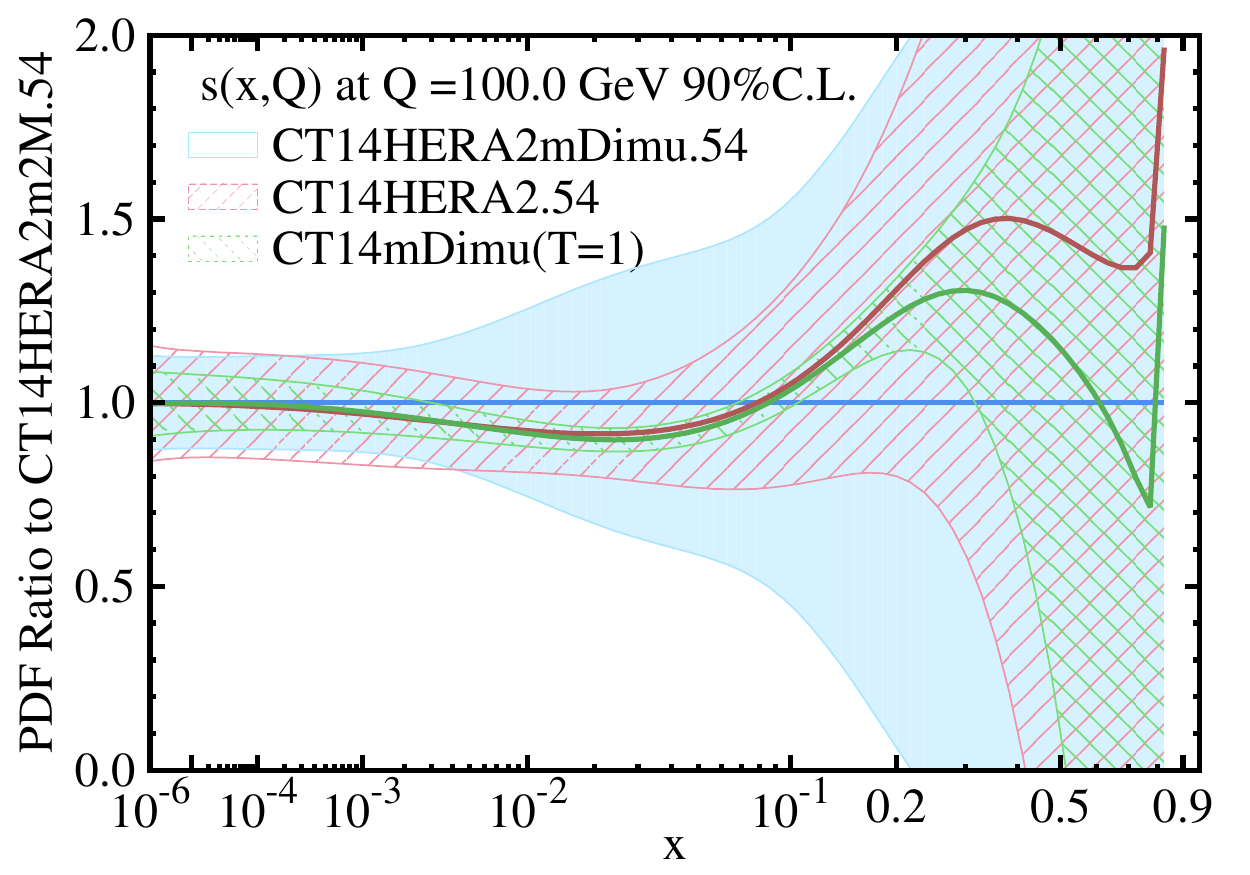}
	\includegraphics[width=0.45\textwidth]{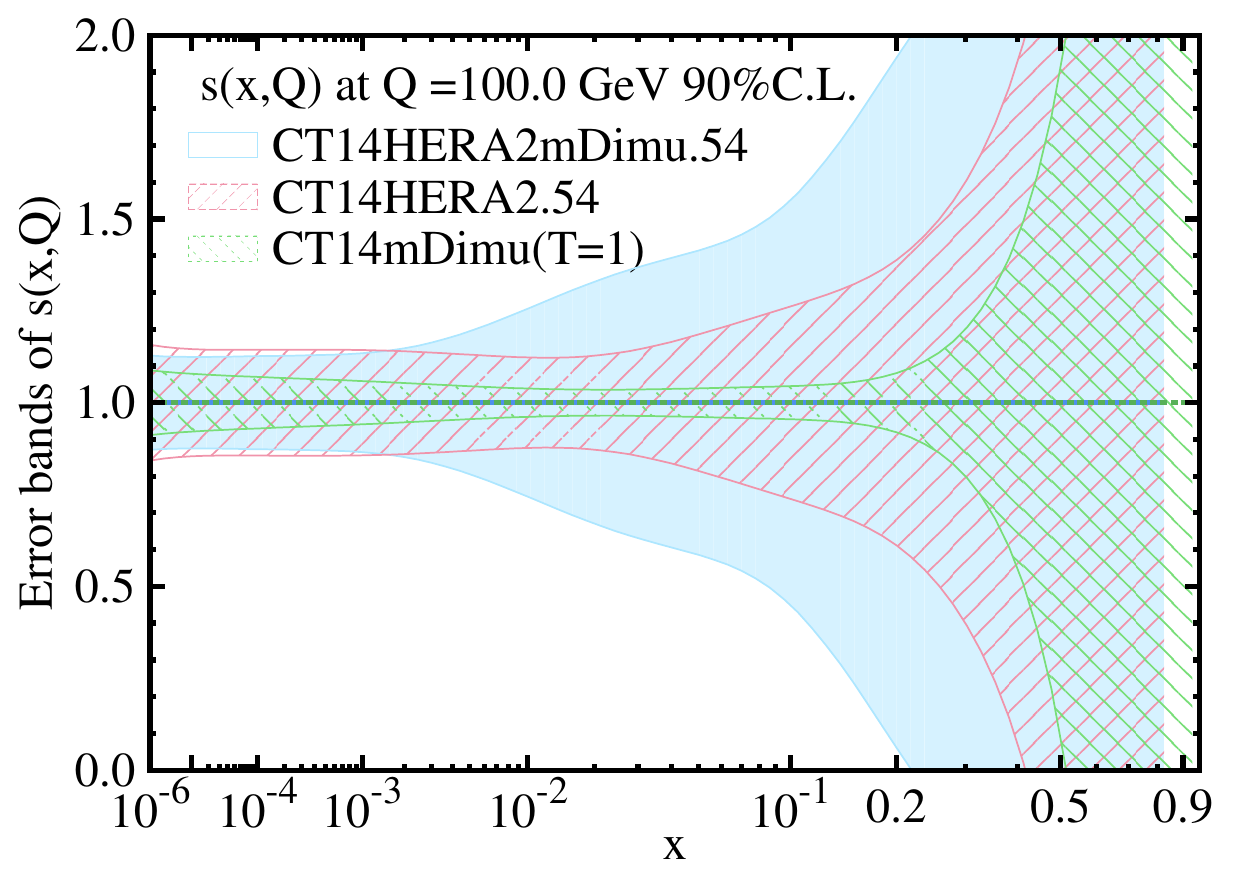}\\
	\includegraphics[width=0.45\textwidth]{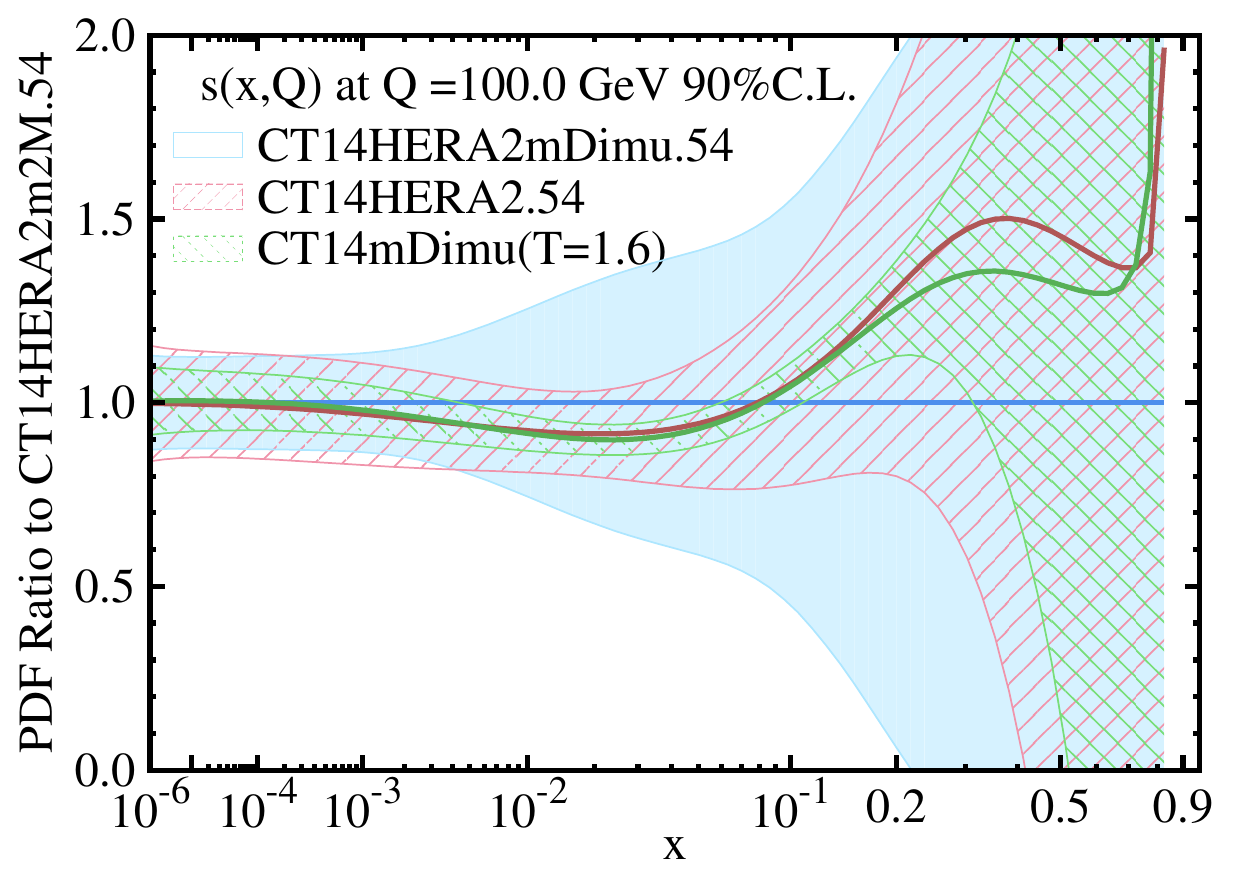}
	\includegraphics[width=0.45\textwidth]{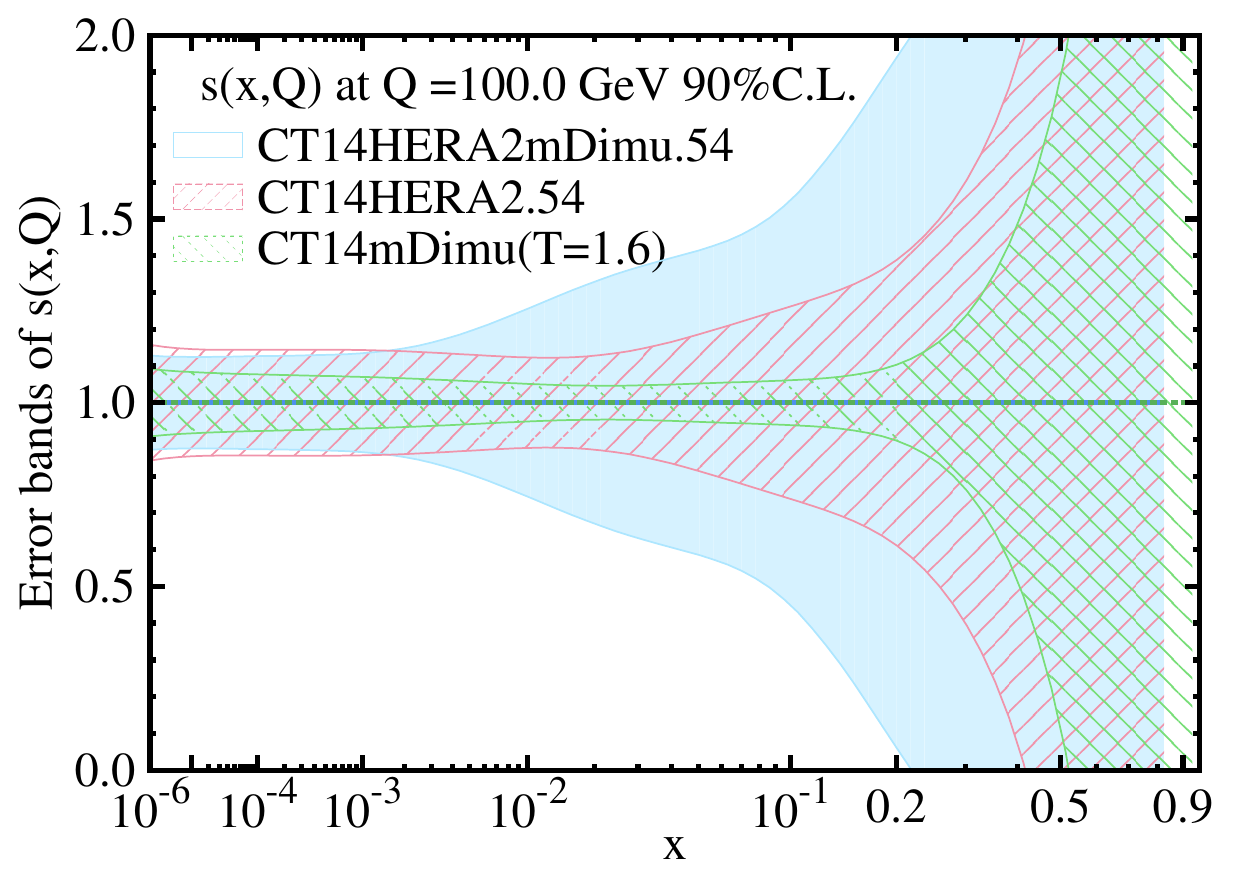}
	\caption{
		Comparison of \texttt{ePump}-updated $s$ PDFs with the global-fit result, at $Q=100~\rm{GeV}$. In the upper two plots, $T^2=1$ was used (labeled as CT14mDimu(T=1)), while in the lower two plots,  and $T^2=1.645^2=2.706$ was assigned (labeled as CT14mDimu(T=1)). Dynamical tolerance was turned off in both two cases. These two results cannot reproduce CT14HERA2, giving too small error bands. Left panel: the PDF ratios over the best-fit of the base CT14HERA2m2M. Right panel: the error bands relative to their own best-fit.}
	\label{Fig:CT14m2DeDimuonT}
\end{figure}

Next we display the results using global tolerance-squares of  $T^2=1$ and $T^2=1.645^2=2.706$. These correspond to the naive use of Gaussian statistics at the 68\% and 90\% C.L.'s, respectively, but are inconsistent with the error PDFs used in this update.
 As explained above, using these small values of $T^2$ with the given error PDFs is equivalent to overweighting this data by a large value. 
The  $s$-PDFs after these updates are shown in Fig.~\ref{Fig:CT14m2DeDimuonT}.  Interestingly, the agreement of the best-fit updates are not too bad when compared with CT14HERA2 (though not as good as when using dynamical tolerances).  This can be understood by the fact that the strange quark PDF is mostly determined by the dimuon data, so overweighting this data does not shift the central value by much.  However, one can see that the $s$-PDF error bands for these two values of the tolerance are much smaller than that of CT14HERA2, so that using $T^2=1$ or $T^2=2.706$ would greatly overestimate the effect of this data on reducing the PDF errors.

In conclusion, in order to best reproduce CT14HERA2 global fit, one should use dynamical tolerance in \texttt{ePump}. Setting tolerance to be 1 will greatly overestimate the impact of a given new data set when updating the existing PDFs in the CT PDF global analysis framework. This conclusion also holds for using MMHT2014~\cite{Harland-Lang:2014zoa} and PDF4LHC15~\cite{Butterworth:2015oua} PDFs in profiling analysis to study the impact of  a new (pseudo-) data on updating the existing PDFs.

\end{document}